\documentclass[twocolumn,superscriptaddress]{revtex4}

\usepackage{graphicx}
\usepackage[T1]{fontenc}
\usepackage{amsmath}
\usepackage{epsfig} 
\usepackage{ulem}
\usepackage{amssymb}
\usepackage[mathscr]{eucal}
\usepackage{hyperref}
\usepackage{color}
\usepackage{bbold}

\begin{document}

\title{From quantum to classical modelling of radiation reaction: \\a focus on stochasticity effects}
\author{F. Niel}\email{fabien.niel@polytechnique.edu}
\author{C. Riconda}                
\author{F. Amiranoff}
\affiliation{LULI, UPMC Universit\'e Paris 06: Sorbonne Universit\'es, CNRS, \'Ecole Polytechnique, CEA, Universit\'e Paris-Saclay, F-75252 Paris cedex 05, France}
\author{R. Duclous}
\affiliation{CEA, DAM, DIF, F-91297 Arpajon, France}
\author{M. Grech}\email{mickael.grech@polytechnique.edu}
\affiliation{LULI, CNRS, \'Ecole Polytechnique, CEA,  Universit\'e Paris-Saclay, UPMC Universit\'e Paris 06: Sorbonne Universit\'es, F-91128 Palaiseau cedex, France}

\begin{abstract}

Radiation reaction in the interaction of ultra-relativistic electrons with a strong external electromagnetic field is investigated using a kinetic approach 
in the non-linear moderately quantum regime.
Three complementary descriptions are discussed considering arbitrary geometries of interaction: 
a deterministic one relying on the quantum-corrected radiation reaction force in the Landau and Lifschitz (LL) form,  
a linear Boltzmann equation for the electron distribution function,
and a Fokker-Planck (FP) expansion in the limit where the emitted photon energies are small with respect to that of the emitting electrons.  
The latter description is equivalent to a stochastic differential equation where the effect of radiation reaction appears in the form of the deterministic term corresponding to the quantum-corrected LL friction force, and by a diffusion term accounting for the stochastic nature of photon emission.

By studying the evolution of the energy moments of the electron distribution function with the three models, 
we were able to show that all three descriptions provide similar predictions on the temporal evolution of the average energy of an electron population in various physical situations of interest, 
even for large values of the quantum parameter $\chi$. 
The FP and full linear Boltzmann descriptions also allow to correctly describe the evolution of the energy variance (second order moment) of the distribution function, while higher order moments 
are in general correctly captured the full linear Boltzmann description only.
A general criterion for the limit of validity of each description is proposed, as well as a numerical scheme for the inclusion of the FP description in Particle-In-Cell codes.
This work, not limited to the configuration of a mono-energetic electron beam colliding with a laser pulse, 
allows further insight into the relative importance of various effects of radiation reaction and in particular of 
the discrete and stochastic nature of high-energy photon emission and its back-reaction in the deformation of the particle distribution function. 

\end{abstract}

\maketitle

\section{Introduction}\label{sec:intro}

Over the last years, high-energy photon emission by ultra-relativistic particles and its back-reaction on the particle dynamics, also known as
radiation reaction, has received a large interest from the strong-field physics, laser-plasma interaction and astrophysics communities. 

The interest of the laser-plasma community is driven by the development of multi-petawatt laser facilities 
such as ELI~\cite{ELI} or APOLLON~\cite{Apollon}. Within the next decade, these laser systems will deliver light pulses with a peak power up to 10 PW 
and durations in the femtosecond regime, thus allowing to reach on-target intensities close to $10^{23}~\rm{W/cm^2}$. 
This opens a novel regime of relativistic laser-matter interaction ruled by both collective and quantum electrodynamics (QED) effects~\cite{Di_Piazza_RMP}. 
Among the latter, high-energy photon emission and electron-positron pair production have attracted 
a lot of attention~\cite{tamburini2010,nerush2011,gonoskov2015,ridgers2012,capdessus2013,blackburn2014,lobet2015,lobet2017,grismayer2017}. 
Some of these QED processes have been observed in recent laser-plasma experiments, some involving pair production in the
Coulomb field of highly-charged ions~\cite{chen2009,sarri2015}, and most recently in link to the problem of radiation reaction~\cite{cole2017,poder2017}. 
This line of study is at the center of various proposals for experiments on forthcoming multi-petawatt laser facilities.

Radiation reaction has also been shown to be of importance in various scenarios relating to relativistic astrophysics.
Kinetic plasma simulations have demonstrated that it can alter the physical nature of radiation-dominated relativistic current sheets
at ultra-high magnetization~\cite{jaroschek2009}. Its importance was also demonstrated for the interpretation and modeling of 
gamma-ray flares in the Crab-Nebulae~\cite{cerutti2014} and of pulsars~\cite{cerutti2016}.

These developments have motivated various theoretical works devoted to the treatment of radiation reaction in both 
classical electrodynamics~\cite{spohn2000,rohrlich2008,sokolov2009,sokolov2010,bulanov2011,burton2014,capdessus2016},
and QED~\cite{MonizSharp,KrivitskiiTsytovich,dipiazza2010,meuren2011,ilderton2013,ilderton2013b} 
(see also Ref.~\cite{Di_Piazza_RMP} for a review).
Radiation reaction, treated either using a radiation friction force in the framework of classical electrodynamics or a Monte-Carlo procedure 
to account for the quantum process of high-energy photon emission, has also recently been implemented in various kinetic simulation codes, 
in particular Particle-In-Cell (PIC) codes~\cite{tamburini2010,nerush2011,duclous2011,arber2015,lobet2016,gonoskov2015}. 
These numerical tools have been used to tackle various problems, from laser-plasma interaction under extreme light conditions 
to relativistic astrophysics.\\

QED effects are negligible (so-called classical regime) when the energy of the emitted photons remains small with respect
to that of the emitting electron, and radiation reaction can then be treated as a continuous friction force acting on the particles, 
as proposed e.g. by Landau and Lifshitz~\cite{Landau_CED}. 
In the quantum regime, photons with energies of the order of the energy of the emitting electrons can be produced~\cite{Di_Piazza_RMP}.
The on-set of QED effects has two important consequences~\cite{uggerhoj,footnote0}: first, the instantaneous power radiated away by an electron is reduced with respect to the "classical" prediction,
second, the discrete and stochastic nature of photon emission impacts the electron dynamics (so-called {\it straggling}) which cannot be treated using the continuous friction force  and
thus motivated the development of Monte-Carlo procedures~\cite{duclous2011, lobet2016}. 

Considering an ultra-relativistic electron beam interacting with a counter-propagating high-intensity laser pulse, 
Neitz and Di Piazza have demonstrated that, even in the limit $\chi \ll 1$, 
the stochastic nature of photon emission cannot always be neglected~\cite{Neitz}. 
Using a Fokker-Planck approach, the authors show that the stochastic nature of high-energy photon emission can lead
to an energy spreading (i.e. effective heating) of the electron beam while a purely classical treatment using the radiation friction force would predict 
only a cooling of the electron beam~\cite{tamburini2011,lehman2012,burton2014}.
Vranic and collaborators have further considered this scenario in Ref.~\cite{vranic2016} to study the competition between this effective
heating and classical cooling of the electron beam distribution.\\

The present study focuses on the effects of this stochastic nature of high-energy photon emission on radiation reaction.
In contrast with previous works~\cite{Neitz,vranic2016}, we extend the study to $\chi \lesssim 1$~\cite{footnoteRidgers} and arbitrary configurations, i.e. we do not
restrict ourselves to the study of an electron beam with a counter-propagating light pulse, and demonstrate the existence
of an intermediate regime, henceforth referred to as the intermediate quantum regime. 
To do so, we rely on a statistical approach of radiation reaction, starting from a linear Boltzmann description of 
photon emission and its back-reaction (from which the Monte-Carlo procedure derives), then studying in detail 
its Fokker-Planck limit. 
This procedure and a systematic comparison with the linear Boltzmann description allow to highlight different effects 
related to the quantum nature of photon emission, among which are the stochastic energy spreading and quantum quenching 
of radiation losses. The appropriate model that needs to be used in different physical situations and the relevant measurable quantities are discussed.

Beyond the theoretical insights offered by the developed descriptions, this  approach is also particularly interesting 
for numerical [in particular Particle-In-Cell (PIC)] simulations as a simple solver (so-called {\it particle pusher}) is obtained 
which can be easily implemented in kinetic simulation tools to account for the on-set of QED effects (namely the reduction
of the power radiated away by the electron, so-called quantum correction, and the straggling following from the stochastic nature of photon emission) 
 in the intermediate quantum regime, without having to rely on the  computationally demanding Monte-Carlo approach. \\

The paper is structured as follows.
In Sec.~\ref{sec:ced}, we summarize the classical treatment of the radiation emission and its back-reaction on the electron dynamics
and show that, in the case of ultra-relativistic electrons, the momentum and energy evolution equations take a simple and intuitive form.
This form of the radiation friction force has the advantage to conserve the on-shell condition while being straightforward to implement numerically. 
High-energy photon emission and its back-reaction as inferred from the quantum approach is then summarized in Sec.~\ref{sec:qed}, which
introduces the key-quantities that appear in the statistical descriptions we develop in the following Sections.
In Sec.~\ref{sec:stoch}, starting from a kinetic master equation, we derive a Fokker-Planck (FP) equation where quantum effects appear both 
as a correction on the friction force (drift term) and in a diffusion term, the latter accounting for the stochastic nature inherent to the quantum emission process.
Interestingly, the leading term of the Landau-Lifshitz equation with a quantum correction naturally appears from the FP expansion.
The domain of validity of the FP expansion is then studied in detail, and for arbitrary conditions of interaction.
In Sec.~\ref{sec:averages}, the equations of evolution for the successive moments of the electron distribution function are discussed considering the classical, 
FP and linear Boltzmann descriptions. These equations allow for some analytical predictions on the average energy and energy dispersion when
considering, but not limited to, an electron beam interacting with a high-intensity laser field. It also sheds light on other processes such as the quantum quenching
of radiation losses observed in recent numerical simulations~\cite{harvey2017}. Most importantly, it allows us to identify the domains of validity of the
various descriptions of radiation reaction. These domains being intimately linked to the relative importance of different effects of radiation reaction, 
our analysis provides new physical insights into how to observe these different effects.
Section~\ref{sec:algo} then presents three complementary numerical algorithms to account for radiation reaction, among which 
the new particle pusher obtained from the FP description.
Section~\ref{sec:numResults} then considers different physical configurations to validate both our theoretical analysis and numerical tools, 
and allows us to further investigate both the domains of validity of the different descriptions as well as the different aspects of radiation reaction.
Finally, conclusions are given in Sec.~\ref{sec:conclusion}.

\section{Dynamics of a radiating electron in classical electrodynamics}\label{sec:ced}

Let us consider a single electron (with charge $-e$ and mass $m$) in an arbitrary external field described by the electromagnetic field tensor $F^{\mu\nu}$.
Classically, its dynamics is determined by the Lorentz equation whose covariant formulation reads
\begin{eqnarray}\label{eq_Lorentz}
\frac{dp^{\mu}}{d\tau} = -\frac{e}{m c}\,F^{\mu\nu}\,p_{\nu}\,,
\end{eqnarray}
where $c$ is the speed of light in vacuum, $\tau$ the proper time, 
and $p^{\mu} = (\gamma m c, {\bf p})$ the electron four-momentum with $\gamma = \sqrt{1+{\bf p}^2/(m c)^2}$ the electron Lorentz factor [SI units will be used throughout this work]. 
Lorentz equation~\eqref{eq_Lorentz}, however, does not account for the fact that, while being accelerated, the electron emits radiation
thus losing energy and momentum.
Accounting for the back-reaction of radiation emission on the electron dynamics has been a long standing 
problem of classical electrodynamics (CED)~\cite{LAD,Jackson}. As such, it has been the focus of many studies, 
and various equations of motion of a radiating charge in an external field have been proposed~\cite{Di_Piazza_RMP}.
In this Section, we briefly discuss the one proposed by Landau and Lifshitz~\cite{Landau_CED}, and apply it to the 
dynamics of an ultra-relativistic electron in an arbitrary external field.

\subsection{The Landau-Lifshitz radiation reaction force}\label{sec:ced1}

A derivation of the equation of motion a radiating electron has been proposed by Landau and Lifshitz (LL)~\cite{Landau_CED}:
\begin{eqnarray}\label{eq_LL0}
\frac{dp^{\mu}}{d\tau} = -\frac{e}{m c}\,F^{\mu\nu}\,p_{\nu} + g^{\mu}\,,
\end{eqnarray}
where the so-called LL radiation reaction force reads:
\begin{eqnarray}
\nonumber
g^{\mu} &=& -\frac{2}{3}\,\tau_e\,\left[ \frac{e}{m^2\,c}\,\partial_{\eta} F^{\mu\nu}\,p_{\nu}\,p^{\eta}   
+ \frac{e^2}{m^2\,c^2}\,F^{\mu\nu}\,F_{\eta\nu}\,p^{\eta} \right. \\
\label{eq_LL} &-& \left. \frac{e^2}{m^4\,c^4}\,(F^{\nu\eta}\,p_{\eta})\,(F_{\nu\alpha}\,p^{\alpha})\,p^{\mu} \right]\,,
\end{eqnarray}
with $\tau_e = r_e/c$ the time for light to travel across the classical radius of the electron $r_e = e^2/(4\pi\epsilon_0\,m c^2)$ and $\epsilon_0$ the permittivity of vacuum.

The first term in Eq.~\eqref{eq_LL}, also known as the Schott term, stands as a four-force (i.e. it is perpendicular to the four-momentum).
Rigorously, the last two terms have to be kept together to form a four-force $f^{\mu}$ such that ${f^{\mu} p_{\mu} = 0}$. 
In addition, upon integration over the particle motion through a given external field (i.e. computing $\int\!d\tau g^{\mu}$), the first two terms in Eq.~\eqref{eq_LL} cancel each other while the last term corresponds to the total four-momentum radiated away by the particle~\cite{footnote1}
\begin{eqnarray}
\Delta p^{\mu} = \frac{P_0}{c^2}\,\int \left\vert \frac{F^{\nu\eta}}{E_{cr}}\,\frac{p_{\eta}}{m c}\right\vert\,dx^{\mu}\,,
\end{eqnarray} 
with $P_0 = 2\,m c^2/(3\,\tau_e)$, $E_{cr} = 4\pi\epsilon_0\,m^2c^4/e^3 \simeq 1.8 \times 10^{20}~{\rm V/m}$ the critical field of CED and $x^{\mu}$ the electron four-position. Therefore, it is not possible, in general, to consider each term of the LL force separately.

\subsection{Radiation friction force acting on an ultra-relativistic electron}\label{sec:ced2}

The time and space components of LL Eqs.~\eqref{eq_LL0} and~\eqref{eq_LL} give the equations of evolution of the energy and 
momentum of an electron with arbitrary $\gamma$, respectively 
\begin{eqnarray}
\nonumber    m c^2\,\frac{d\gamma}{dt} &=& -e c\,{\bf u} \cdot {\bf E} - \frac{2}{3}\,e c \tau_e \gamma\, \dot{\bf E} \cdot {\bf u}   \\    
\nonumber    &+& \frac{2}{3}\,\frac{e c}{E_{cr}}\,{\bf E} \cdot ({\bf E} + {\bf u}\times {\bf H}) \\
\label{eq_A1} &-& \frac{2}{3}\,\frac{e c}{E_{cr}}\,\gamma^2\,\left[ ({\bf E} + {\bf u} \times {\bf H})^2 - ({\bf u} \cdot {\bf E})^2 \right]\,,                                              \\
\nonumber    \frac{d{\bf p}}{dt} &=& -e\,({\bf E} + {\bf u} \times {\bf H}) - \frac{2}{3}\,e\,\tau_e\,\gamma\,\left(\dot{\bf E} + {\bf u} \times \dot{\bf H}\right) \\
\nonumber    &+& \frac{2}{3}\,\frac{e}{E_{cr}}\,\left[ ({\bf u} \cdot {\bf E})\, {\bf E} - {\bf H} \times ({\bf E} + {\bf u \times {\bf H}}) \right] \\
\label{eq_A2} &-& \frac{2}{3}\,\frac{e}{E_{cr}}\,\gamma^2\left[({\bf E} + {\bf u }\times {\bf H})^2 - ({\bf u} \cdot {\bf E})^2\right]\,{\bf u}\,,
\end{eqnarray}
where ${\bf u}={\bf p}/(\gamma m c)$ is the normalized electron velocity,
${\bf E}$ and ${\bf H}=c {\bf B}$ are the electric and magnetic fields, respectively, and dotted fields are (totally) differentiated with respect to time $t$.

It is interesting to develop the radiation reaction force ${\bf f}_{\rm rad}$ [three last terms in Eq.~\eqref{eq_A2}] as a longitudinal force 
(acting in the direction of the electron velocity) and a transverse force (acting in the direction normal to the electron velocity).
One obtains ${\bf f}_{\rm rad} = {\bf f}^{\Vert}_{\rm rad} + {\bf f}^{\perp}_{\rm rad}$ with
\begin{eqnarray}
\nonumber {\bf f}^{\Vert}_{\rm rad} &=& -\frac{2}{3}\,e \tau_e \gamma (\dot{\bf E}\cdot{\bf u})\,{\bf u}/{{\bf u}^2}\\
\nonumber &+&\frac{2}{3}\,\frac{e}{E_{cr}}\,{\bf E}\cdot({\bf E} + {\bf u} \times {\bf H})\,{\bf u}/{{\bf u}^2}\\
\label{eq_fpar} &-& \frac{2}{3}\,\frac{e}{E_{cr}}\,\gamma^2\,\left[({\bf E} + {\bf u} \times {\bf H})^2 - ({\bf u}\cdot{\bf E})^2 \right]\,{\bf u}/{{\bf u}^2}\,,\\
\nonumber {\bf f}^{\perp}_{\rm rad} &=& -\frac{2}{3}\,e \tau_e \gamma\,\left[(\dot{\bf E})_{\perp} + {\bf u} \times \dot{\bf H}\right] \\
\label{eq_fper} &+& \frac{2}{3}\,\frac{e}{E_{cr}}\,\left[ ({\bf u}\cdot{\bf E})\,{\bf E}_{\perp} + ({\bf u}\cdot{\bf H})\,{\bf H}_{\perp} + ({\bf E}\times{\bf H})_{\perp}\right],\quad
\end{eqnarray}
where $\Vert$ ($\perp$) denotes the vector component parallel (perpendicular) to the electron velocity ${\bf u}$.

Let us stress at this point that the last term in Eq.~\eqref{eq_fpar} contains not only the contribution of the last term of the LL radiation reaction force Eq.~\eqref{eq_LL}, but also part of its second term. This has interesting implications, in particular in terms of conservation of the on-shell condition, as will be further
discussed at the end of this Section.

For an ultra-relativistic electron ($\gamma \gg 1$), the last terms in Eqs.~\eqref{eq_A1} and~\eqref{eq_fpar} give the important contributions,
and all other terms together with the perpendicular component of the radiation reaction force [Eq.~\eqref{eq_fper}] can be neglected.
One then obtains the equations of evolution for the ultra-relativistic electron energy and momentum
\begin{eqnarray}
\label{eq_ener}  m c^2 \, \frac{d\gamma}{dt} &=& -e c\ {\bf u} \cdot {\bf E} - P_{\rm cl}\,,\\
\label{eq_force}  \frac{d{\bf p}}{dt} &=& -e\,({\bf E} + {\bf u} \times {\bf H}) - P_{\rm cl}\,{\bf u}/(c {\bf u}^2) \,,
\end{eqnarray}
where $P_{\rm cl}$ denotes the classical instantaneous power radiated away by the electron
\begin{eqnarray}\label{eq_Prad_CED}
P_{\rm cl} = P_0\,\eta^2\,,
\end{eqnarray}
for which we have introduced the Lorentz invariant:
\begin{eqnarray}
\eta = \left\vert \frac{F^{\mu\nu}}{E_{cr}}\,\frac{p_{\nu}}{m c} \right\vert = \frac{\gamma}{E_{cr}}\,\sqrt{({\bf E} + {\bf u} \times {\bf H})^2 - ({\bf u} \cdot {\bf E})^2}\,.
\end{eqnarray}

In contrast with previous works, we point out that the radiation reaction force, last term in Eq.~\eqref{eq_force}, takes the form of a friction force
${\bf f}_{\rm rad} = \nu {\bf u}$ [with a nonlinear friction coefficient $\nu$] that also
has the nice property of conserving the on-shell condition $p^{\mu}p_{\mu}=m^2 c^2$. 
This can be seen by taking the scalar product of Eq.~\eqref{eq_force} by $c {\bf u}$, which turns out to be consistent with the 
energy conservation Eq.~\eqref{eq_ener}. 
This would not have been the case had we retained only the leading ($\propto \gamma^2$) terms in Eqs.~\eqref{eq_A1} and~\eqref{eq_A2}.
Our formulation hence contrasts with that proposed, e.g. by Tamburini {\it et al.}~\cite{tamburini2010}, where the authors choose to retain both last two terms 
in Eq.~\eqref{eq_A2} to preserve the on-shell condition.
Beyond its simple and intuitive form, the radiation force given by Eq.~\eqref{eq_force} is also straightforward to implement in 
numerical tools. Indeed, in the formulation by Tamburini {\it et al.}, the two components kept in the radiation reaction force 
have huge ($\propto \gamma^2$) differences in their magnitude that may lead to accumulation of round-off errors 
(see, e.g., Ref~\cite{Accurate_LL} for an accurate treatment of this problem). The simple form of the radiation friction force derived
here allows to avoid this issue.

\subsection{High-energy synchrotron-like radiation emission}\label{sec:ced3}

Let us first briefly discuss, in the framework of classical electrodynamics, the radiation emission of an ultra-relativistic electron 
in an arbitrary external field.
It is well known that, for an ultra-relativistic - otherwise arbitrary - motion, the radiation emission is approximately the same as 
that of an electron moving instantaneously along a circular path~\cite{Jackson}. Hence, it is well approximated by the so-called 
synchrotron emission. The corresponding emitted power distribution as a function of the frequency $\omega$ of the emitted 
photons reads~\cite{Jackson} 
\begin{eqnarray}\label{eq_Pspec_CED}
\frac{dP}{d\omega} = \frac{9\sqrt{3}}{8\pi}\,\frac{P_{\rm cl}}{\omega_c}\,\frac{\omega}{\omega_c}\,\int_{\omega/\omega_c}^{+\infty}\!\!\!dy\,{\rm K}_{5/3}(y)\,,
\end{eqnarray}
with ${\rm K}_{\nu}(z)$ the modified Bessel function of the second kind, $\omega_c = 3\,\gamma \eta/(2\tau_e)$ the so-called critical frequency for synchrotron emission, and where the instantaneous power radiated away is given by Eq.~\eqref{eq_Prad_CED}.
The classical approach requires the emitted photon energy $\hbar\omega$ to be much smaller than that of the emitting electron~\cite{Di_Piazza_RMP}.
This is ensured for $\hbar\omega_c \ll \gamma m c^2$, which translates to the condition on the electron parameter: $\eta/\alpha = \chi \ll 1$, with $\alpha = e^2/(4\pi \epsilon_0 \hbar c)$ the fine-structure constant.

\section{Radiation reaction and high-energy photon emission in quantum electrodynamics}\label{sec:qed}

In the regime where quantum effects are not negligible, determining the spectral properties of the emission radiated away 
by an electron in an arbitrary external field is greatly simplified when considering ultra-relativistic electrons in the presence of a 
(i) slowly varying compared to the formation time of the radiated photon (this is the so-called local constant field approximation~\cite{footnoteLCFA}), 
and (ii) undercritical (as defined below), otherwise arbitrary, field~\cite{Di_Piazza_RMP}. 

Condition (i) is fulfilled when the electromagnetic field has a relativistic strength~\cite{footnote2}
\begin{eqnarray}\label{eq_cond1}
a_0 = \frac{e\,\vert A^{\mu}\vert}{m c^2} \gg 1\,,
\end{eqnarray}
where $A^{\mu}$ is the four-potential corresponding to the electromagnetic field tensor $F^{\mu\nu} = \partial^{\mu}\! A^{\nu} - \partial^{\nu}\! A^{\mu}$. 

Condition (ii) requires that both Lorentz invariants of the electromagnetic fields are small with respect to the corresponding invariants of the critical field of QED [$E_s = \alpha\, E_{cr} \simeq 1.3 \times 10^{18}~{\rm V/m}$]:
\begin{eqnarray}
\label{eq_cond2a}\zeta_1 = F^{\mu\nu}\,F_{\mu\nu}/E_s^2 = ({\bf H}^2 - {\bf E}^2)/E_s^2 \ll 1\,,\\
\label{eq_cond2b}\zeta_2 = \epsilon^{\mu\nu\eta\alpha}\,F_{\mu\nu}\,F_{\eta\alpha}/E_s^2 = ({\bf E} \cdot {\bf H})/E_s^2 \ll 1\,,
\end{eqnarray}
where $\epsilon^{\mu\nu\eta\alpha}$ is the completely antisymmetric unit tensor with $\epsilon^{0123} = 1$.

Let us now introduce the Lorentz invariant quantum parameter for the electron:
\begin{eqnarray}
\chi = \left\vert \frac{F^{\mu\nu}}{E_s}\,\frac{p_{\nu}}{m c} \right\vert = \frac{\eta}{\alpha}\,.
\end{eqnarray}
In addition to conditions (i) and~(ii), the following condition (iii) imposes $\chi$ to be much larger than both field invariants $\zeta_1$ and $\zeta_2$:
\begin{eqnarray}
\label{eq_cond3} \chi \gg {\rm max}(\zeta_1,\zeta_2)\,,
\end{eqnarray}
For completeness, following Ref.~\cite{dipiazza2010,Di_Piazza_RMP}, we further restrict our study to the so-called 
{\it non-linear moderately quantum regime} corresponding to $\chi \lesssim 1$ and $a_0 \gg 1$, for which radiation reaction in the QED framework has been 
identified as the overall electron energy and momentum loss due to the emission of many photons consecutively, and incoherently~\cite{footnote3}.

Under these assumptions, the (Lorentz invariant) production rate of high-energy photons emitted by the electron can be written as~\cite{Ritus}:
\begin{eqnarray}\label{eq_QEDprodrate}
\label{eq_N}\frac{d^2N}{d\tau d\chi_{\gamma}} = \frac{2}{3} \frac{\alpha^2}{\tau_e}\,\frac{G(\chi,\chi_{\gamma})}{\chi_{\gamma}}\,,
\end{eqnarray}
where 
\begin{eqnarray}\label{eq_G}
G(\chi,\chi_{\gamma}) = \frac{\sqrt{3}}{2\pi}\frac{\chi_{\gamma}}{\chi}\left[\int_{\nu}^{+\infty} \!\!\!\!\!\!{\rm K}_{5/3}(y)\,dy 
+ \frac{3}{2}\chi_{\gamma}\nu\, {\rm K}_{2/3}(\nu)\right]\,\quad
\end{eqnarray}
is the so-called quantum emissivity, and $\nu = 2\,\chi_{\gamma}/[3\chi\,(\chi-\chi_{\gamma})]$. 

\begin{figure}
\begin{center}
\includegraphics[width=7.25cm]{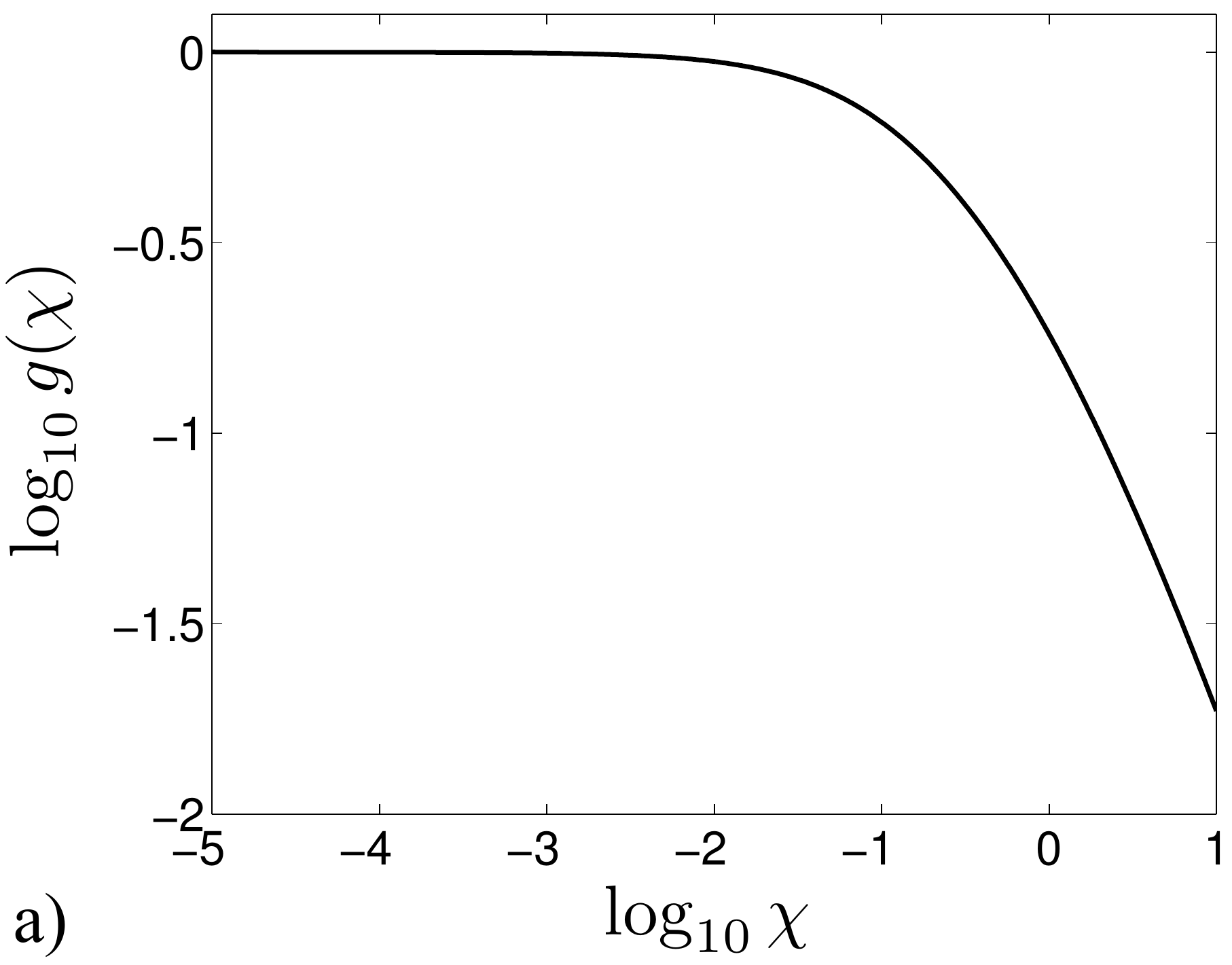}
\includegraphics[width=7.5cm]{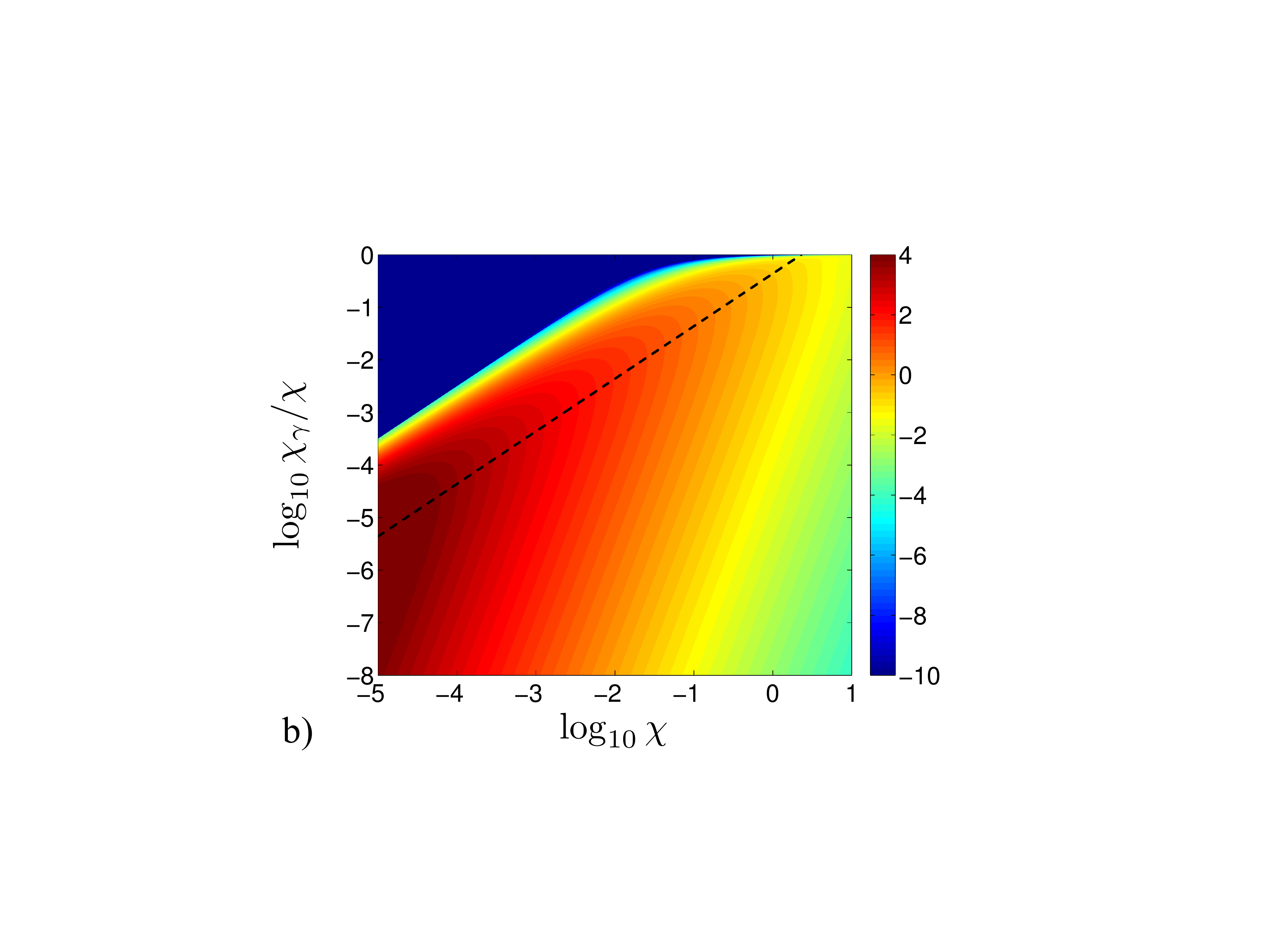}
\includegraphics[width=7.5cm]{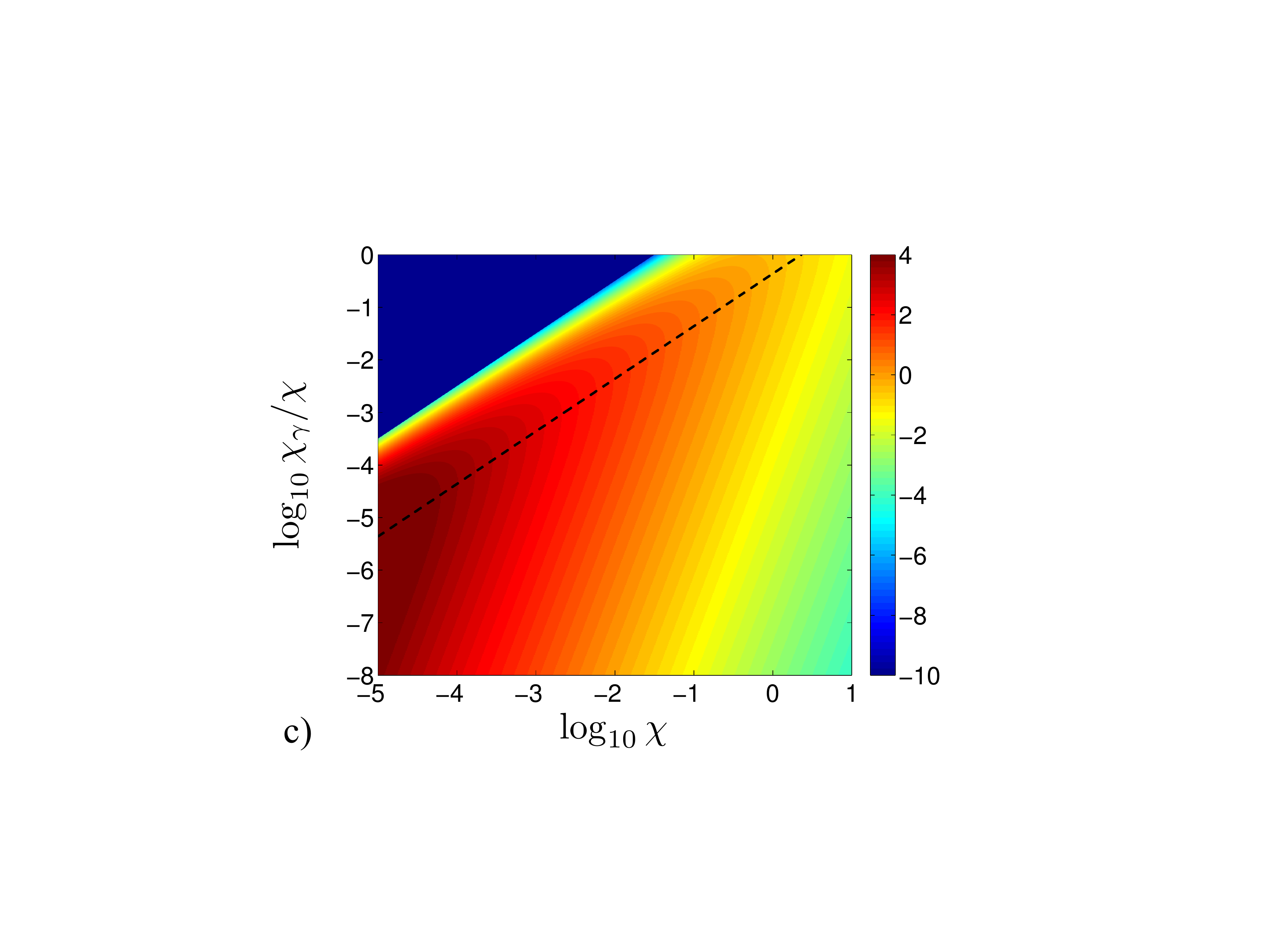}
\caption{(a) Dependence of $g(\chi)$ on the electron quantum parameter $\chi$ leading to a reduction of the 
emitted power due to quantum effects. (b) Quantum emissivity $G(\chi/\chi_{\gamma})/\chi^2$ and (c) its classical limit 
as a function of $\chi$ and $\chi_{\gamma}/\chi = \gamma_{\gamma}/\gamma$. 
Dashed lines in panel (b) and (c) show $\chi_{\gamma} \simeq 0.435\,\chi^2$ for which the classical limit of $G(\chi,
\chi_{\gamma})$ is maximum.} 
\label{fig:qed}
\end{center}
\end{figure}

The production rate Eq.~\eqref{eq_QEDprodrate} only depends on the electron quantum parameter $\chi$ and on the (Lorentz invariant)
quantum parameter for the emitted photon:
\begin{eqnarray}
\chi_{\gamma} = \left\vert \frac{F^{\mu\nu}}{E_s}\,\frac{\hbar k_{\nu}}{m c} \right\vert\,,
\end{eqnarray}
where $k^{\nu} = (\hbar\omega/c,\hbar {\bf k})$ is the four-momentum of the emitted photon.
Considering an ultra-relativistic electron, the photon quantum parameter $\chi_{\gamma}$ can be expressed in terms 
of the electron quantum parameter $\chi$ and the electron and photon energies as
\begin{eqnarray}\label{eq:linkChi}
\chi_{\gamma} = \frac{\gamma_{\gamma}}{\gamma}\,\chi\,.
\end{eqnarray}
Another Lorentz invariant can be derived from Eq.~\eqref{eq_N}
\begin{eqnarray}
\label{eq_Pspec}  \frac{d^2 \mathcal{E}}{d\tau d\gamma_{\gamma}} = P_0\, \alpha^2\, G(\chi, \chi_{\gamma})\,,
\end{eqnarray}
which denotes the emitted power distribution in terms of the photon normalized energy.
The instantaneous power radiated away by the electron is another Lorentz invariant. It is obtained by integrating
 Eq.~\eqref{eq_Pspec} over all photon energies giving
\begin{eqnarray}
\label{eq_PradQED}  P_{\rm rad} = \int_0^{+\infty}\!\!\!d\gamma_{\gamma} \, \frac{1}{\gamma} \frac{d^2 \mathcal{E}}{d\tau 
d\gamma_{\gamma}} = P_0\, \alpha^2 \chi^2\,g(\chi) \,,
\end{eqnarray}
where 
\begin{eqnarray}
\nonumber g(\chi) &=& \int_0^{+\infty} \!\!\!\!\! d\chi_{\gamma}\,\frac{G(\chi,\chi_{\gamma})}{\chi^3} = \frac{9\sqrt{3}}{8\pi}\!\!
 \int_0^{+\infty}\!\!\!\!\! d\nu\,\left[ \frac{2\nu^2\,{\rm K}_{5/3}(\nu)}{(2+3\nu\chi)^2}  \right. \\
\label{eq_h_chie} &+& \left. \frac{4\nu\, (3\nu\chi)^2}{(2+3\nu\chi)^4}\, {\rm K}_{2/3}(\nu)\right]\,.
\end{eqnarray}
Figure~\ref{fig:qed}(a) shows $g(\chi)$ for $\chi$ ranging from $10^{-5}$ to $10$.

Let us now stress that Eq.~\eqref{eq_PradQED} is nothing but the classical instantaneous power radiated away by the electron 
[Eq.~\eqref{eq_Prad_CED}] multiplied by $g(\chi)$  (we recall that $\chi = \eta/\alpha$):
\begin{eqnarray}
P_{\rm rad} = P_{\rm cl}\,g(\chi).
\end{eqnarray}
The classical limit is recovered when the emitted photon energies remain much smaller than the emitting electron energy, i.e. 
by taking the limit $\chi_{\gamma} \ll \chi \ll 1$ [correspondingly $\nu \sim 2\chi_{\gamma}/(3 \chi^2)$]. 
In this limit, $g(\chi) \sim 1$, and Eqs.~\eqref{eq_Pspec} and~\eqref{eq_PradQED}
reduce to their classical forms Eqs.~\eqref{eq_Pspec_CED} and~\eqref{eq_Prad_CED}, respectively.

Therefore, to take into account the difference between the classical and quantum radiated spectrum 
in our classical equation of motion [Eqs~\eqref{eq_ener} and~\eqref{eq_force}], 
we can replace phenomenologically $P_{\rm cl}$  
by its quantum expression $P_{\rm rad}$, 
$g(\chi)$ giving a so-called quantum correction (see, e.g.,~\cite{ridgers2014,Erber_RevModPhys.38.626})
If this approach is here mainly heuristic, we will see in Sec.~\ref{sec:stoch} 
that it is actually correct, the statistical average of the quantum description providing
the quantum correction naturally.

Finally, to highlight QED effects on the emitted radiation properties, 
we have plotted $G(\chi,\chi_{\gamma})/\chi^2$
and its classical limit in Fig.~\ref{fig:qed}(b) and~\ref{fig:qed}(c), respectively. 
As can be seen, quantum effects mainly tend to decrease
the photon emission rate at high-energies. In particular, emission of photons
with an energy larger than the emitting photon
energy (i.e. for $\chi_{\gamma} > \chi$) is prevented. As a result, the overall 
emitted power is reduced [see also  Fig.~\ref{fig:qed}(a)].

\section{From quantum to classical radiation reaction for ultra-relativistic electrons}\label{sec:stoch}

In this Section, we propose a statistical description of high-energy photon emission and its back-reaction,
starting from the quantum point of view, and getting toward the classical regime.

\subsection{Basic assumptions}\label{sec:stoch:assumptions}

We now consider not a single electron but a set (population) of electrons interacting with an arbitrary 
external electromagnetic field, with the only provision that this field satisfies conditions (i) - (iii) in Sec.~\ref{sec:qed} 
[correspondingly Eqs.~\eqref{eq_cond1} - \eqref{eq_cond3}] and that of the non-linear moderately quantum regime 
where pair production and higher-order coherent processes are neglected. 
This statistical approach also neglects all collisions between electrons,
and considers that the electron population can be described by its 
distribution function $f_e(t,\boldsymbol{r},\gamma,\boldsymbol{\Omega})$, where $\gamma$ and $\bf\Omega$ denote 
the phase-space Lorentz factor and velocity direction, respectively, so that they define uniquely the phase-space momentum
 ${\bf p} = m c \gamma{\bf u}$. 

For the sake of completeness, we also point out that all electrons are considered to emit high-energy radiation 
in an incoherent way, that is the radiation emission by an electron is not influenced by neighbor electrons. 
This is justified whenever the high-energy photon emission has a wavelength much shorter than the typical distance 
between two electrons $\propto n_e^{-1/3}$, with $n_e$ the characteristic density of the electron population~\cite{schlegel_POP_2009}.

\subsection{Kinetic point of view: the linear Boltzmann equation}\label{sec:stoch:linearBoltzmann}

The equation of evolution for the electron distribution function $f_e(t,\boldsymbol{r},\gamma,\boldsymbol{\Omega})$ 
accounting for the effect of high energy photon emission
and the corresponding photon distribution function $f_{\gamma}(t,\boldsymbol{r},\gamma,\boldsymbol{\Omega})$ 
can be written in the form:
\begin{eqnarray}
\nonumber\!\!\!\!\frac{d}{dt} f_e &=& \,\int_0^{+\infty}\!\!\!\!d\gamma_{\gamma}\,w_{\chi} (\gamma+\gamma_{\gamma},\gamma_{\gamma})  \, f_e(t,{\bf x,}\gamma + \gamma_{\gamma},{\bf\Omega}) \\
\label{eq:Master1} &-& f_e(t,{\bf x,}\gamma,{\bf\Omega})\,\int_0^{+\infty}\!\!\!\!d\gamma_{\gamma} w_{\chi} (\gamma,\gamma_{\gamma})\,,\\
\label{eq:Master2}\!\!\!\! \frac{d}{dt} f_{\gamma} &=& \int_1^{+\infty}\!\!\!\!d\gamma\, w_{\chi}(\gamma+\gamma_{\gamma},\gamma_{\gamma})\,f_e(t,{\bf x,}\gamma + \gamma_{\gamma},{\bf\Omega}),
\end{eqnarray}
where it has been assumed that radiation emission (and its back-reaction) is dominated by
the contribution of ultra-relativistic electrons (for which ${\bf p}\simeq mc \gamma {\bf\Omega}$),
and that such ultra-relativistic electrons emit radiation in the direction $\bf\Omega$ of their velocity,
and the total time derivatives in Eqs.~\eqref{eq:Master1} and~\eqref{eq:Master2} will be detailed in Sec.~\ref{sec:stoch}.

Equation~\eqref{eq:Master1} is a linear Boltzmann equation, and its right-hand-side (rhs), henceforth denoted $\mathcal{C}[f_e]$, acts as a {\it collision} operator
(see also Refs.~\cite{sokolov2010,Elkina,Neitz,ridgers2017}).
It accounts for the effect of high-energy photon emission on the dynamics of an electron radiating in the electromagnetic fields $\bf E$ and $\bf H$, 
that is for radiation reaction.
It depends on $w_{\chi}(\gamma,\gamma_{\gamma})$ which denotes the rate of emission of a photon with energy $m c^2\gamma_{\gamma}$ 
by an electron with energy $mc^2\gamma$ and quantum parameter $\chi$. Note that the dependency on $\chi$ implicitly states that the emission 
rate is computed locally in space and time, i.e. taking the local value of the electromagnetic field at time $t$ and position $\bf x$, 
for a given electron momentum direction $\bf\Omega$. 
Under the assumptions previously introduced (Sec.~\ref{sec:stoch:assumptions}), this emission rate reads:
\begin{eqnarray}\label{eq:Kernel}
w_{\chi} (\gamma,\gamma_{\gamma}) = \left. \frac{d^2N}{dt d \gamma_{\gamma}} \right|_{\chi}\!\!\!(\gamma_{\gamma},\gamma)=\frac{2}{3} \frac{\alpha^2}{\tau_e} 
\frac{\widetilde{G}(\chi,\gamma_{\gamma}/\gamma)}{\gamma \gamma_{\gamma}} \, ,
\end{eqnarray}
where:
\begin{eqnarray}
\nonumber \widetilde{G}(\chi, \xi) = \frac{\sqrt{3}}{2\pi} \xi \Bigg[ \int_{\nu}^{+\infty} K_{5/3}(y)dy + \frac{\xi^2}{1-\xi}K_{2/3}(\nu)\Bigg] \, ,
\end{eqnarray}
with $\xi = \chi_{\gamma}/\chi = \gamma_{\gamma}/\gamma$ and $\nu=2\xi/[3\chi (1-\xi)]$.

It is complemented by Eq.~\eqref{eq:Master2} that describes the temporal evolution of the photon distribution function.
In this work, photons are simply created and then propagate freely. The rhs of Eq.~\eqref{eq:Master2} thus stands as a source term
and will be denoted $\mathcal{S}[f_e]$ in the rest of this work.

In Sec.~\ref{sec:averages}, we will show that Eq.~\eqref{eq:Master1} conserves the total number of electrons,
while Eq.~\eqref{eq:Master2} predicts a total number of photons increasing with time as more and more photons are radiated away. 
It will also be demonstrated that the total energy lost by electrons due to radiation emission is indeed transferred to high-energy photons,
that is, the system of Eqs.~\eqref{eq:Master1} and~\eqref{eq:Master2} does conserve the total energy in the system.

\subsection{High-energy photon emission as a random process}\label{sec:stoch:PoissonProcess}

The system of Eqs.~\eqref{eq:Master1} and~\eqref{eq:Master2} stands as a Master equation.
It describes a discontinuous jump process with $w_{\chi}(\gamma,\gamma_{\gamma})$ 
giving the rate of jump from a state of electron energy $mc^2\gamma$ to the state of energy 
$mc^2(\gamma-\gamma_{\gamma})$, via the emission of a photon of energy $mc^2\gamma_{\gamma}$.

Considering the dynamics of a single electron, this process can be described by three different (but related) random variables: 
(i) the electron energy itself, (ii) the number $N_t$ of photon emission events in a time interval $[0,t]$,
and (iii) the time $T_n$ of the $n^{th}$ emission event.
The  last two variables are of course equivalent since $T_n \geq t \iff N_t \leq n$, 
both denoting that there are at least $n$ emissions in the time interval $[0,t]$. 
It is possible to show that $N_t$ follows a Poisson process of parameter
\begin{eqnarray}
\tau(\chi,\gamma, t) = \int_0^t W(\chi,\gamma) dt'
\end{eqnarray}
which is usually referred to as the {\it optical depth}, and where:
\begin{eqnarray}\label{eq:PhotonRate}
W(\chi,\gamma) =
 \frac{2}{3} \frac{\alpha^2}{\tau_e \gamma} \, \int_0^{+\infty} \!\!\! d\xi\, \widetilde{G}(\chi,\xi)/\xi \, 
\end{eqnarray}
is the instantaneous rate of photon emission.
Hence, the probability for the electron to emit $n$ photons during a time interval $t$ is given by
 \begin{eqnarray}\label{Poisson_eq}
P[N(\gamma, t) = n] = e^{-\tau(\chi,\gamma, t)}\frac{\tau(\chi,\gamma, t)^n}{n!} \, ,
\end{eqnarray}
while the cumulative probability of the random variable $T_n$ is given by
\begin{eqnarray}\label{final_opt_depth}
P[T_{n \geq 1} < t] = 1 - e^{-\tau(\chi,\gamma, t)} \, .
\end{eqnarray}

A discrete stochastic formulation of these discontinuous jumps can be rigorously deduced~\cite{lapeyre},
leading to a Monte-Carlo description (see Sec.~\ref{sec:algoMC} and Refs.~\cite{duclous2011,lobet2016} for more details).
While it allows to fully model high-energy photon emission and its back-reaction as depicted by the linear 
Boltzmann Eq.~\eqref{eq:Master1} and Eq.~\eqref{eq:Master2}, the Monte-Carlo procedure has some limitations.
Indeed, in regimes of intermediate $\chi$ parameters, numerous discrete events of small energy content may occur,
giving rise to computational cost overhead.
These events may however have a non-negligible cumulative effect.
As will be shown in what follows, this case is precisely the operating regime of the Fokker-Planck 
approximation (a by-product of the master equation). In the following, we show that a Fokker-Planck approach 
can be used to treat many discrete events at once.

\subsection{Toward the classical limit: the Fokker-Planck approach}\label{sec:stoch:FokkerPlanck}

Let us now focus on the linear Boltzmann Eq.~\eqref{eq:Master1} which we rewrite in the form:
\begin{eqnarray}\label{eq:Master}
\partial_t f_e &+& \nabla\cdot\left[c u{\bf\Omega} f_e\right]-\frac{1}{m c^2}\partial_\gamma\!\left[e c u({\bf\Omega}\cdot{\bf E})f_e\right]\\
\nonumber&-&\frac{e}{p}\nabla_{\tiny\bf\Omega}\cdot\left[(\mathbb{1}-{\bf\Omega}\otimes{\bf\Omega})\cdot({\bf E}+u{\bf\Omega}\times{\bf H})f_e\right] = \mathcal{C}\left[f_e\right]\,,
\end{eqnarray}
where $u=\sqrt{\gamma^2-1}/\gamma$, $p=mc\sqrt{\gamma^2-1}$, $\nabla_{\bf\Omega}$ 
denotes the derivative with respect to ${\bf \Omega}$, $\mathbb{1}$ the rank 2 unit tensor and $\otimes$ stands for the dyadic product.
While the left hand side of Eq.~\eqref{eq:Master} is the standard Vlasov operator written for the energy-direction distribution function
$f_e(t,\boldsymbol{r},\gamma,\boldsymbol{\Omega})$ (see e.g. Ref.~\cite{touati2014}), $\mathcal{C}[f_e]$ is the {\it collision} operator given
by the rhs side of Eq.~\eqref{eq:Master1}.

Rewriting the integrand in the first integral of $\mathcal{C}[f_e]$ as a Taylor series in $\gamma_\gamma/\gamma$,
the {\it collision} operator can be formally casted in the form of the Kramers-Moyal expansion:
\begin{eqnarray}\label{eq:KramersMoyal}
 \mathscr{C}\left[f_e\right] = \sum_{n=1}^{\infty} \frac{1}{n!}\,\partial^n_{\gamma}\left[ A_n(\chi,\gamma)\,f_e\right]\,,
 \end{eqnarray}
 with $A_n(\chi,\gamma) = \int d\gamma_{\gamma}\,\gamma_{\gamma}^n\,w_{\chi}(\gamma,\gamma_{\gamma})$ the $n^{th}$ moment
associated to the kernel $w_{\chi}(\gamma,\gamma_{\gamma})$. For the particular kernel given by Eq.~\eqref{eq:Kernel}, we get for the 
associated moments:
\begin{eqnarray}\label{eq:An}
A_n(\chi,\gamma) = \frac{2}{3}\frac{\alpha^2}{\tau_e}\, \gamma^{n-1}\,a_n(\chi)\,
\end{eqnarray}
with:
\begin{eqnarray}
a_n(\chi) = \int_{0}^{+\infty}\!\!\!\! d\xi\,\xi^{n-1}\,\widetilde{G}(\chi,\xi)\,.
\end{eqnarray}
Note that the first moment depends on $\gamma$ only through $\chi$ and reads $A_1(\chi,\gamma)= \frac{2}{3}\frac{\alpha^2}{\tau_e}\, \chi^2\,g(\chi)$ 
[where $\chi^2 g(\chi) = a_1(\chi)$ is the quantum correction given by Eq.~\eqref{eq_h_chie}].

In general, using expansion~\eqref{eq:KramersMoyal} for the operator in the linear Boltzmann Eq.~\eqref{eq:Master} would require to solve an infinite order partial differential equation. Therefore, it is common to truncate Eq.~\eqref{eq:KramersMoyal}. This truncation cannot however be done properly by a finite and larger than two number of terms~\cite{pawula1967}. 

The truncation using the first two terms in Eq.~\eqref{eq:KramersMoyal} is actually justified in the limit $\gamma_{\gamma}\ll\gamma$.
This is ensured for $\chi\ll1$, i.e. in the classical regime of radiation emission, and the resulting truncation corresponds to a Fokker-Planck expansion.
In this limit, the {\it collision} operator reduces to:
\begin{eqnarray}\label{eq:FPoperator}
\mathcal{C}_{\mbox{\tiny\rm FP}}\left[f_e\right] = \partial_{\gamma}\!\left[S(\chi)f_e\right] + \frac{1}{2}\partial_{\gamma}^2\!\left[R(\chi,\gamma)f_e\right] ,
\end{eqnarray}
the first term being referred to as the {\it drift} term, and the second one as the {\it diffusion} term, and where we have introduced:
\begin{eqnarray}
\label{eq_Sgamma} \!\!\!\!S(\chi) \!&=&\!\! \!\! \int_0^{+\infty}\!\!\!\!\!\! d\gamma_{\gamma} \, \gamma_{\gamma} \, w_{\chi} (\gamma,\gamma_{\gamma}) = \frac{2}{3}\frac{\alpha^2}{\tau_e} \,\chi^2\, g(\chi)\\
\label{eq_Rgamma} \!\!\!\!R(\chi,\gamma) \!&=&\!\! \int_0^{+\infty}\!\!\!\!\!\! d\gamma_{\gamma} \, \gamma_{\gamma}^2 \, w_{\chi} (\gamma,\gamma_{\gamma}) = \frac{2}{3}\frac{\alpha^2}{\tau_e}\,\gamma\, h(\chi), 
\end{eqnarray}
where $h(\chi)=a_2(\chi)$ reads:
\begin{eqnarray}
\nonumber  h(\chi) &=&  \frac{9 \sqrt{3}}{4\pi}\!\!  \int_0^{+\infty}\!\!\!\!\! d\nu \left[ \frac{2 \chi^3\nu^3}{(2+3\nu\chi)^3}\rm{K}_{5/3}(\nu) \right. \\
\label{eq:g}&+& \left. \frac{54\chi^5\nu^4}{(2+3\nu\chi)^5}\rm{K}_{2/3}(\nu) \right] \, .
\end{eqnarray}

Equation~\eqref{eq:Master} rewritten using the operator Eq.~\eqref{eq:FPoperator} is a Fokker-Planck equation.
Mathematically, it is equivalent to the Ito stochastic differential equation for the random process $\gamma(t)$~\cite{KP}:
\begin{eqnarray}
\nonumber mc^2 d\gamma &=& -ec\,({\bf u}\cdot{\bf E})\,dt-mc^2S(\chi) dt \\
\label{eq:stochEnergy}&+& mc^2\sqrt{R(\chi,\gamma)}\,dW\, ,
\end{eqnarray}
together with the equation on the electron momentum direction $\bf\Omega$:
\begin{eqnarray}
\frac{d{\bf \Omega}}{dt} = -\frac{e}{p}(\mathbb{1}-{\bf\Omega}\otimes{\bf\Omega})\cdot({\bf E}+u{\bf\Omega}\times{\bf H})\,.
\end{eqnarray}
Note that Eq.~\eqref{eq:stochEnergy} on the electron energy contains both deterministic (first two terms in its rhs) and stochastic (last term) increments,
the latter being modeled using $dW$, a Wiener process of variance $dt$.
As high-energy photon emission does not modify the direction $\bf\Omega$ of the emitting ultra-relativistic electron, 
the equation on the momentum direction ${\bf \Omega}$ is found to be given by the Lorentz force only.

It follows that the electron momentum satisfies the stochastic differential equation:
\begin{eqnarray}\label{eq:stochMomentum}
\nonumber d{\bf p} &=& -e({\bf E}+{\bf u}\times{\bf H}) dt - mc^2 S(\chi) \,{\bf u}/(c {\bf u}^2) dt\\ 
&+& mc^2\sqrt{R(\chi,\gamma)}\,dW \,{\bf u}/(c {\bf u}^2)\,.
\end{eqnarray}

Derived from the framework of quantum electrodynamics in the limit $\gamma_{\gamma}\ll\gamma$, 
Eqs.~\eqref{eq:stochEnergy} and~\eqref{eq:stochMomentum} are the generalization 
of the purely deterministic equations of motion Eqs.~\eqref{eq_ener} and~\eqref{eq_force} 
derived in the framework of classical electrodynamics (CED).
The first terms in the rhs of Eqs.~\eqref{eq:stochEnergy} and \eqref{eq:stochMomentum} 
correspond to the effect of the Lorentz force. 
The second terms follow from the {\it drift} term of the
Fokker-Planck operator [Eq.~\eqref{eq:FPoperator}] and account for the deterministic effect of radiation
reaction on the electron dynamics. It is important to point out that, from Eq.~\eqref{eq_Sgamma}, we get
\begin{eqnarray}\label{quantum_correction}
P_{\rm rad} = P_{\rm cl} \, g(\chi) = m c^2\,S(\chi) \, .
\end{eqnarray}
The deterministic terms thus turn out to be {\it the leading terms of the LL equation 
including the quantum correction} introduced phenomenologically in Sec.~\ref{sec:qed},
and here rigorously derived from the quantum framework.

Note that this procedure fully and rigorously justifies the use of the quantum corrected LL friction force in Particle-In-Cell codes.

Finally, the last terms in Eqs.~\eqref{eq:stochEnergy} and~\eqref{eq:stochMomentum},
which follow from the {\it diffusion} term of the Fokker-Planck operator [Eq.~\eqref{eq:FPoperator}],
account for the stochastic nature of high-energy photon emission and its back-reaction on the
electron dynamics. It is a purely quantum effect, which is not present in the framework of CED.
As a result, Eqs.~\eqref{eq:stochEnergy} and~\eqref{eq:stochMomentum} extend the validity of
Eqs.~\eqref{eq_ener} and~\eqref{eq_force} from the classical regime of radiation reaction ($\chi \ll 1$) 
to the intermediate quantum regime ($\chi \lesssim 1$) by accounting for both
the deterministic radiation friction force, and the  stochastic nature of radiation emission. 
The domain of validity of this Fokker-Planck description and the extension of validity to the weakly quantum
regime $\chi \lesssim 1$ will now be discussed in more details.\\

Before doing so, however, we want to briefly discuss how our findings fit in with respect to previous works.
In contrast with the work of Elkina {\it et al.}~\cite{Elkina} (in which the Master equation approach is applied 
to the cascades in circularly polarized laser fields), and that by Neitz and Di Piazza~\cite{Neitz} (in which the
authors provide a Fokker-Planck based analytical description of an electron beam colliding with an ultra-relativistic 
light pulse), our approach is more general. No assumption is here done on the electron and field configuration,
which led us to treat the full Vlasov operator in Eq.~\eqref{eq:Master} and allowed us to derive Eq.~\eqref{eq:stochMomentum},
valid for arbitrary geometries.
As we will show in the next Sections, this will allow us to bring new useful insights and predictions, 
and opens new opportunities in the numerical treatment of radiation reaction in arbitrary geometries.

\subsection{Domain of validity of the Fokker-Planck and quantum-corrected Landau-Lifshitz descriptions}\label{sec:stoch:FPValidity}

Let us now study the domain of validity of the previously derived Fokker-Planck and quantum-corrected Landau-Lifshitz descriptions.
We use as a starting point the Kramers-Moyal expansion [Eq.~\eqref{eq:KramersMoyal}] for the {\it collision} operator.
If the high-order moments of the kernel in Eq.~\eqref{eq:KramersMoyal} do not give a proper description of the {\it collision} operator 
unless all accounted for, computing them still allows us to infer 
the limit of validity of the Fokker-Planck and corrected Landau-Lifshitz descriptions, and of Eqs.~\eqref{eq:stochEnergy} 
and~\eqref{eq:stochMomentum}, in particular.

Here we derive an estimate of the relative importance of the successive terms in the Kramers-Moyal expansion~\eqref{eq:KramersMoyal},
by computing: 
\begin{eqnarray}
B^{n+1}_{n} = \frac{n!}{(n+1)!}\frac{\partial_{\gamma}^{n+1}[A_{n+1}f_e]}{\partial_{\gamma}^{n}[A_{n}f_e]} \sim \frac{A_{n+1}}{\gamma(n+1)A_n}\,.
\end{eqnarray}
For the particular kernel Eq.~\eqref{eq:Kernel}, we can use Eq.~\eqref{eq:An}, and the ratio of the $(n+1)^{th}$ to the $n^{th}$ contribution is found to depend on the $\chi$ parameter only:
\begin{eqnarray}\label{estimationKM}
B^{n+1}_{n} = \frac{1}{n+1}\frac{a_{n+1}(\chi)}{a_{n}(\chi)}\,.
\end{eqnarray}

In the limit $\chi \ll 1$, which ensures $\gamma_{\gamma} \ll \gamma$, the previous ratio reduces to $B^{n+1}_{n} \rightarrow b_n \chi$,
where (introducing the Gamma function $\Gamma$)
\begin{eqnarray}
b_n = \frac{3}{n+2}\,\frac{\Gamma(\frac{n}{2}+\frac{2}{3})\Gamma(\frac{n}{2}+\frac{7}{3})}{\Gamma(\frac{n}{2}+\frac{1}{6})\Gamma(\frac{n}{2}+\frac{11}{6})}
\end{eqnarray}
slowly increases with $n$ from $b_1 \simeq 1.07$, up to its asymptotic value ${b_{\infty}=3/2}$ for $n\gg1$.
 This ordering $\propto \chi^{n}$ confirms that the Fokker-Planck expansion~\eqref{eq:FPoperator}  and therefore Eqs.~\eqref{eq:stochEnergy} and~\eqref{eq:stochMomentum} are valid for small $\chi$. 
As $\chi \rightarrow 1$, all terms in the Kramers-Moyal expansion~\eqref{eq:KramersMoyal}
become of the same order, no truncation can be made in the {\it collision} operator, 
and the full linear Boltzmann Eq.~\eqref{eq:Master} needs to be considered. 
It is however well known that this limit also corresponds to the onset of various other QED processes such as
electron-positron pair production, and the present approach is not satisfying anymore.

\begin{figure}
\begin{center}
\includegraphics[width=8cm]{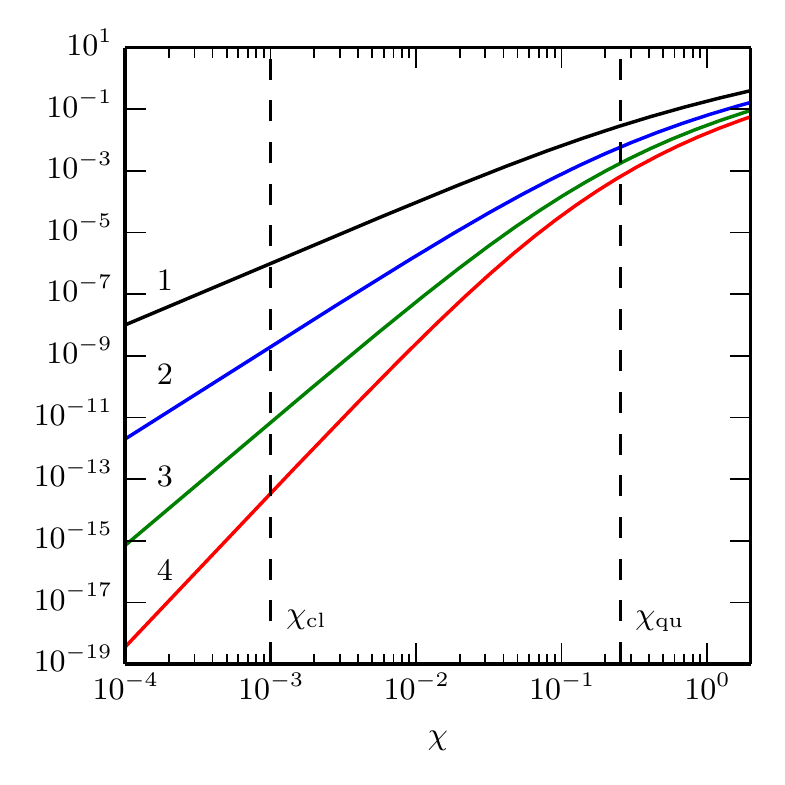}
\caption{Dependence with $\chi$ of the functions $a_n(\chi)/n!$ for $n=1$ to $4$, in black, blue, green and red (respectively). The left vertical line at $\chi = \chi_{\rm{cl}} = 1\times 10^{-3}$ indicates the threshold of the classical to the intermediate regime and the right vertical line at $\chi = \chi_{\rm{qu}} = 2.5\times 10^{-1}$ the limit of the intermediate to the quantum regime (see explicit definitions in the text).} 
\label{fig:stoch:ordering}
\end{center}
\end{figure}

To be more quantitative on the domain of applicability of the Fokker-Planck and quantum-corrected Landau-Lifshitz descriptions, 
a first general  criterion can be obtained by comparing the relative importance of the various terms in the Kramers-Moyal expansion as given by the approximated expression Eq.~\eqref{estimationKM}. 
For this reason,
we have plotted in Fig.~\ref{fig:stoch:ordering} the functions $a_n(\chi)/n!$ for $n=1$ to 4.
The classical regime of radiation reaction can now be defined as the region $\chi<\chi_{\rm cl} \simeq 1\times 10^{-3}$ 
for which the second term in the Kramers-Moyal expansion (diffusion term) is at least three order of magnitude 
below the first (drift) term.
The intermediate quantum regime is defined as the region $\chi_{\rm cl}<\chi<\chi_{\rm qu} \simeq 2.5\times 10^{-1}$
where the diffusion term contribution is not negligible but the higher-order terms in the Kramers-Moyal expansion are.
Finally, the quantum regime is the region $\chi > \chi_{\rm qu}$ for which the third term in the Kramers-Moyal expansion becomes larger than a tenth of the diffusion term. 

This criterion is based on a statistical description of the 
system and is consistent with the one proposed in the literature,
which is solely based on the argument that the recoil of a single
emitting photon can be neglected.
This first criterion, that depends only on the quantum
parameter $\chi$, gives a first approximation for the validity 
of the deterministic (quantum corrected LL) or FP description when 
considering an arbitrary electron distribution function. In the
next Section~\ref{sec:averages:domainval}, we will derive a more 
general criterion for the domain
of validity of the different models that accounts for the 
general energy distribution of an arbitrary distribution function
by analyzing into more details the temporal evolution of its
successive moments.

\section{Temporal evolution of integrated quantities}\label{sec:averages}

In this Section, we discuss the temporal evolution of the 
successive moments of the electron distribution function
as inferred from the different descriptions discussed in Sec.~\ref{sec:stoch}. 
These are defined as 
\begin{eqnarray}
\langle \gamma \rangle_{\alpha}(t) = \frac{\int\!d^3x\,d^2\Omega\,d\gamma\, \gamma f_e}{\int\!d^3x\,d^2\Omega\,d\gamma\,f_e} \, ,
\end{eqnarray}
and for $n > 1$, 
\begin{eqnarray}
\mu_{n}\vert_{\alpha}(t) = \frac{\int\!d^3x\,d^2\Omega\,d\gamma\, (\gamma - \langle \gamma \rangle_{\alpha}(t))^n f_e}{\int\!d^3x\,d^2\Omega\,d\gamma\,f_e} \, .
\end{eqnarray}
where the index $\alpha$ indicates the method used to compute  the distribution function [$\alpha={\mbox{\rm cLL}}, {\rm FP}, {\rm MC}$  for the deterministic
 (quantum corrected Landau-Lifshitz friction force),  Fokker-Planck and linear Boltzmann (Monte-Carlo) approach, respectively].
 The deterministic description here corresponds to 
Eq.~\eqref{eq:Master} accounting for the drift term only in the {\it collision} operator that is equivalent to the LL formalism 
including the quantum correction $g(\chi)$ in the radiated power, hereafter indicated as "corrected Landau-Lifshitz" (cLL). 
The Fokker-Planck description corresponds to Eq.~\eqref{eq:Master} with the collision operator given by Eq.~\eqref{eq:FPoperator},
and the linear Boltzmann description corresponds to Eq.~\eqref{eq:Master} with the full collision operator. 

\subsection{Energy moments of the collision operators}\label{sec:averages:CollOpMoments}

Let us start by discussing briefly the first moments, in energy, of the {\it collision} operators.
Computing the order $0$ and $1$ moments for all three descriptions leads to:
\begin{eqnarray}
\int\!d^2\Omega\! \int_{1}^{+\infty}\!\!\!\!\!\!\!\! d\gamma\,\mathcal{C}_{\alpha}[\hat{f}_e] &=& 0\,,\\
\int\!d^2\Omega\! \int_{1}^{+\infty}\!\!\!\!\!\!\!\! d\gamma\, \gamma\, \mathcal{C}_{\alpha}[\hat{f}_e] &=& \overline{S(\chi)}_{\alpha}\,,
\end{eqnarray}
where $\overline{q}_{\alpha}(t,{\bf x)} = \int\!d^2\Omega\,d\gamma\,q\,\hat{f}_e$ denotes for a given quantity $q$ its local average 
over the normalized electron distribution function $\hat{f}_e=f_e/n_e$ taken at a time $t$ and position $\bf x$, 
with $n_e(t,{\bf x}) =  \int\!d^2\Omega\,d\gamma\,f_e$ the electron density at this time and position. 

Concerning the second order moment of the {\it collision} operators, only the linear Boltzmann and Fokker-Planck methods give a similar form:
\begin{eqnarray}
\!\int\!\!d^2\Omega\,d\gamma \big(\gamma\!-\!\overline{\gamma}_{\alpha}\big)^2\mathcal{C}_{\alpha}[\hat{f}_e] \!=\!- 2\overline{ (\gamma\!-\!\overline{\gamma}_{\alpha}) S(\chi) }_{\alpha}\!+\!\overline{ R(\chi,\gamma)}_{\alpha}\,\,\,
\end{eqnarray}
while the deterministic description simply leads to:
\begin{eqnarray}
\int\!\!d^2\Omega\,d\gamma \big(\gamma\!-\!\overline{\gamma}_{\mbox{\rm\tiny cLL}}\big)^2\mathcal{C}_{\mbox{\rm\tiny cLL}}[\hat{f}_e] \!=\!- 2\overline{ (\gamma\!-\!\overline{\gamma}_{\mbox{\rm\tiny cLL}}) S(\chi) }_{\mbox{\rm\tiny cLL}}\,.
\end{eqnarray}

The third order moment is the first for which all three descriptions lead to different equations of evolution:
\begin{eqnarray}\label{eq:thirdMomentCollOp}
&&\int\!d^2\Omega\,d\gamma\, \big(\gamma\!-\!\overline{\gamma}_{\alpha}\big)^3\,\mathcal{C}_{\alpha}[\hat{f}_e] =\\
&=&\left\lbrace
\begin{array}{lcc}
\nonumber \!\!\! -3 \overline{(\gamma-\overline{\gamma}_{\alpha})^2\,S(\chi)}_{\alpha}  \hfill\,\,\alpha={\mbox{\tiny cLL}}\,\,\\ \\ 
\nonumber \!\!\!-3\overline{(\gamma-\overline{\gamma}_{\alpha})^2\,S(\chi)}_{\alpha}+3\overline{(\gamma-\overline{\gamma}_{\alpha})R(\chi,\gamma)}_{\alpha} \hfill \,\,\alpha={\rm FP}\\ \\
\nonumber \!\!\!-3\overline{(\gamma-\overline{\gamma}_{\alpha})^2\,S(\chi)}_{\alpha}+3\overline{(\gamma-\overline{\gamma}_{\alpha})R(\chi,\gamma)}_{\alpha} \\
\nonumber \!\!\!-\overline{A_3(\chi,\gamma)}_{\alpha} \hfill\,\,\alpha={\rm MC}
\end{array}\right.  
\end{eqnarray}

Finally, the $n^{th}$ order moment of the {\it collision} operator for any order $n$ reads:
\begin{eqnarray}\label{eq:nthMomentCollOp}
&& \int\!d^2\Omega\,d\gamma\, \big(\gamma\!-\!\overline{\gamma}_{\alpha}\big)^n\,\mathcal{C}_{\alpha}[\hat{f}_e] =\\
&=&\left\lbrace
\begin{array}{lcc}
\nonumber - n \overline{(\gamma-\overline{\gamma}_{\alpha})^{n-1}\,S(\chi)}_{\alpha}  \hfill\,\,\alpha={\mbox{\tiny cLL}}\,\,\\ \\
\nonumber - n\overline{(\gamma-\overline{\gamma}_{\alpha})^{n-1}\,S(\chi)}_{\alpha} \\
+\frac{n}{2} (n-1)\overline{(\gamma-\overline{\gamma}_{\alpha})^{n-2}\,R(\chi,\gamma)}_{\alpha}\,\, \hfill \,\,\alpha={\rm FP}\\ \\
\nonumber \sum_{k = 1}^n (-1)^k \binom{n}{k}  \overline{(\gamma-\overline{\gamma}_{\alpha})^{n-k}\,A_k(\chi,\gamma})_{\alpha}  \\
\nonumber  \hfill\,\,\,\,\alpha={\rm MC}
\end{array}\right.  
\end{eqnarray}
and we can verify that for $n \ge 3$, the three description give different results.

Note that in the complete (MC) description, the evolution of the energy momentum of order $n$ involves the first $n^{th}$ moments associated to the kernel of the Kramers-Moyal expansion, so that only the equation for the first two moments are formally the same for the FP and MC description.

In Appendix~\ref{app:Conservation}, the first two moments of the {\it collision} operators are used to demonstrate that
all three descriptions correctly conserve both the number of electrons and the total energy in the system. 
In what follows, we will further use these results to discuss the temporal evolution of the successive moments 
of the electron distribution.

\subsection{Temporal evolution of average quantities}\label{sec:averages:timeEvolution}

In this Section, we first give the equations of evolution for the successive moments of the electron distribution function and then discuss their implications on both
the different (cLL, FP, MC) descriptions, and the physics they describe.

\subsubsection{Equations of evolution}

Let us first discuss the temporal evolution of the electron average energy.
Dividing Eq.~\eqref{eq:TotalEnergy} by the total (constant) number of electrons $N_e=\int\!d^3x\,n_e(t,{\bf x})$ (see Appendix~\ref{app:Conservation}), 
one finds that all three descriptions lead to the same equation of evolution for the average electron energy:
\begin{eqnarray}\label{eq:eonMeanEnergy}
mc^2 \frac{d\langle\gamma\rangle_{\alpha}}{dt} = -e c\,\langle{\bf u}\!\cdot\!{\bf E}\rangle_{\alpha} - mc^2\,\langle S(\chi)\rangle_{\alpha}\,,
\end{eqnarray}
where $\langle q\rangle_{\alpha}(t) = \int\!d^3x\,d^2\Omega\,d\gamma\,q f_e/\int\!d^3x\,d^2\Omega\,d\gamma\,f_e$ 
stands as the total (including spatial) average of $q$ over the distribution function.
The first term in the rhs of Eq.~\eqref{eq:eonMeanEnergy} stands for the average work rate of the external field,
and the second term denotes the power radiated away averaged over the whole distribution function.

It is important to stress that the distribution functions considering the different descriptions are not necessarily the same.
Therefore, Eq.~\eqref{eq:eonMeanEnergy} does not, in general, predicts that all three approaches will give similar results on the average electron energy.
We will later quantify this difference in Section~\ref{sec:ElectronMeanEnergy}.\\

Let us now turn to the derivation of the equation of evolution for the variance in energy. 
To do so, we focus on the radiation reaction effect and will
thus neglect the effect of the external field on the energy 
dispersion which cannot be treated for arbitrary configurations~\cite{footnote4}.
This is justified in particular for the case of an electron
population evolving in a constant magnetic field or interacting
with a plane wave. In those cases indeed, the additional Vlasov
term is either zero or negligible and the numerical simulations
presented in Sec.~\ref{sec:numResults} will show very good agreement with the
present analysis.

The equations of evolution of the energy variance are obtained by multiplying the Master equations for the electron distribution by $(\gamma - \langle \gamma \rangle)^2$,
and integrating over $\gamma$, $\bf\Omega$ and space. 
In contrast with the previous case (mean energy), only the linear Boltzmann  and Fokker-Planck descriptions formally give
 the same equation for the time evolution of 
$\sigma_{\gamma}^2 = \left\langle(\gamma - \langle \gamma \rangle)^2\right\rangle_{\alpha}$ [here $\alpha={\rm FP}, {\rm MC}$]:
\begin{eqnarray}\label{eq:variance}
\frac{d\sigma_{\gamma}^2}{dt} \Big \vert_{\alpha} &=& \langle R(\chi,\gamma)\rangle_{\alpha} - 2\,\big\langle (\gamma -\langle \gamma \rangle_{\alpha} )S(\chi)\big\rangle_{\alpha}  \, ,
\end{eqnarray}
while the deterministic description gives:
\begin{eqnarray}\label{eq:variancecLL}
\frac{d\sigma_{\gamma}^2}{dt} \Big \vert_{\mbox{\tiny cLL}} &=& - 2\,\big\langle (\gamma -\langle \gamma \rangle_{\mbox{\tiny cLL}} )S(\chi)\big\rangle_{\mbox{\tiny cLL}}   \, .
\end{eqnarray}
Present in all three descriptions, the term $- 2\,\langle (\gamma -\langle \gamma \rangle )S(\chi)\rangle_{\alpha}$ is, 
in most cases~\cite{footnote5}, negative since high-energy photon emission and its back-reaction
is dominated by electrons at the highest energies.
It will therefore lead to a decrease of $\sigma_{\gamma}$, i.e. to a cooling of the electron population.

In contrast, the term $\langle R(\chi,\gamma)\rangle$, which pertains to the stochastic nature of high-energy photon emission in the QED framework, 
is a purely quantum term, and as such is absent from the deterministic description.
This quantum term is always positive and leads to a spreading of the energy distribution, i.e. to an effective heating of the electron population. 

In the following, we will further discuss the relative importance of the deterministic (cooling) and quantum (energy spreading/heating) terms and their impact on the 
electron population.\\

Using Eq.~\eqref{eq:thirdMomentCollOp} for the third moment of the {\it collision} operators, an equation of evolution
for $\mu_3(t) = \langle (\gamma-\langle\gamma\rangle_{\alpha})^3 \rangle_{\alpha}$ can be derived. 
As the three descriptions  lead to a  different form for the third moment of the {\it collision} operator, they will also lead to 
different equations of evolution for $\mu_3$ (note that the contribution from the Vlasov operator~\cite{footnote4} is not considered, 
and one focuses on the radiation reaction contribution only).
The equation of evolution in the deterministic description is given by 
\begin{eqnarray}\label{eqCLmu3}
\nonumber\!\!\frac{d \mu_3}{dt} \Big \vert_{\mbox{\tiny cLL}}\!\!\!\!\!\!\!\! &=& 3 \left\langle S(\chi)\right\rangle\!\left\langle(\gamma-\langle\gamma\rangle)^2\right\rangle_{\mbox{\tiny cLL}} \!\!- 3 \big\langle (\gamma -\langle \gamma \rangle_{_{\rm{cLL}}} )^2 S(\chi)\big\rangle_{_{\rm{cLL}}},\\ 
\end{eqnarray}
in the Fokker-Planck description by
\begin{eqnarray}\label{eqFPmu3}
\nonumber \!\! \frac{d \mu_3}{dt} \Big \vert_{\mbox{\tiny FP}} \!\!\!\!&=& 3 \left\langle S(\chi)\right\rangle\,\left\langle(\gamma-\langle\gamma\rangle)^2\right\rangle_{\mbox{\tiny FP}} \!\!- 3 \big\langle (\gamma -\langle \gamma \rangle_{\mbox{\tiny FP}} )^2 S(\chi)\big\rangle_{\mbox{\tiny FP}} \\
&+& 3 \big\langle (\gamma -\langle \gamma \rangle_{\mbox{\tiny FP}} ) R(\chi, \gamma)\big\rangle_{\mbox{\tiny FP}}\,, 
\end{eqnarray}
and in the linear Boltzmann (MC) description by
\begin{eqnarray}\label{eqMCmu3}
\nonumber \!\!\frac{d \mu_3}{dt} \Big \vert_{\mbox{\tiny MC}} \!\!\!\!\!\!\!\!&=& 3 \left\langle S(\chi)\right\rangle\,\left\langle(\gamma-\langle\gamma\rangle)^2\right\rangle_{\mbox{\tiny MC}}- 3 \big\langle (\gamma -\langle \gamma \rangle_{\mbox{\tiny MC}} )^2 S(\chi)\big\rangle_{\mbox{\tiny MC}} \\
\!\!\!\!\!  \label{eq:mu3_MC}&+& 3 \big\langle (\gamma -\langle \gamma \rangle_{\mbox{\tiny MC}} ) R(\chi, \gamma)\big\rangle_{\mbox{\tiny MC}}
\!\!\!\!\! -\langle A_3(\chi,\gamma) \rangle_{\mbox{\tiny MC}}.
\end{eqnarray}
As will be further discussed in Sec.~\ref{sec:averages:mu3Discussion}, this third order moment relates to the skewness (asymmetry) of the electron distribution function.
In particular, we will show that it provides a link to the so-called {\it quenching} of radiation losses, as introduced and discussed in Ref.~\cite{harvey2017}.

Finally, for any order $n$, we get for the $n^{th}$ moment $\mu_{n}\vert_{\alpha}(t) = \left\langle (\gamma-\langle\gamma\rangle_{\alpha})^n \right\rangle_{\alpha}$:
\begin{eqnarray}\label{eq:mun}
 \frac{d \mu_{n}}{dt}\Big\vert_{\mbox{\tiny cLL}} &=&  n \left\langle S(\chi) \right\rangle_{\mbox{\tiny cLL}}\,\mu_{n-1}\vert_{\mbox{\tiny cLL}} \\
\nonumber &-&n \left\langle (\gamma - \langle \gamma \rangle_{\mbox{\tiny MC}})^{n-1} S(\chi) \right\rangle_{\mbox{\tiny cLL}} \, , \\
\frac{d \mu_{n}}{dt}\Big\vert_{\mbox{\tiny FP}} &=& n \left\langle S(\chi) \right\rangle_{\mbox{\tiny FP}}\,\mu_{n-1}\vert_{\mbox{\tiny FP}} \\
\nonumber &-& n \left\langle (\gamma - \langle \gamma \rangle_{\mbox{\tiny FP}})^{n-1} S(\chi) \right\rangle_{\mbox{\tiny FP}} \\
\nonumber &+&\!\! \frac{n(n-1)}{2} \left\langle (\gamma - \langle \gamma \rangle_{\mbox{\tiny FP}})^{n-2} R(\chi,\gamma) \right\rangle_{\mbox{\tiny FP}}\!, \\
\frac{d \mu_{n}}{dt}\Big\vert_{\mbox{\tiny MC}} &=& n \left\langle S(\chi) \right\rangle_{\mbox{\tiny MC}}\,\mu_{n-1}\vert_{\mbox{\tiny MC}} \\
\nonumber&+&\sum_{k = 0}^{n-1} (-1)^{n-k}\!\binom{n}{k}\!\!  \left\langle (\gamma - \langle \gamma \rangle_{\mbox{\tiny MC}})^{k} A_{n-k}(\chi, \gamma) \right\rangle_{\mbox{\tiny MC}} .
\end{eqnarray}

Before investigating into more details the evolution of the first two moments, let us discuss more closely these equation of evolution
for an arbitrary order $n$.

At any order $n$, the purely deterministic term $\propto S(\chi)$ comes in the form 
$-n \langle (\gamma - \langle \gamma \rangle)^{n-1} S(\chi) \rangle$,
while the $n^{th}$-order moment of the kernel $w_{\chi}$ comes in the form
$(-1)^n \langle A_n(\chi,\gamma) \rangle$. 
This implies that, while the purely deterministic term always tends to decrease any moment of the electron distribution function 
(except in special geometries where the external field would take small enough values whenever $\gamma > \langle \gamma \rangle$, see also~\cite{footnote5}), 
the purely quantum term $\propto A_n(\chi,\gamma)$ tends to increase even moments and decrease odd moments. Concerning the 
other terms, it is in general difficult to state the behavior of 
the $n^{th}$ moment since it will involve all moments from $0$ to $n$ [see Eq.~\eqref{eq:approx_mun} in 
Appendix~\ref{app:ApproximatedEqEvolutionMoments}].

\subsubsection{Electron mean energy}\label{sec:ElectronMeanEnergy}

Let us now further discuss the temporal evolution of the electron mean energy as inferred from all three descriptions.
As previously discussed, we will not consider here the additional Vlasov term and focus on the effect of radiation reaction~\cite{footnote4}.

Obviously, the fact that the quantum-corrected leading term of the LL friction force naturally appears by taking 
the FP limit of the linear Boltzmann description already leads us to stress that, whenever the diffusion term [in Eq.~\eqref{eq:stochMomentum}] can be neglected,
all three descriptions will lead to the same evolution of the distribution function, and therefore to the same mean energy prediction [as seen from Eq.~\eqref{eq:eonMeanEnergy}].
This can be easily understood as all three models have been proved to share the same deterministic (drift) term.

Yet, even in a regime where the diffusion term is not negligible and where the three models result in sensibly different distribution functions,
simulations relying on either the quantum-corrected friction force only or full MC simulations have been shown to lead to similar predictions 
on the electron mean energy (see, e.g., Ref.\cite{ridgers2014}). In Sec.~\ref{sec:numResults}, we will actually see that all three approaches
lead to very similar predictions on the electron mean energy even when the overall electron distribution is very different from one model to another.
In what follows, we explain this result.

To do so, we need to quantify the difference on the mean electron energy predictions by the different approaches.
Our first step is to formally expand $S(\chi)$ around the average value $\langle\chi\rangle_{\alpha}$ in  the last term of the rhs of Eq.~\eqref{eq:eonMeanEnergy}
\begin{eqnarray}\label{eq:expansion}
\langle S(\chi)\rangle_{\alpha} \simeq S\left(\langle\chi\rangle_{\alpha}\right)+ \frac{1}{2} \sigma_{\chi}^2\,S''\!\left(\langle \chi \rangle_{\alpha}\right) \,,
\end{eqnarray}
where $S''\!(\chi)$ is the second derivative of $S(\chi)$ with respect to $\chi$, and
\begin{eqnarray}
\sigma^2_{\chi}=\big\langle (\chi-\langle\chi\rangle_{\alpha})^2 \big\rangle_{\alpha}
\end{eqnarray}
measures the variance of the distribution in $\chi$ of the electron population. 
From this, one expects all three descriptions to predict similar average electron energies whenever the first term in Eq.~\eqref{eq:expansion} dominates.
 
\begin{figure}
\begin{center}
\includegraphics[width=8cm]{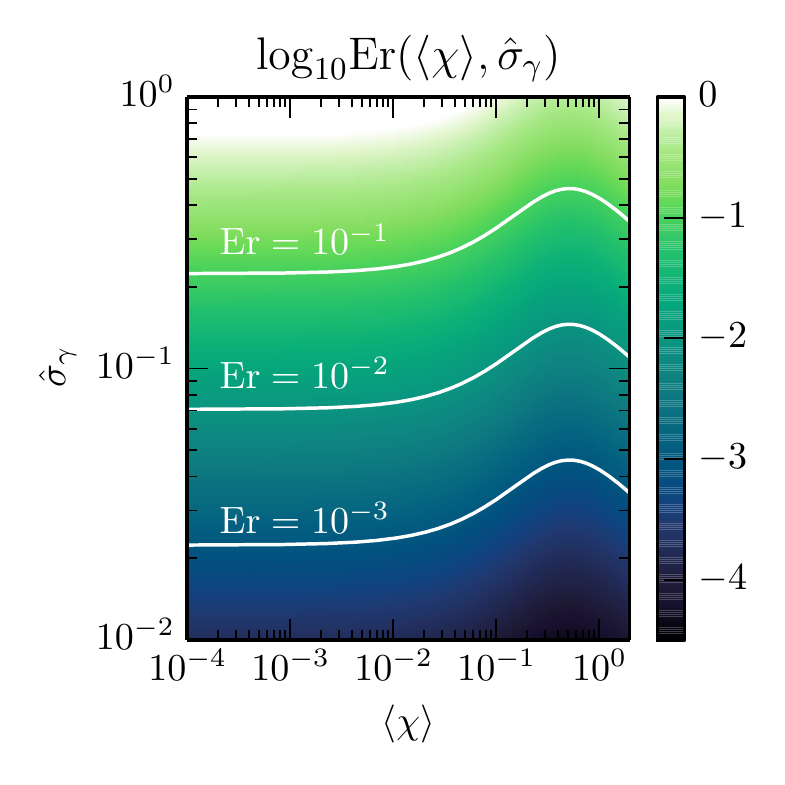}
\caption{Dependence with $\langle \chi \rangle$ and $\hat{\sigma}_{\gamma}$ of $\rm{Er}(\langle \chi \rangle, \hat{\sigma}_{\gamma})$ which represents the relative difference between $d_t \langle \gamma \rangle_{\rm{MC}}$ and $d_t \langle \gamma \rangle_{\rm{cLL}}$. The curve $\rm{Er} = 10^{-3}, 10^{-2}$ and $10^{-1}$ are plotted in white lines.} 
\label{fig:averages:errorMeanEnergyMap}
\end{center}
\end{figure}

Then, we introduce the error on the rate of change of the electron energy:
\begin{eqnarray}\label{eq:maxErrorEnergyRate}
\nonumber&&\!\!\!\!\frac{d_t \langle \gamma \rangle_{\rm{MC}} - d_t \langle \gamma \rangle}{d_t \langle \gamma \rangle}
\simeq \frac{F^2 \hat{\sigma}_{\gamma}^2 \langle \chi \rangle^2 \widetilde{S}''(\langle \chi \rangle)}{\widetilde{S}(\langle \chi \rangle)} \equiv \widetilde{\rm{Er}}(\langle \chi \rangle, \hat{\sigma}_{\gamma},F) \, ,\\
\end{eqnarray}
where we have introduced $F = \hat{\sigma}_{\chi}/\hat{\sigma}_{\gamma}$, 
$\hat{\sigma}_{\gamma}=\sigma_{\gamma}/\langle\gamma\rangle$ and $\hat{\sigma}_{\chi}=\sigma_{\chi}/\langle\chi\rangle$. 
Considering situations where all particles radiate in a similar field (e.g. localized electron bunch and/or uniform electromagnetic field),
$\hat{\sigma}_{\chi} \sim \hat{\sigma}_{\gamma}$ ($F = 1$) and we introduce 
$\rm{Er}(\langle \chi \rangle, \hat{\sigma}_{\gamma}) \equiv \widetilde{\rm{Er}}(\langle \chi \rangle, \hat{\sigma}_{\gamma},1)$ 
that now depends only on $\langle\chi\rangle$ and $\hat{\sigma}_{\gamma}$, as presented in Fig.~\ref{fig:averages:errorMeanEnergyMap}.
We can see that, for a given $\hat{\sigma}_{\gamma}$, $\rm{Er}(\langle \chi \rangle, \hat{\sigma}_{\gamma})$ is only weakly dependent on $\langle \chi \rangle$,
while it depends more strongly on $\hat{\sigma}_{\gamma}$ at fixed $\langle\chi\rangle$.  

\begin{figure}
\begin{center}
\includegraphics[width=8cm]{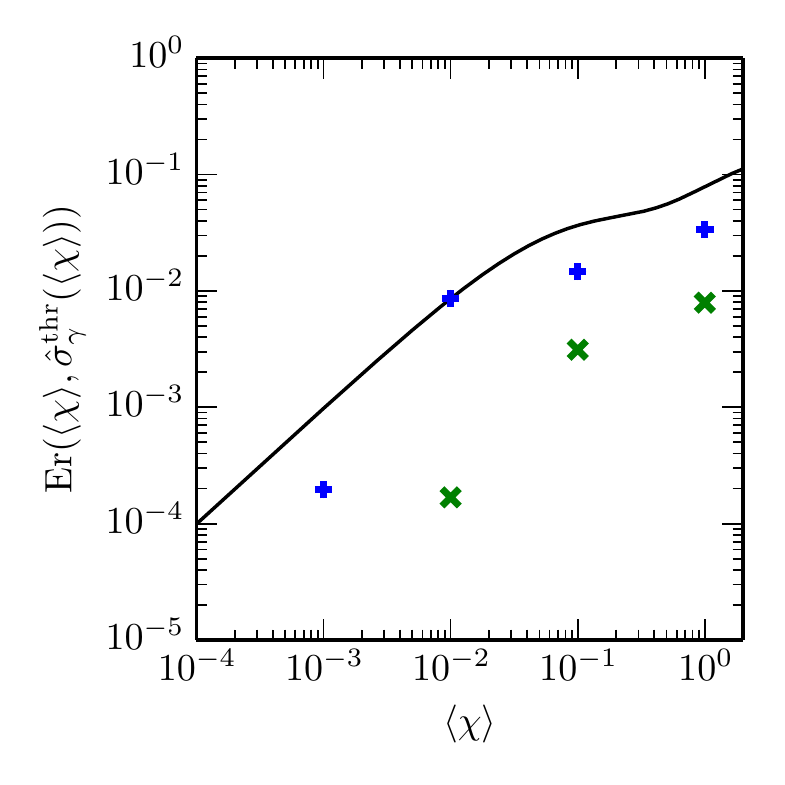}
\caption{Dependence with $\langle \chi \rangle$ of $\rm{Er}(\langle \chi \rangle, \sigma_{\gamma}^{\rm thr})$, which represents the relative difference between $d_t \langle \gamma \rangle_{\rm{MC}}$ and $d_t \langle \gamma \rangle_{\rm{cLL}}$ using $\hat{\sigma}_{\gamma}^{\rm thr}(\langle\chi\rangle)$. The green and blue crosses represent the value of $(\langle \gamma \rangle_{\rm{MC}}(t_{\rm{heat}}) - \langle \gamma \rangle_{\rm{cLL}}(t_{\rm{heat}}))/\langle \gamma \rangle_{\rm{cLL}}(t_{\rm{heat}})$ for the plane-wave field and the constant-uniform magnetic field (respectively) and for $\chi_0 = 10^{-2}, 10^{-1}$ and $1$, $t_{\rm{heat}}$ being the time at which $\sigma_{\gamma}$ stops to increase.} 
\label{fig:averages:errorMeanEnergy}
\end{center}
\end{figure}

In next Sec.~\ref{sec:averages:varianceDiscussion}, we will show that whenever the {\it initial} electron distribution is such that $\hat{\sigma}_{\gamma} < \hat{\sigma}_{\gamma}^{\rm thr}$,
with $\hat{\sigma}_{\gamma}^{\rm thr}$ a threshold energy dispersion [given by Eq.~\eqref{eq:sigmaThreshold}] that depends only on $\langle\chi\rangle$, 
the energy dispersion of the electron distribution can increase up to, but never exceed $\hat{\sigma}_{\gamma}^{\rm thr}$.
Replacing $\hat{\sigma}_{\gamma}$ by this threshold value $\hat{\sigma}_{\gamma}^{\rm thr}$ in our previous estimate
thus provides us with an estimate of the error on the rate of change Eq.~\eqref{eq:maxErrorEnergyRate} that depends on $\langle\chi\rangle$ only.
It is plotted (solid line) in Fig.~\ref{fig:averages:errorMeanEnergy} for mean quantum parameters
in the range $10^{-3} < \langle\chi\rangle < 2$, and does not exceed a few percents (at $\langle\chi\rangle \simeq 1$).
It increases with $\langle\chi\rangle$ as the threshold energy dispersion increases with $\langle\chi\rangle$.
We also report in Fig.~\ref{fig:averages:errorMeanEnergy} the relative discrepancy (at the end of the simulation) in between Monte-Carlo 
and deterministic modeling, measured as $\Delta\gamma = (\langle\gamma\rangle_{\mbox{\tiny MC}}-\langle\gamma\rangle_{\mbox{\tiny cLL}})/\gamma_0$ [with $\gamma_0=\langle\gamma\rangle(t=0)$].
This discrepancy follows the same trend as $\rm Er$ (solid line) and the latter is found to provide an upper-bound to $\Delta\gamma$.
For the two cases presented here [initially narrow electron bunch interacting with a constant magnetic field (blue crosses) and 
linearly polarized plane-wave (green crosses)], the two methods predict similar average energies with a relative discrepancy 
of a few percents, maximum at initially large quantum parameters.

\subsubsection{Variance in energy: radiative cooling vs energy spreading}\label{sec:averages:varianceDiscussion}

We now get back to the equation of evolution of the variance $\sigma_{\gamma}^2$ and discuss the relative importance of radiative cooling 
and `stochastic' energy spreading. To do so, we rewrite (exactly) Eq.~\eqref{eq:variance} in the form
 \begin{eqnarray}\label{eq:variance2}
\nonumber \left(\frac{3\tau_e}{2\alpha^2}\right)\frac{d\sigma_{\gamma}^2}{dt}  &=& \langle \gamma \rangle_{\alpha} \langle h(\chi)\rangle_{\alpha} \\
&-& \,\big\langle (\gamma -\langle \gamma \rangle_{\alpha} )(2\widetilde{S}(\chi) - h(\chi))\big\rangle_{\alpha},\,\,
\end{eqnarray}
where $\widetilde{S}(\chi)=\chi^2\,g(\chi)$.
There are now two possible situations : either (i) the energy distribution of the electron population is initially broad and the standard deviation
 $\sigma_{\gamma}$ is of the same
 order than the average energy, or (ii) it is initially narrow and $\sigma_{\gamma}$ is small with respect to the average energy. 

In the first case, the second term in Eq.~\eqref{eq:variance2} will be dominant at all times, and since for $\chi \lesssim 1$, $2\widetilde{S}(\chi) - h(\chi) > 0$ there will be cooling of the electron population 
even for initially large values of $\chi$.
In the second case, the first term in Eq.~\eqref{eq:variance2} dominates and results in an energy spreading of the electron population. 
As the variance increases, so does the second term that will eventually become dominant:
a phase of cooling will then take place.\\

\begin{figure}
\begin{center}
\includegraphics[width=8cm]{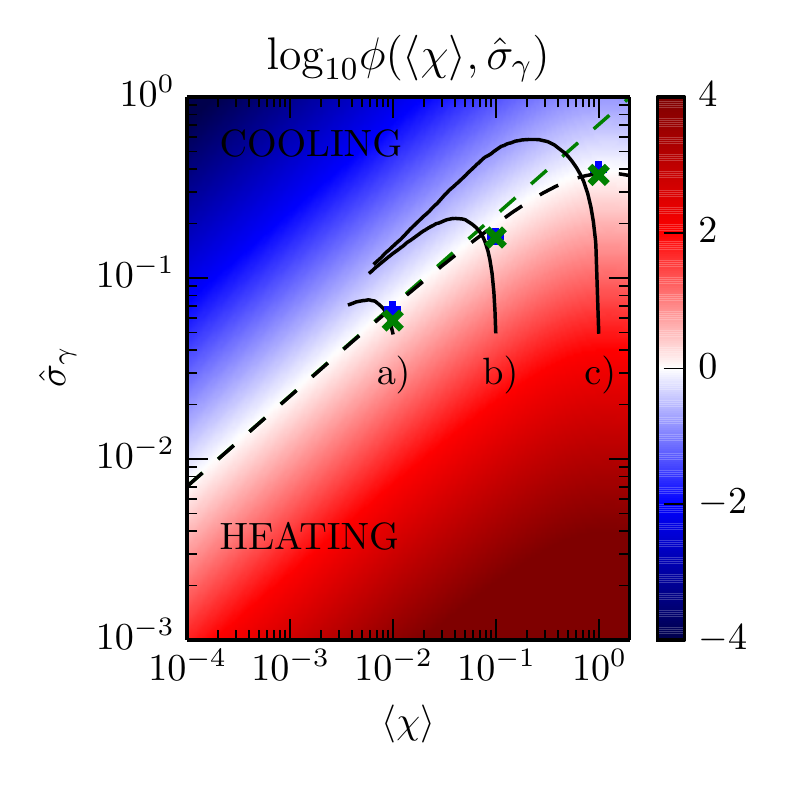}
\caption{Dependence with $\langle \chi \rangle$ and $\hat{\sigma}_{\gamma}$ of $\phi(\langle \chi \rangle, \hat{\sigma}_{\gamma})$. When $\phi > 1$, the electron population is predicted to experience energy-spreading (heating), while it is predicted to cool down when $\phi < 1$. The black dashed line corresponds to $\phi = 1$ and represents the threshold $\hat{\sigma}^{\rm{thr}}_{\gamma}(\langle \chi \rangle)$ between the regions of energy spreading (heating) and cooling. The green dashed line shows the first order expansion in $\langle \chi \rangle$ of the previous equation and corresponds to the prediction of Ref.~\cite{vranic2016}. 
The black lines represent the trajectories $\hat{\sigma}_{\gamma}(\langle \chi \rangle)$ for the interaction of an ultra-relativistic electron bunch
 with different constant-uniform magnetic fields corresponding to a) $\chi_0 = 10^{-2}$, b)~$\chi_0 = 10^{-1}$ and c) $\chi_0 = 1$ (the corresponding simulations are discussed in Sec.~\ref{sec:numResults}). The green and blue crosses represent the value of $\hat{\sigma}_{\gamma}$ extracted from the simulations considering the plane-wave field and the constant-uniform magnetic field (respectively) for $\chi_0 = 10^{-2}, 10^{-1}$ and $1$.} 
\label{fig:averages:sigmamax}
\end{center}
\end{figure}

To be more quantitative, let us consider the latter case ($\sigma_{\gamma} \ll \langle \gamma\rangle$) in more details. 
Expanding  Eq.~\eqref{eq:variance} at first order in $\chi$  around $\langle \chi \rangle_{\alpha}$, we get:
 \begin{eqnarray}
\left(\frac{3\tau_e}{2\alpha^2}\right)\frac{d\sigma_{\gamma}^2}{dt}  &\simeq& \langle \gamma \rangle_{\alpha} h(\langle \chi \rangle_{\alpha}) \\
\nonumber&-& \rm{Cov}(\gamma,\chi)_{\alpha} \big[ 2\widetilde{S}'(\langle \chi \rangle_{\alpha}) - h'(\langle \chi \rangle_{\alpha})\big] \, ,
\end{eqnarray}
where $\rm{Cov}(\gamma, \chi)_{\alpha} = \langle (\gamma - \langle \gamma \rangle_{\alpha})(\chi - \langle \chi \rangle_{\alpha}) \rangle_{\alpha}$.

Whether one should expect heating or cooling depends on the sign of the rhs of the previous equation.
In particular, heating is expected whenever this rhs is positive, which arises for
\begin{eqnarray}\label{eq:mg}
\frac{\langle\gamma\rangle_{\alpha}\,h(\langle \chi \rangle_{\alpha})}{\rm{Cov}(\gamma,\chi)_{\alpha} \big[ 2\widetilde{S}'(\langle \chi \rangle_{\alpha}) - h'(\langle \chi \rangle_{\alpha})\big]} > 1,
\end{eqnarray}
with $2\widetilde{S}'(\chi) - h'(\chi)>0$ whenever $\chi\lesssim 1$.

Introducing $F=\hat{\sigma}_{\chi}/\hat{\sigma}_{\gamma}$ and the correlation factor $\rho_{\gamma\chi} = \rm{Cov}(\gamma, \chi)/(\sigma_{\gamma} \sigma_{\chi})$,
where both coefficients $F$ and $\rho_{\chi \gamma}$ depend on the geometry of the interaction, one can rewrite
 $\rm{Cov}(\gamma, \chi)_{\alpha} = \rho_{\gamma \chi} F \langle \chi \rangle \langle \gamma \rangle \hat{\sigma}_{\gamma}^2$.
 The previous condition Eq.~\eqref{eq:mg} can then be rewritten in the form:
\begin{eqnarray}\label{eq:mg2}
\nonumber\tilde{\phi}\big(\langle\chi\rangle_{\alpha},\hat{\sigma}_{\gamma}, F, \rho_{\gamma\chi}\big)\!\!\! &=&\!\!\! \frac{h(\langle\chi\rangle_{\alpha})/\langle \chi \rangle_{\alpha}}{\rho_{\gamma\chi} F\,\hat{\sigma}_{\gamma}^2\,\big[ 2\widetilde{S}'(\langle \chi \rangle_{\alpha}) - h'(\langle \chi \rangle_{\alpha})\big]} > 1\,, \\
& &
\end{eqnarray}
with $\hat{\sigma}_{q} = \sigma_{q}/\langle q \rangle_{\alpha}$.

This new functional parameter $\tilde{\phi}$ allows to account for the overall properties (mean energy and energy spread) of the electron population, 
and shows that electron heating is {\it not only} correlated to large values of the quantum parameter $\chi$, 
as will be further demonstrated in Sec.~\ref{sec:numResults}.

Let us now focus on the case where all particles radiate in a similar external field 
(e.g. a localized electron bunch and/or uniform external field) for which  
$\hat{\sigma}_{\chi} \simeq \hat{\sigma}_{\gamma}$ ($F \simeq 1$) and $\rho_{\gamma\chi} \simeq 1$.
Equation~\eqref{eq:mg2} then reduces to:
\begin{eqnarray}
\phi \big(\langle\chi\rangle_{\alpha},\hat{\sigma}_{\gamma}) = \frac{h(\langle\chi\rangle_{\alpha})/\langle \chi \rangle_{\alpha}}{\hat{\sigma}_{\gamma}^2\,\big[ 2\widetilde{S}'(\langle \chi \rangle_{\alpha}) - h'(\langle \chi \rangle_{\alpha})\big]} > 1\,.
\end{eqnarray}
Comparing the value of $\phi\big(\langle\chi\rangle_{\alpha},\hat{\sigma}_{\gamma}\big)$ with respect to 1,
one can deduce whether heating ($\phi>1$) and by extension cooling ($\phi<1$) will take place.  
A threshold value $\hat{\sigma}_{\gamma}^{\rm thr}$ for the standard deviation can be derived considering $\phi  = 1$, 
that reads:
\begin{eqnarray}\label{eq:sigmaThreshold}
\hat{\sigma}_{\gamma}^{\rm thr} = 
\sqrt{\frac{h(\langle\chi\rangle_{\alpha})}{\langle \chi \rangle_{\alpha}[2\widetilde{S}'(\langle\chi\rangle_{\alpha}) - h'(\langle\chi\rangle_{\alpha})]}} \, .
\end{eqnarray}
This threshold value gives the maximal standard deviation (in energy) that can be reached starting
from an initially narrow electron energy distribution. 
Once the energy spread has reached this threshold, the cooling phase will take over. 

Figure~\ref{fig:averages:sigmamax} presents, in color scale, $\phi(\langle\chi\rangle_{\alpha},\hat{\sigma}_{\gamma}\big)$
as a function of the normalized standard deviation in energy $\hat{\sigma}_{\gamma}$ 
and the average quantum parameter $\langle\chi\rangle$. 
The dashed line corresponds to the threshold value $\sigma_{\gamma}^{\rm thr}$ [Eq.~\eqref{eq:sigmaThreshold}] delimiting the regions
in parameter space where cooling and heating are expected. The first order expansion of Eq.~\eqref{eq:sigmaThreshold} is plotted as a green dashed line 
and corresponds to the prediction (derived in the limit $\chi \ll 1$) by Vranic {\it et al.}~\cite{vranic2016} (see also~\cite{footnoteRidgers}). 
Also reported are the measures of the maximal standard deviation extracted
from simulations (discussed in Sec.~\ref{sec:numResults}) of an initially narrow electron bunch interacting with a constant magnetic field (blue crosses) and linearly polarized plane-wave 
(green crosses). These measures show the maximum (normalized) standard deviation
obtained in the simulations as a function of the initial
average quantum parameter $\chi_0=\langle\chi\rangle(t=0)$. Values reported here
correspond to the results of either Monte-Carlo or Fokker-Planck simulations that lead to the same predictions, even for $\chi_0 =  1$.
The evolution of the normalized standard deviation as a function of $\langle \chi \rangle$ (as it evolves with time) in the simulations considering a constant uniform magnetic field is also reported (solid lines). During the energy spreading (heating) phase (for $t<t_{\rm heating}$ as defined in Sec.~\ref{sec:numResults}), the average $\chi$ is approximately constant, justifying the fact that in this phase, we can make the approximation $\langle \chi \rangle(t_{\rm{heat}}) \simeq \chi_0$ when plotting the maximum standard deviation $\hat{\sigma}_{\gamma}^{\rm{thr}}$ (blue and green crosses in figure \ref{fig:averages:sigmamax}). Moreover, we note that all curves end on the same line which acts as an attractor.\\

Note that, when the electron distribution function is initially broad, the hypothesis leading to the calculation of 
$\sigma_{\gamma}^{\rm thr}$ are not valid. The general reasoning nevertheless holds and simulations presented in Sec.~\ref{sec:numResults} 
indicate that $\hat{\sigma}_{\gamma}^{\rm thr}$ can still be interpreted as a threshold: whenever the standard deviation of the considered electron 
distribution initially exceeds this threshold, only cooling of the electron population will be observed. \\

\subsubsection{Third order moment and link to the quenching of radiation losses}\label{sec:averages:mu3Discussion}
In this section, we wish to show that the evolution of the third order moment can be used in order 
to interpret some interesting phenomenon associated to the quantum 
behavior of the system. As previously mentioned, we wish to link $\mu_3$ to the so-called {\it quenching} of radiation losses.
This process follows from the discrete nature of photon emission by the radiating electron in a quantum framework.
As shown in Ref.~\cite{harvey2017} for an electron bunch interaction with a high-amplitude electromagnetic wave, a deformation
of the distribution function can appear on short time scales as some electrons did radiate high-energy photons away thus decreasing their energy,
while other electrons have not yet emitted any photon thus conserving their initial energy. 
As a result, the electron distribution function becomes asymmetric, showing a low energy tail, associated to a negative third order moment (negative skewness).

It is therefore interesting to investigate under which conditions the third order moment assumes negative values.
As its evolution is correctly described in general in the linear Boltzmann (MC) approach {\it {only}}, we will consider Eq.~\eqref{eq:mu3_MC} and 
derive the conditions under which $d \mu_3/dt < 0$.
This would require to compute the average over the electron distribution function of the different quantities appearing in Eq.~\eqref{eq:mu3_MC}, 
which implies solving the full linear Boltzmann equation. 
Instead, we proceed as previously done and expand at first order the different functions in the rhs of Eq.~\eqref{eq:mu3_MC} that 
depend on $\chi$ around $\langle \chi \rangle$.
We find that $d \mu_3/dt < 0$ whenever
\begin{eqnarray}\label{eq:psi_tilde}
\tilde{\psi}(\langle \chi \rangle_{\alpha}, \hat{\sigma}_{\gamma}, F, \rho_{\gamma,\chi},\tilde{C}_{\gamma,\chi}) < 1\,
\end{eqnarray}
 where we have introduced:
 \begin{eqnarray}
\tilde{C}_{\gamma,\chi} = \left\langle (\gamma - \langle \gamma \rangle_{\alpha})^2 (\chi - \langle \chi \rangle_{\alpha})\right\rangle_{\alpha}/\langle \gamma\rangle_{\alpha}^2
 \end{eqnarray}
 and
\begin{eqnarray}\label{eqmu3negative}
\!\!\! \tilde{\psi} = \hat{\sigma}_{\gamma}^2\,\left[ f_1(\langle\chi\rangle) + F \rho_{\gamma,\chi} \langle\chi\rangle\  f_2(\langle\chi\rangle) \right] -\tilde{C}_{\gamma,\chi}\,f_3(\langle\chi\rangle)
\end{eqnarray}
with:
\begin{eqnarray}
f_1(\chi) &=& \frac{3 h(\chi)-a_3(\chi)}{a_3(\chi)}\,,\\
f_2(\chi) &=& \frac{3 h'(\chi)-2 a'_3(\chi)}{a_3(\chi)}\,,\\
f_3(\chi) &=& \frac{3 S'(\chi) + a'_3(\chi) - 3 h'(\chi)}{a_3(\chi)}\,.
\end{eqnarray}

Interestingly, Eq.~\eqref{eq:psi_tilde} simplifies when considering a well localized (in space) electron population (e.g. narrow electron bunch and/or uniform electromagnetic field)
so that $F \simeq 1$
and $\tilde{C}_{\gamma,\chi} \simeq \langle\chi\rangle\,\hat{\mu}_3$.
The condition given by Eq.~\eqref{eq:psi_tilde} then reduces to:
\begin{eqnarray}
\psi(\langle \chi \rangle_{\alpha}, \hat{\sigma}_{\gamma}, \hat{\mu}_3) < 1\,,
\end{eqnarray}
where:
 \begin{eqnarray}\label{eqmu3negative}
 \psi = \hat{\sigma}_{\gamma}^2 \left[ f_1(\langle\chi\rangle) + \langle\chi\rangle\ f_2(\langle\chi\rangle) \right] - \hat{\mu}_3\,\langle\chi\rangle\,f_3(\langle\chi\rangle)\,.
\end{eqnarray}
 
\begin{figure}
\begin{center}
\includegraphics[width=8cm]{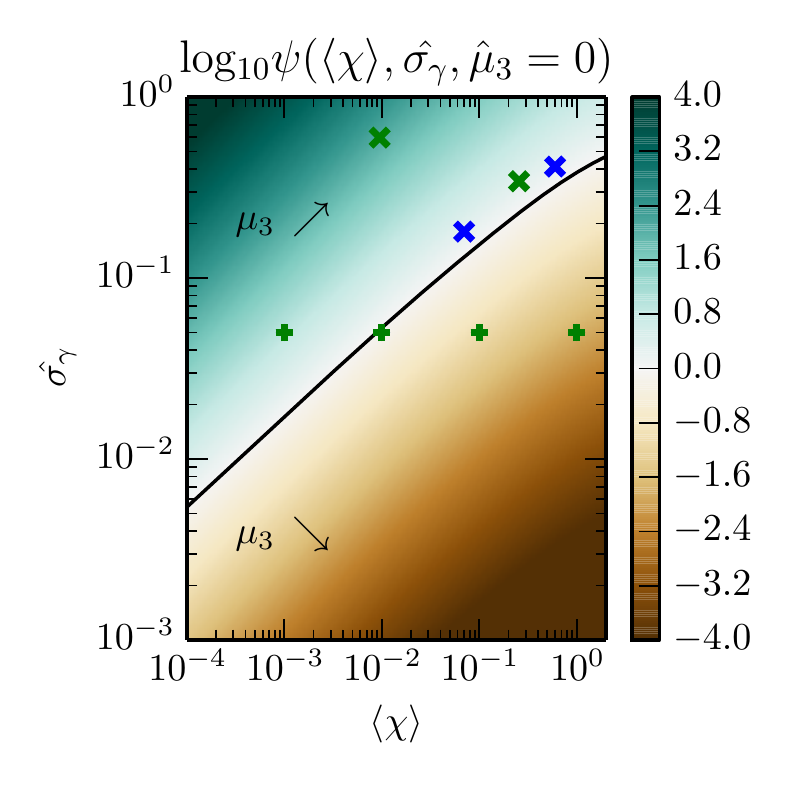}
\caption{Dependence with $\langle \chi \rangle$ and 
$\hat{\sigma}_{\gamma}$ of $\psi(\langle \chi \rangle,
\hat{\sigma}_{\gamma},\hat{\mu}_3 = 0)$. When $\psi > 1$, the 
electron 
population is predicted to acquire an asymmetry toward high energies
($\mu_3$ increases, positive skewness) while it is predicted to acquire an asymmetry
toward low energies ($\mu_3$ decreases, negative skewness) when $\psi < 1$. The black dashed line
 correspond to $\psi = 1$ and represents the threshold 
 $\hat{\sigma}^{\rm{lim}}_{\gamma,0}(\langle \chi \rangle)$ [Eq.~\eqref{sigma_threshold_mu3_0}]. 
 The blue and green crosses represent the value of 
 $\hat{\sigma}_{\gamma}$ when $\mu_3 = 0$ in the interaction of an
 ultra-relativistic electron bunch with 
different constant-uniform magnetic field and plane-waves
(respectively) and initial quantum parameter a) $\chi_0 = 10^{-2}$, b) 
$\chi_0 = 10^{-1}$ and c) $\chi_0 = 1$. Vertical crosses 
correspond to the initial values of $\hat{\sigma}_{\gamma}$ and
$\chi$. Diagonal crosses correspond to the values 
of $\hat{\sigma}_{\gamma}$ and $\chi$ when $\mu_3$ goes back to 
$0$ (see, the 6th panel of Figs.~\ref{fig:numResults:feB} and \ref{figfePW}).
These simulations are discussed in Sec.~\ref{sec:numResults}.}
\label{fig:psi:mu3}
\end{center}
\end{figure}

Considering the situation of an initially symmetric (in energy) electron beam, $\mu_3 = 0$, the condition for $\mu_3$ to decrease (and thus assume negative values) can then be 
reduced to a condition on the normalized energy variance $\hat{\sigma}_{\gamma}^2$ and average quantum parameter only $\langle\chi\rangle$.
Figure~\ref{fig:psi:mu3} presents, in color scale, the function $\psi(\langle \chi \rangle_{\alpha}, \hat{\sigma}_{\gamma}, \hat{\mu}_3=0)$
as a function of both $\langle\chi\rangle$ and $\hat{\sigma}_{\gamma}$.
The solid line corresponds to the condition $\psi=1$ and defines, for a given initial value of $\langle\chi\rangle$, a limiting energy variance for the electron bunch:
\begin{eqnarray}\label{sigma_threshold_mu3_0}
\hat{\sigma}_{\gamma,0}^{\rm lim}(\langle \chi \rangle) = \left[ f_1(\langle\chi\rangle) + \langle\chi\rangle\,f_2(\langle\chi\rangle) \right]^{-1/2}\,.
\end{eqnarray}
This value of the initial electron bunch energy dispersion delimits the regions in parameter space where $\mu_3$ is expected to increase or decrease.

Also reported are the measures of the standard deviation extracted from the simulations (discussed in Sec.~\ref{sec:numResults}) 
of an initially spatially narrow and energy symmetric [$\mu_3(t=0)=0$] electron bunch interacting with a constant magnetic field (blue crosses) and linearly polarized plane-wave 
(green crosses) considering different initial values for the average quantum parameter $\langle\chi\rangle$.
Vertical ($+$) crosses report the corresponding values of $\langle\chi\rangle$ and $\hat{\sigma}_{\gamma}$ at time $t=0$.
Crosses for $\langle\chi\rangle = 10^{-3}$ and $10^{-2}$ are either above or close to the delimiting solid line. 
In this simulation, one could thus expect an increase of $\mu_3$. In our simulations, however, (and as expected from the scaling in $\chi^n$ of the successive moments 
discussed in Sec.~\ref{sec:stoch:FPValidity}), $\mu_3$ assumes very small values and it was not possible to confirm this prediction.
Nevertheless, for the case $\langle\chi\rangle = 10^{-1}$ and $1$, where $\vert\mu_3\vert$ reaches larger values, all crosses are found to be below the limiting 
$\hat{\sigma}^{\rm lim}_{\gamma,0}$ (solid) line, and the third order moment is found to decrease in the corresponding simulations, consistently with the present analysis.
Also reported as "diagonal" ($\times$) crosses are the corresponding values of $\langle\chi\rangle(t^{\star})$ and $\hat{\sigma}_{\gamma}(t^{\star})$ extracted from the simulations
at the first time $t^{\star}$ for which $\mu_3$ is found to vanish. In every cases, $\mu_3$ crosses $0$ again when increasing, and finding all these points above the $\hat{\sigma}_{\gamma,0}^{\rm lim}$
limit further confirms the present analysis.\\

For completeness, we study the influence of the initial value of $\hat{\mu}_3$ on its evolution, and we compute the limiting value of $\hat{\sigma}_{\gamma}$ at which $d \mu_3/dt$ changes sign for an initial non-zero value of $\hat{\mu}_3$
\begin{eqnarray}\label{sigma_threshold_mu3}
\hat{\sigma}_{\gamma}^{\rm{lim}}(\langle \chi \rangle, \hat{\mu}_3) = \hat{\sigma}^{\rm{lim}}_{\gamma,0}(\langle \chi \rangle) \sqrt{1 +  \hat{\mu}_3 \langle \chi \rangle f_3(\langle \chi \rangle)} \, .
\end{eqnarray}

\begin{figure}
\begin{center}
\includegraphics[width=8cm]{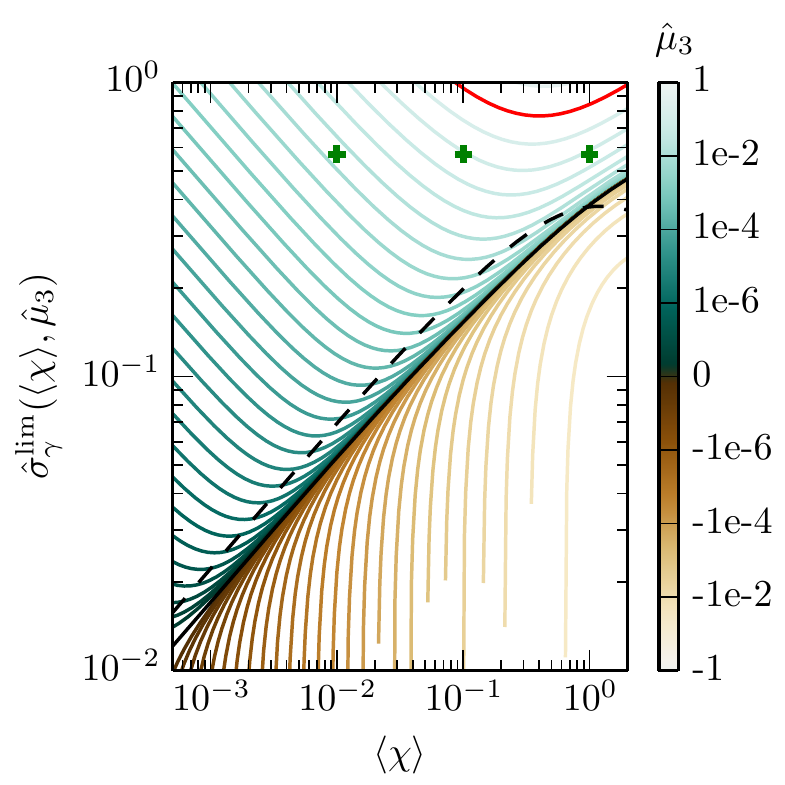}
\caption{Dependence of $\hat{\sigma}_{\gamma}^{\rm{lim}}$ [Eq~\eqref{sigma_threshold_mu3}] with $\langle \chi \rangle$ for 
different values of $\hat{\mu}_3$. The red line represents $\hat{\sigma}_{\gamma}^{\rm{lim}}$ for the case $\hat{\mu_3} = 0.2$ used in Sec.~\ref{sec:numResultsMJ}. 
The black solid line corresponds to $\hat{\mu}_3 = 0$ while the black dotted line corresponds to $\hat{\sigma}_{\gamma}^{\rm thr}$ [Eq.~\eqref{eq:sigmaThreshold}]. 
The green crosses correspond to the three initial situations considered in Sec.~\ref{sec:numResultsMJ} considering a broad Maxwell-J\"uttner energy distribution.} 
\label{fig:sigmalim:mu3}
\end{center}
\end{figure}

Figure~\ref{fig:sigmalim:mu3} presents, in color scale $\hat{\sigma}_{\gamma}^{\rm{lim}}(\langle \chi \rangle, \hat{\mu}_3)$ as a function of the average quantum parameter 
$\langle\chi\rangle$ and for different values of $\hat{\mu}_3$. The black dashed line represents the value of $\hat{\sigma}_{\gamma}^{\rm thr}$ [Eq.~\eqref{eq:sigmaThreshold}], while the straight 
black line represents the value of $\hat{\sigma}^{\rm{lim}}_{\gamma,0}$ [Eq.~\eqref{sigma_threshold_mu3_0}]. The other lines following the colorscale are the values of 
$\hat{\sigma}^{\rm{lim}}_{\gamma}$ for different values of $\hat{\mu}_3$. For a given initial $\hat{\mu}_3$, as for $\hat{\sigma}^{\rm{lim}}_{\gamma,0}$, 
if the initial values of $\hat{\sigma}_{\gamma}$ and $\langle \chi \rangle$ are above the limiting curve, the skewness will increase, and if below it will decrease.  

We recall the meaning of the quantity  
$\hat{\sigma}_{\gamma}^{\rm thr}$ (dashed line): if $\hat{\sigma}_{\gamma}$ is above this curve there will be cooling, and below, heating. For a system starting from initial non zero positive $\hat{\mu}_3$ (blue-shaded curves), the higher the value of $\hat{\mu}_3$, the broader the range of parameter in which we have cooling and decrease of $\mu_3$, especially for small $\langle \chi \rangle$. On the contrary, for a system starting from an 
initially negative $\hat{\mu}_3$ (brown-shaded curves), we have the possibility of a range of parameters for which we can have both heating and an increase of $\mu_3$. 

The situation is different if we start from a symmetrical distribution function (or quasi-symmetrical distribution function, with very small values of $\mu_3$). 
As we can see the lines  $\hat{\sigma}^{\rm{lim}}_{\gamma,0}$ and $\hat{\sigma}_{\gamma}^{\rm thr}$ are very close, and in this case values of $\hat{\sigma}_{\gamma}$ above this line correspond to an electron beam 
acquiring an asymmetry toward the left (negative skewness). Because of the proximity of these two lines, we can see in Fig.~\ref{fig:sigmalim:mu3}, that in most cases (for $\chi < 1$), cooling is 
 accompanied by an increase of $\hat{\mu}_3$ and heating by a decrease of $\hat{\mu}_3$. In Sec.~\ref{sec:numResultsMJ}, we will examine in more detail the case of a Maxwell-J\"uttner distribution that has a non-zero initial value of $\mu_3$. The corresponding limiting line is marked in red in figure \ref{fig:sigmalim:mu3} and the initial conditions in $\langle\chi\rangle$ and $\hat{\sigma}_{\gamma}$ of the three simulations considered in that Section are reported in the figure as green crosses. According to our prediction, the system will cool down while its skewness diminishes (all points are below the red line), as will indeed be demonstrated in all these simulations.

\subsection{Domain of validity of the different descriptions \& Physical insights}\label{sec:averages:domainval}

The analysis of the evolution of the successive moments allows us to infer more precisely the domain of validity of the three descriptions
by taking into account the properties of the electron distribution function (in particular its first three moments). 
Most importantly, these domains of validity also provide us with a deeper insight into the various aspects of radiation reaction 
from which can be drawn new guidelines for designing experiments, as will be further discussed in Sec.~\ref{sec:PhysicalImplications}.

In what follows, we discuss this, first considering an initially symmetric distribution function for the electron 
(most interesting when considering, e.g., an electron beam),
then considering an asymmetric electron distribution function (as can be expected considering a hot electron plasma, see, e.g., Sec.~\ref{sec:numResultsMJ}).\\

\subsubsection{Case of an initially symmetric energy distribution}

In the case of an initially symmetric (or nearly symmetric) distribution function, the deterministic (cLL) description is expected to hold whenever: 
(i) the purely quantum term in the equation of evolution for the variance [first term in the rhs of Eq.~\eqref{eq:variance}] 
is negligible with respect to the classical one [second term in the rhs of Eq.~\eqref{eq:variance}],
and (ii) the third order moment remains negligible compared to the second order one 
(i.e. it does not play any role in the description of the distribution function).

The first condition (i) leads:
\begin{eqnarray}\label{condition_cLL}
\hat{\sigma}^2_{\gamma} \gg \frac{h(\langle\chi\rangle)+\langle\chi\rangle h'(\langle\chi\rangle)\,\hat{\sigma}^2_{\gamma}}{2\langle\chi\rangle\,S'(\langle\chi\rangle)}\,,
\end{eqnarray}
while the second condition (ii) can be computed estimating the extremum value of $\hat{\mu}_3$ [obtained canceling the rhs of Eq.~\eqref{eqMCmu3}] leading to:
\begin{eqnarray}\label{condition_mu3_min}
\frac{\hat{\sigma}_{\gamma}^2 \langle \chi \rangle f_3(\langle \chi \rangle)}{\left\vert \hat{\sigma}_{\gamma}^2[f_1(\langle \chi \rangle) + \langle \chi \rangle f_2(\langle \chi \rangle)] - 1 \right\vert} \gg 1 \,.
\end{eqnarray}

\begin{figure*}
\begin{center}
a) \includegraphics[width=0.4\textwidth]{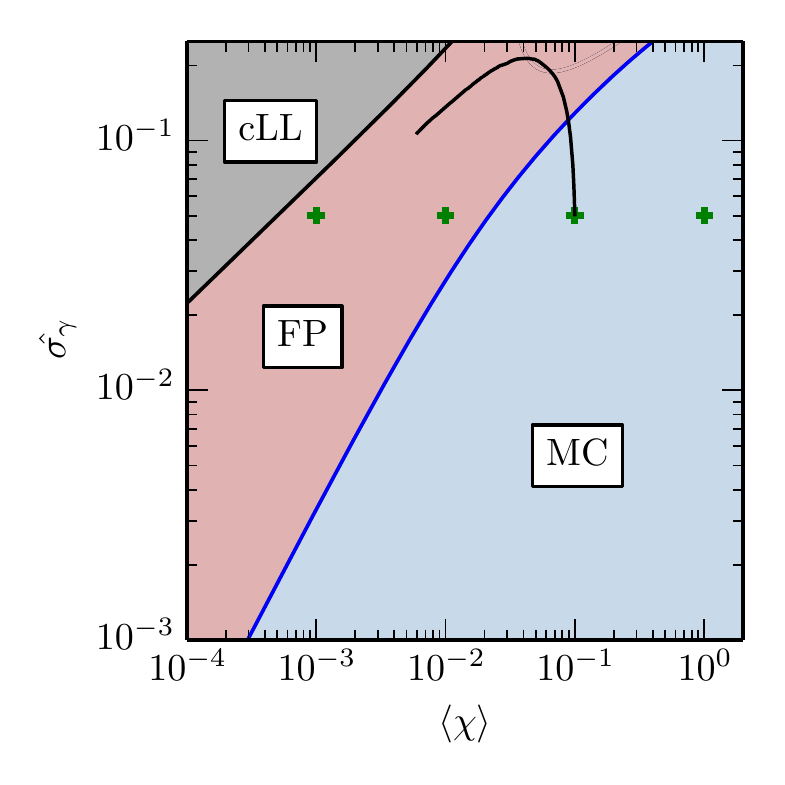}\quad
b) \includegraphics[width=0.4\textwidth]{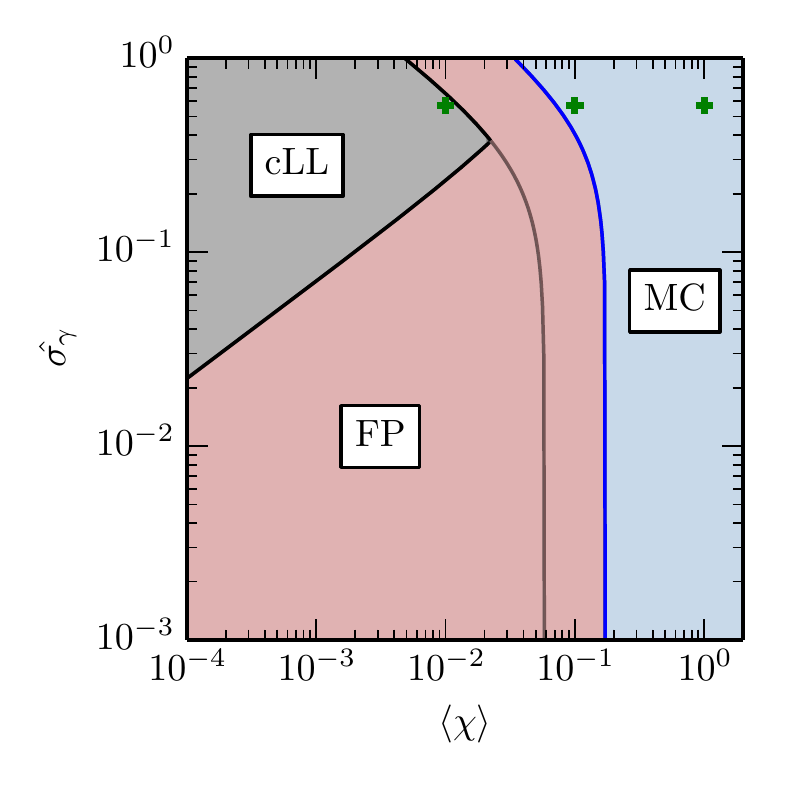}
\caption{Domain of validity of the different approaches. a) In the case of an initially symmetric electron energy distribution; as given by Eqs.~\eqref{condition_cLL} and \eqref{condition_mu3_min} for a symmetric distribution function ($\hat{\mu}_3 = 0$). b) In the case of an initially asymmetric distribution function;  as given by Eqs.~\eqref{condition_dmu3_dt_cLL} and \eqref{condition_dmu3_dt_FP}. Here presented for $\hat{\mu}_3 \simeq 0.2$ corresponding to the Maxwell-J\"uttner distribution presented in Sec.~\ref{sec:numResultsMJ}.}
\label{fig:validity_mu3}
\end{center}
\end{figure*}

In contrast, the FP model providing the correct description for the energy variance, its validity is constrained by the second condition (ii) [Eq.~\eqref{condition_mu3_min}] only.

The derived conditions, Eqs.~\eqref{condition_cLL} and~\eqref{condition_mu3_min}, depend on both, the average quantum parameter $\langle\chi\rangle$
and the electron distribution normalized energy variance $\hat{\sigma}_{\gamma}$. They provide us with a path to define the domain of validity of the different approaches (deterministic/cLL, FP or MC) in the ($\langle\chi\rangle$,$\hat{\sigma}_{\gamma}$) parameter-space.
Our results are summarized in Fig.~\ref{fig:validity_mu3}. 

The blue shaded regions correspond to the parameter-space for which condition (ii) [Eq.~\eqref{condition_mu3_min}] is not satisfied (more exactly the lhs of Eq.~\eqref{condition_mu3_min} is $\le 10$~\cite{footnote6}). In this region, the third order moment of the distribution function can assume large values, so that only the MC procedure provides a correct description. 

Finally, outside of this domain (grey and red shaded regions), the FP description holds while 
the gray shaded region corresponds to the region where the deterministic (cLL) description is valid, i.e. both conditions (i) and (ii) [Eqs.~\eqref{condition_cLL} 
and~\eqref{condition_mu3_min}, respectively] are satisfied [for condition (i), we specify here that the lhs of Eq.~\eqref{condition_cLL} is $\ge 10$
which defines the region above the black-dashed line].

\subsubsection{Case of an initially non-symmetric energy distribution}

A similar analysis can be performed with an initially asymmetric distribution function.
In that case however, even though condition (i) [Eq.~\eqref{condition_cLL}] is unchanged, one has to reconsider the condition on $\hat{\mu}_3$ that is now not negligible, but should be correctly handled by the various description.
For the deterministic (cLL) description to hold, one thus requires that all terms appearing in Eq.~\eqref{eqMCmu3} (for $d\mu_3/dt$ in the MC description) 
and not in Eq.~\eqref{eqCLmu3} (for $d\mu_3/dt$ in the deterministic description) are small with respect to the rhs of Eq.~\eqref{eqCLmu3}.
This leads: 
\begin{widetext}
\begin{eqnarray}\label{condition_dmu3_dt_cLL}
\frac{3 \hat{\mu}_3 \langle \chi \rangle S'(\langle \chi \rangle) / a_3(\langle \chi \rangle)}{\left\vert -\hat{\mu}_3 \langle \chi \rangle f_3(\langle \chi \rangle) + 3 \hat{\mu}_3 \langle \chi \rangle \tilde{S}'(\langle \chi \rangle) /a_3(\langle \chi \rangle) + \hat{\sigma}^2_{\gamma} [f_1(\langle \chi \rangle) + \langle \chi \rangle f_2(\langle \chi \rangle)] -1 \right\vert} \gg 1\, .
\end{eqnarray}
\end{widetext}
Proceeding in a similar way but for the FP description [i.e. using Eqs.~\eqref{eqMCmu3} and ~\eqref{eqFPmu3}], one obtains the condition of validity for the FP approach in the form:
\begin{eqnarray}\label{condition_dmu3_dt_FP}
& &\!\!\!\!\!\!\!\! \nonumber \frac{\left\vert 3 \hat{\sigma}^2_{\gamma} [h(\langle \chi \rangle) + \langle \chi \rangle h'(\langle \chi \rangle)] + 3 \hat{\mu}_3 \langle \chi \rangle [h'(\langle \chi \rangle) - \tilde{S}(\langle \chi \rangle)]  \right\vert}{\left\vert a_3(\langle \chi \rangle) + \hat{\sigma}^2_{\gamma} [a_3(\langle \chi \rangle) + 2 \langle \chi \rangle a_3'(\langle \chi \rangle)] + \hat{\mu}_3 \langle \chi \rangle a_3'(\langle \chi \rangle)  \right\vert} \!\! \gg \!\! 1 \, , \\
\end{eqnarray}

The derived conditions of validity, Eqs.~\eqref{condition_cLL} and~\eqref{condition_dmu3_dt_cLL} for the deterministic (cLL) description 
and  Eqs.~\eqref{condition_cLL} and~\eqref{condition_dmu3_dt_FP} for the Fokker-Planck description 
now depend not only on $\langle\chi\rangle$ and $\hat{\sigma}_{\gamma}$, but also on the normalized third order moment $\hat{\mu}_3$.
In figure~\ref{fig:sigmalim:mu3}, we have projected them on the ($\langle\chi\rangle,\hat{\sigma}_{\gamma}$) parameter-space assigning 
for $\hat{\mu}_3$ its initial value ($\hat{\mu}_3 = 0.2$) for the particular case (broad Maxwell-J\"uttner distribution) studied in Sec.~\ref{sec:numResultsMJ}.

The regions of validity of the different models are shown in Fig.~\ref{fig:validity_mu3}b following the same lines then for Fig.~\ref{fig:validity_mu3}a.
While the region of validity of the deterministic (cLL) description is mainly unchanged (note that the range of accessible $\hat{\sigma}_{\gamma}$ is larger
for asymmetric functions), one finds that the region validity of the FP description is significantly increased.
As will be discussed in Sec.~\ref{sec:numResultsMJ}, this is indeed what is observed in our simulations.

\subsubsection{Physical implications}\label{sec:PhysicalImplications}

Our analysis of the successive (and in particular second and third) moments has allowed us to identify more clearly
the domain of validity of the different descriptions in terms of both the initial average quantum parameter and shape
of the energy distribution of the electron population.
As each of these descriptions account for different physical effects of radiation reaction,
these domains of validity provide us with some insights on which processes play a dominant role under given conditions
of the external field and properties of the electron population, as well as some guidelines for future experiments to
observe various aspects of radiation reaction.

Indeed, should our physical conditions (e.g. experimental set-up or astrophysical scenario) lie in the regime where the
deterministic (cLL) description is valid, radiation reaction acts as a friction force 
and one can expect to observe both a reduction of the electron population mean energy as well as a narrowing (cooling) 
of the electron energy distribution function.
In the case where our physical conditions are characteristic of the regime where the deterministic description is not valid anymore 
but the FP approach is, the stochastic nature of high-energy photon emission starts to play an important role. 
Under such conditions, and in particular for short (with respect to the so-called {\it heating} time that will be further 
discussed in Sec.~\ref{sec:numResults}) times, one can expect to observed a measurable broadening of the electron energy distribution.
Finally, should our experimental (or astrophysical) conditions lie in the regime where only the linear Boltzmann (MC) description is valid, 
not only the stochastic nature of photon emission, but its discrete nature will play a role.
In particular, this defines the experimental regime for which the {\it quenching} of radiation losses can be observed, and diagnosed as a negative skewness
of the electron distribution function. In this regime, and for short enough times, the diffusion approximation supporting the FP approach is not 
valid anymore and rare emission events, that can be described only as a Poisson process (i.e. by the MC procedure), are of outmost importance.

\section{Numerical algorithms}\label{sec:algo}

In what follows we detail the numerical algorithms, so-called pushers, developed to treat the dynamics of ultra-relativistic electrons radiating
in an external electromagnetic field. These pushers will be used and compared to each other in the next Sec.~\ref{sec:numResults}.

\subsection{Deterministic pusher (friction force with quantum correction)}\label{sec:algo:cLL}

The first and simplest pusher is the deterministic radiation reaction pusher that allows one to describe the 
radiating electron dynamics in the framework of classical electrodynamics, and accounting for the quantum correction
on the power radiated away by the particle. 
Its implementation closely follows that proposed by Tamburini {\it et al.}~\cite{tamburini2010} with the difference
that it relies on the equations of motions~\eqref{eq_ener} and~\eqref{eq_force}, and uses the quantum correction 
given by Eq.~\eqref{eq_h_chie}.

This pusher (and all pushers discussed here) are based on the {\it leap-frog} technique and assume that 
forces and momenta are known at integer ($n$) and half-integer ($n-\tfrac{1}{2}$) timesteps, respectively.
Following Ref.~\cite{tamburini2010}, we first compute the effect of the Lorentz and radiation reaction force separately. 
Starting from momentum ${\bf p}^{(n-\tfrac{1}{2})}$ and considering 
the Lorentz force ${\bf f}_L^{(n)}$, the first step is performed using the standard Boris pusher~\cite{Boris_pusher} giving:
\begin{eqnarray}
{\bf p}_L = {\bf p}^{(n-\tfrac{1}{2})} + {\bf f}_L^{(n)}\,\Delta t\,,
\end{eqnarray}
with $\Delta t$ the time-step. In a second step, we compute the effect of the radiation reaction force:
\begin{eqnarray}\label{eq:num_p_R}
{\bf p}_R = {\bf p}^{(n-\tfrac{1}{2})} + {\bf f}_{\rm rad}^{(n)}\,\Delta t\,,
\end{eqnarray}
where ${\bf f}_{\rm rad}^{(n)}=P_{\rm cl}\,g(\chi)\,{\bf u}/(c{\bf u}^2)$ [Eqs.~\eqref{eq_force} and~\eqref{eq_Prad_CED} with quantum correction] 
is computed using the particles properties at time step $(n-\tfrac{1}{2})$ and current value of the fields to estimate $\chi$,
and the quantum correction $g(\chi)$ can be either tabulated or approximated by a fit.
Finally, the momentum at time-step $(n+\tfrac{1}{2})$ is computed as:
\begin{eqnarray}
{\bf p}^{(n+\tfrac{1}{2})} = {\bf p}_L + {\bf p}_R - {\bf p}^{(n-\tfrac{1}{2})}\,.
\end{eqnarray}

This pusher has been validated (not shown) against analytical solutions for the cases of a radiating electron in a constant, 
homogeneous magnetic field~\cite{Accurate_LL} and a plane-wave~\cite{Di_Piazza_sol}.

\subsection{Stochastic (Fokker-Planck) pusher}\label{sec:algo:fp}

The stochastic pusher we now describe is based on the Fokker-Planck treatment developed in Sec.~\ref{sec:stoch}.
It follows the very same step as described earlier for the deterministic pusher with the difference that the radiation reaction
force in Eq.~\eqref{eq:num_p_R} now contains an additional stochastic term:
\begin{eqnarray}\label{eq:fradFP}
{\bf f}_{\rm rad}^{(n)}=\left[-P_{\rm rad}\Delta t+mc^2\sqrt{R(\chi,\gamma)}\,dW\right] \,{\bf u}/(c {\bf u}^2)\,,
\end{eqnarray}
where ${\bf u} = {\bf p}^{(n-\tfrac{1}{2})}/(mc\gamma)$, and $dW$ is a random number generated using a normal
distribution of variance $\Delta t$. Both functions $g(\chi)$ and $h(\chi)$ [the latter appearing when evaluating $R(\chi,\gamma)$ using Eq.~\eqref{eq_Rgamma}]
can be either tabulated or estimated from a fit.\\

Let us now note that Eq.~\eqref{eq:fradFP} can in some cases (when its rhs is positive) lead to an electron gaining energy.
This up-scattering is not physical, and is a well-known short-cut of the Fokker-Planck approach.
It may become problematic only in cases where $\chi \rightarrow 1$ for which the stochastic term can become of the 
order of the drift term. 
However, if this may be a problem should one consider only a single particle dynamics, this problem
is alleviated when using this kind of pusher in Particle-In-Cell (PIC) codes. In that case indeed, one deals not with real particles
but with so-called macro-particles that actually represent discrete element of a distribution function (see, e.g., Ref.~\cite{smilei}),
and up-scattering is then in average suppressed.
This will be discussed into more details in the next Sec.~\ref{sec:numResults}.\\

For the sake of completeness, we also note that Wiener
process involve sample paths that are non-differentiable~\cite{KP}.
This requires much care when the issue comes to the numerical treatment of these random
discrete increments. Here, as a first tentative, we introduce the simplest possible scheme, know as Euler-Maruyama~\cite{KP}.
Of course more sophisticated and accurate schemes exist, that have not been tested in this work.
An importance issue lies in the stiffness of the Stochastic Differential Equation (SDE). 
This stiffness can be quantified with use of the SDE Lyapunov exponents, which
basically indicates the presence of different scales in the
solution~\cite{KP}. {\it A priori}, the stiffness of the rate of photon emission is avoided by the SDE
because it is precisely the operating regime of the Monte-Carlo pusher.

\subsection{Monte-Carlo pusher}\label{sec:algoMC}\label{sec:algo:mc}

We finally introduce the Monte-Carlo pusher as it provides a discrete formulation of the linear Boltzmann Eq.~\eqref{eq:Master}. 
It is therefore valid for a wide range of electron quantum parameter $\chi$,
and will thus be used in the next Sec.~\ref{sec:numResults} to infer the validity of the two previous pushers.

Our implementation closely follows that presented in Refs.~\cite{duclous2011,arber2015,lobet2016} and is described in Appendix~\ref{AppendixMC}.

\section{Numerical results}\label{sec:numResults}

In this Section, we confront the various numerical algorithms 
(pushers) introduced above against each other as well as against 
our theoretical predictions (Section~\ref{sec:averages}). We consider an electron beam, 
first in a constant magnetic field, then in a counter-propagating linearly polarized plane-wave. The case of an electron bunch
with a broad (Maxwell-J\"uttner) energy distribution in a constant magnetic field is also discussed.
Note that, throughout this Section, the Monte-Carlo (MC) simulations will be used as a reference as they 
provide a more general description equivalent to the full linear-Boltzmann description.

\subsection{Constant-uniform magnetic field}\label{sec:numResultsB0}

We start by simulating the interaction of a Gaussian electron beam with mean energy $\gamma_0 =1800$ and standard deviation 
$\sigma_0 = 90$ (corresponding to approximately 920 $\pm$ 46 MeV) with different constant-uniform magnetic fields of 
magnitude corresponding to $\chi_0 = \langle\chi\rangle(t=0) = 10^{-3}$, $10^{-2}$, $10^{-1}$ and $1$ (correspondingly, $B = 2.5~\rm{kT}$,  
$25~\rm{kT}$, $250~\rm{kT}$ and $2.5~\rm{MT}$). 
The end of the simulation is taken when the energy decrease becomes very slow (i.e. we approach a regime in which radiation losses are not important)
except for the case $\chi_0 = 10^{-3}$, where the energy loss is always small and we stop arbitrarily at $t_{\rm end} = 20/\omega_c$,
with $\omega_c = eB/(m \gamma)$ the synchrotron frequency. 
For $\chi_0 = 1$ the simulation ends at $t_{\rm end} = 3/\omega_c$, for $\chi_0 = 10^{-1}$
at $t_{\rm end} = 5/\omega_c$, and for $\chi_0 = 10^{-2}$ and $10^{-3}$ at $t_{\rm end} = 20/\omega_c$. 
In all cases, we used 10 000 test particles. 
 
 \begin{figure*}
\begin{center}
\includegraphics[width=16cm]{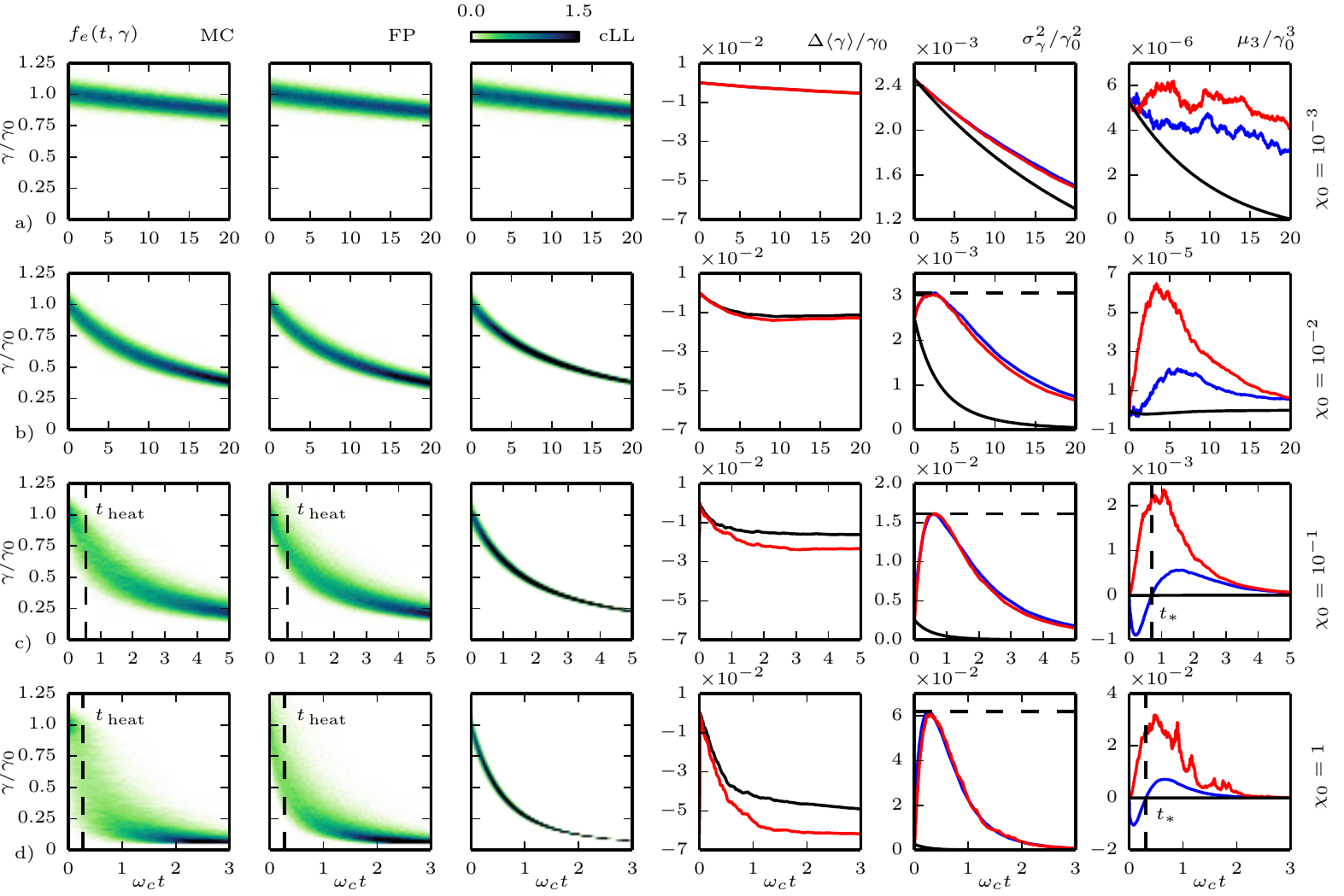}
\caption{Simulations of an ultra-relativistic electron beam in a constant uniform magnetic field for 
a) $\chi_0 = 10^{-3}$, b) $\chi_0 =10^{-2}$, c) $\chi_0 =10^{-1}$ and d) $\chi_0 =1$. 
The first three panels of each row show the electron distribution functions from the Monte-Carlo simulations (MC, first panels),
stochastic (Fokker-Planck) simulations (FP, second panels) and quantum-corrected deterministic simulations (cLL, third panels).
The fourth panel shows the difference in the prediction of the mean electron energy in between the MC simulation and
 the deterministic (black line) and FP (red line) simulations. 
  The two last panels (in each row) correspond to the moments of order 2 (energy variance) and 3 for the MC (blue line), 
  FP (red line) and deterministic (black line) simulations.} \label{fig:numResults:feB}
\end{center}
\end{figure*}

\begin{figure}
\begin{center}
\includegraphics[width=8cm]{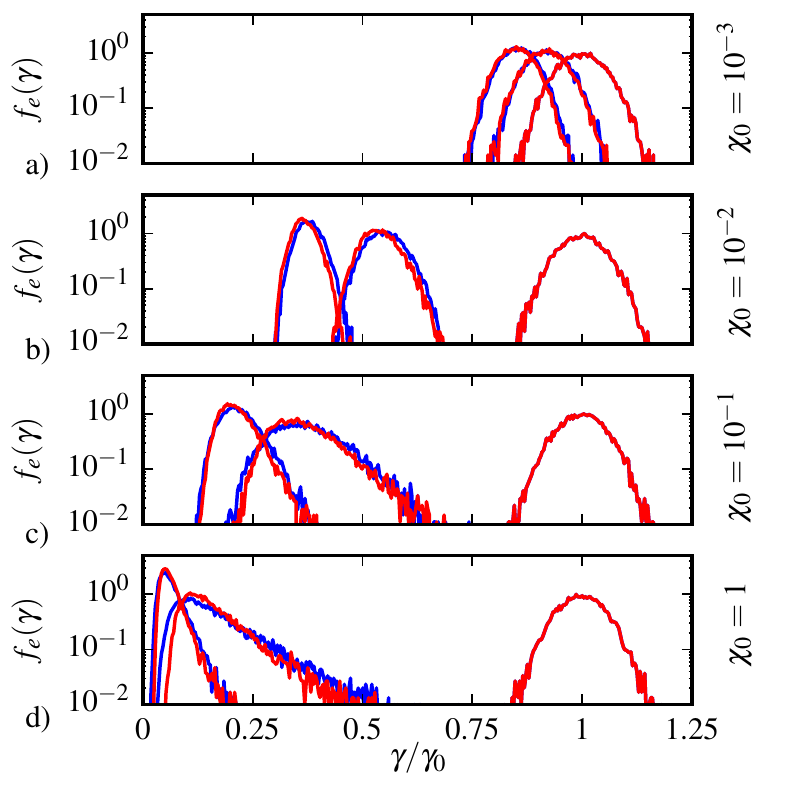}
\caption{Simulations of an ultra-relativistic electron beam in a constant uniform magnetic field for 
a) $\chi_0 = 10^{-3}$, b) $\chi_0 =10^{-2}$, c) $\chi_0 =10^{-1}$ and d) $\chi_0 =1$ electron distribution functions  at  times $t = 0$, $t = t_{\rm end}/2$ and $t = t_{\rm end}$ (from right to left). 
The red lines correspond to FP simulations, the blue one to MC simulations.} \label{figfeBgaussian}
\end{center}
\end{figure}

 The results are summarized in Fig.~\ref{fig:numResults:feB}. 
 The first row a)~corresponds to $\chi_{0} = 10^{-3}$,  the second b) to $\chi_{0} = 10^{-2}$, the third c) to 
$\chi_{0} = 10^{-1}$ and the last one d) to $\chi_{0} = 1$. 
The first three columns correspond to the evolution of the distribution function 
$f_e(t,\gamma)$ respectively in the case of the Monte-Carlo simulation (MC), the stochastic (Fokker-Planck) pusher (FP) 
and the deterministic (cLL) radiation reaction pusher [including the quantum correction $g(\chi)$]. 
The fourth column corresponds to the (normalized) difference between the average energy extracted from the Monte-Carlo simulations 
and the average energy obtained from the stochastic pusher (red line), and that obtained using the deterministic pusher (black line). 
Both are normalized to the initial mean energy $\gamma_0$:
$\Delta \gamma_{\alpha} /\gamma_0 = (\langle \gamma \rangle_{\mbox{\tiny MC}} -  \langle \gamma \rangle_{\alpha})/\gamma_0$, with $\alpha = {\mbox{ cLL}}$ or FP.
Finally the last two rows correspond to the normalized variance $\sigma_{\gamma}^2 /\gamma_0^2 = 
\langle (\gamma-\langle \gamma \rangle)^2\rangle/\gamma_0^2$ and to the normalized moment of order 3, $\mu_3/\gamma_0^3
= \langle (\gamma-\langle \gamma \rangle)^3\rangle/\gamma_0^3$ (in all plots, the blue line corresponds to the Monte-Carlo
 simulation, the red line to the stochastic pusher and the black line to the deterministic pusher [with the quantum 
 correction $g(\chi)$].\\

Let us first consider the case $\chi_{0} =10^{-3}$.
As $\chi_0 < \chi_{\rm cl}$ , one could expect all three descriptions to lead to similar results, and indeed,
there is a very good agreement in the evolution of the distribution function as calculated by the three models 
[see first three panels of Fig.~\ref{fig:numResults:feB}a].
Small differences eventually appear that are as predicted by the analysis performed in the previous sections. 
In particular we see in the variance [fifth panel of Fig.~\ref{fig:numResults:feB}a] 
that cooling is slightly overestimated by the deterministic (cLL) model. 
This is expected as we are close but not exactly into the deterministic domain of validity (see Fig.~\ref{fig:validity_mu3}a) 
so that the quantum terms leading to energy spreading are not completely negligible.
Moreover, there is a small difference between models in the third order moment [last panel of Fig.~\ref{fig:numResults:feB}a], 
as expected from Sec.~\ref{sec:averages:timeEvolution}. 
Yet this moment remains 3 orders of magnitude smaller than the variance, as expected from the scaling $\propto \chi$ of the various moments (see Sec.~\ref{sec:stoch:FPValidity}), and it is thus negligible (see also Fig.~\ref{figfeBgaussian}a). \\

\begin{figure*}
\begin{center}
\includegraphics[width=16cm]{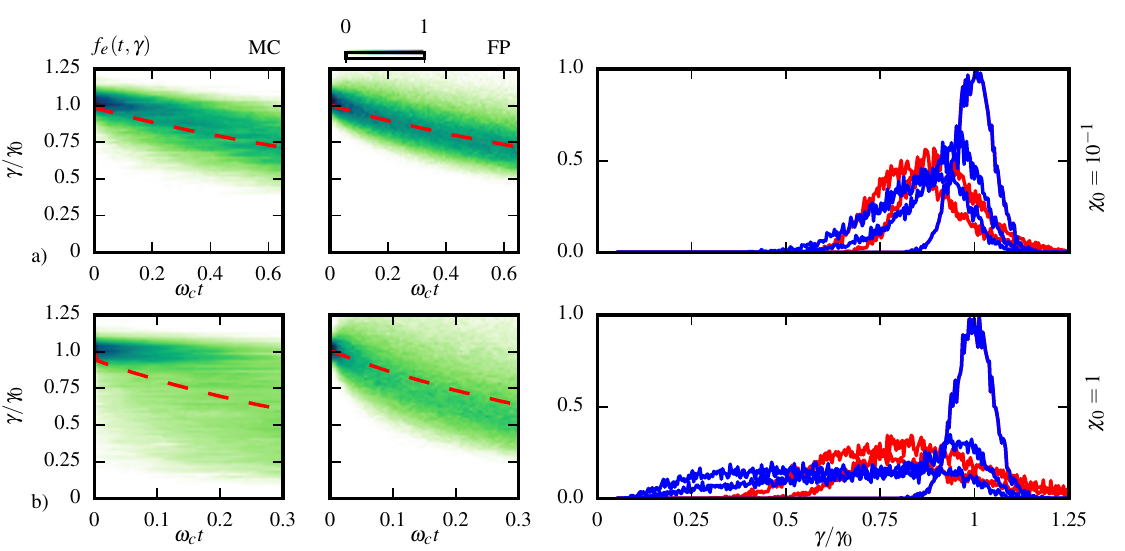}
\caption{Simulations of an ultra-relativistic electron beam in a constant, uniform magnetic field for 
a) $\chi_0 = 10^{-1}$, b) $\chi_0 =1$. This figure focuses on the early times of interaction ($t\le t_{\rm heating}$ during which 
the energy dispersion increases).
The first column corresponds to MC simulations, the second to the FP ones. 
The last row shows snap-shots of the electron distribution functions at different times $t=0$, $t=t_{\rm heating}/4$ and $t=t_{\rm heating}/2$ (from right to left).
The red lines correspond to FP simulations, the blue one to MC simulations.} \label{figfeBheating}
\end{center}
\end{figure*}

We now examine the cases $\chi_0 = 10^{-2}$ and $10^{-1}$. 
We are now in what we called the intermediate quantum regime  ($\chi_{\rm cl} < \chi < \chi_{\rm qu}$), 
and one expects the deterministic (cLL) description to fail in describing the evolution of the electron population.
It turns out to be the case as, while the evolution of the distribution function obtained from the stochastic (FP) pusher is in good agreement with the one obtained using the MC procedure, both are very different from the deterministic one. 
Still, and as expected from Sec.~\ref{sec:averages:timeEvolution}, all three models give similar results concerning the evolution of the average energy
which is found to be very close to the evolution of a single classical particle with initial energy equal to the average energy of the initial population (see the fourth panels of Fig.~\ref{fig:numResults:feB}b and~\ref{fig:numResults:feB}c.). 

The main difference between the deterministic model and the quantum (FP and MC) models is in the variance (fifth panels of Figs.~\ref{fig:numResults:feB}b 
and~\ref{fig:numResults:feB}c). 
The quantum models exhibit an energy spreading (effective heating) phase, with $\sigma_{\gamma}$ increasing up to a maximum value, and a later phase of cooling, while the deterministic model predicts only cooling. As predicted, both quantum models (MC and FP) predict the same evolution of the energy variance, and the maximum
value of $\hat{\sigma}_{\gamma}$ is found to be in perfect 
agreement with the predicted value $\hat{\sigma}^{\rm thr}_{\gamma}$ given by Eq.~\eqref{eq:sigmaThreshold}.
The good agreement between both quantum models with respect to the temporal evolution of the distribution function can be clearly 
seen in Fig.~\ref{figfeBgaussian}b and~\ref{figfeBgaussian}c, where we superimposed the electron distribution functions obtained from the stochastic (FP) pusher (blue line) and Monte Carlo approach (red line) at different times $t = 0$, $t = t_{\rm end}/2$ and $t = t_{\rm end}$.

As expected, differences in between the FP and MC models only appear in the evolution of the third order moment (last panels in ~\ref{fig:numResults:feB}b 
and~\ref{fig:numResults:feB}c). Yet, for both cases, this third order moment is found to be $\chi$ times smaller than the second order moment.
For the case $\chi_0=10^{-2}$, this discrepancy is negligible and both FP and MC are equivalent.
For the case $\chi_0=10^{-1}$, one can argue that for short times, this difference of the order of 10\% is not so negligible.
This is exactly what one could have expected from the extended analysis of the domain of validity presented in Sec.~\ref{sec:averages:domainval},
and Fig.~\ref{fig:validity_mu3}a, where the case $\chi_0=10^{-1}$ is found to lie close to, yet outside of the domain of validity of the FP description. Furthermore, at longer times 
($t > t_{\rm heat}$), the slight decrease of $\langle \chi \rangle$
and significant increase of $\hat{\sigma}_{\gamma}$ bring the 
simulation conditions back into the FP domain of validity as shown
by the solid black line in Fig~\ref{fig:validity_mu3}
Note that this finding supports our claim that not only $\langle \chi \rangle$ but also $\hat{\sigma}_{\gamma}$ are relevant 
to discriminating which physical processes of radiation reaction are important, in particular at short times.\\

Finally, for $\chi_0 = 1 > \chi_{\rm qu}$, we are in the quantum regime and we start to see some differences in the global shape of the distribution function among the two different quantum models (see first two panels of Fig.~\ref{fig:numResults:feB}d and~\ref{figfeBgaussian}d), 
in particular during the early stage of interaction.  
This is expected as, for $\chi \sim 1$, the higher order moment ($n\ge 3$) contributions are not negligible anymore (see Sec.~\ref{sec:stoch:FPValidity})
and leads to different predictions whether one considers the FP or linear Boltzmann approach (see Sec.~\ref{sec:averages}), as clearly seen
in the last panel of Fig.~\ref{fig:numResults:feB}d.
This particular case also clearly lies outside of the deterministic and FP domains of validity (as shown in Fig.~\ref{fig:validity_mu3}a and discussed in Sec.~\ref{sec:averages:domainval}).
Let us note however that the prediction of the average electron energy (fourth panel of Fig.~\ref{fig:numResults:feB}d) and
energy dispersion (fifth panel of Fig.~\ref{fig:numResults:feB}d) is consistent in between all three approaches.

In addition, the clear difference between the FP and MC description is in the third order moment.
Figure~\ref{figfeBheating} shows the evolution of the distribution function in the FP and MC models focusing on the 
initial stage of interaction (heating phase, corresponding to an increase  of the variance) at three different times 
$t = 0$, $t = t_{\rm heating}/4$ and $t = t_{\rm heating}/2$.  
Let us first note that the FP simulation, in contrast with the MC one, exhibits a non negligible amount of particles gaining energy.
This unphysical behavior follows from what we earlier introduced as particle up-scattering (see Sec.~\ref{sec:algo:fp}).
As $\chi \rightarrow 1$, the contribution of the diffusion term becomes of the same order of the drift term, and clearly the FP model reaches his limit. 

Furthermore, the energy spreading in the MC simulation is, for such large values of $\chi$, strongly asymmetric (see also Fig.~\ref{figfeBgaussian}d). 
As the variance increases, the moment of order 3 (that becomes of the same order than the variance as $\chi \rightarrow 1$, see Fig.~\ref{figfeBgaussian}d last two panels) is negative.
This corresponds to a tail towards the low energies, with the distribution still being peaked at high-energy (similar to the time $t=0$ peak).
As the variance reaches its threshold value [still correctly predicted by Eq.~\eqref{eq:sigmaThreshold}, see also Fig.~\ref{fig:averages:sigmamax} (blue crosses)],
we reach $t=t_{\rm heating}$ and our simulation shows that the sign of the third order moment changes at this time. 
This is coherent with the fact that $\hat{\sigma}_{\gamma}^{\rm thr} \simeq \hat{\sigma}_{{\gamma}_0}^{\rm lim}$ as explained in more detail
in Section~\ref{sec:averages:mu3Discussion}.
 
This function peaked at high (close to initial) energy can be interpreted as the result of the {\it quenching} of radiation losses~\cite{harvey2017}. 
This quantum quenching, which is not accounted for in the deterministic (even 
quantum corrected) and FP approaches, follows from the discrete nature of photon emission. 
As a result, in the regime where {\it quenching} is important, each electron trajectory can be modelled only considering
the discrete nature of the emission process, i.e. it requires the use of the full Monte-Carlo approach.

Yet, when one follows the mean energy only, all three descriptions provide similar results.
That is, even in this regime of quantum quenching, the mean energy of the overall electron population is reduced and 
still closely follows deterministic radiation reaction (quantum corrected) and FP predictions. 
This has consequences on future experiments, where only a careful measurement of the electron
energy spectra (and in particular their symmetry) will allow to observe this quenching process.
We will get back to this particular point in more detail at the end of next Sec.~\ref{sec:numResultsPW}.

\subsection{Linearly polarized plane-wave}\label{sec:numResultsPW}

We now consider the interaction of this same Gaussian electron beam (mean energy $\gamma_0 =1800$ and 
standard deviation $\sigma_0 = 90$) with counter-propagating (linearly polarized) electromagnetic plane-waves with different 
amplitudes with $\chi_0 = \langle\chi\rangle(t=0) = 10^{-2}$, $10^{-1}$ and $1$ (corresponding
to the wave normalized vector potentials $a_0 = 1.14$,  $11.4$ and $114$, respectively). 
The duration of each simulation is chosen so that we get the interesting features of the interaction. 
For $\chi_0 = 10^{-2}$, the duration of the simulation is $t_{\rm end} = 2000/\omega_0$, for $\chi_0 = 10^{-1}$  $t_{\rm end} = 200/\omega_0$ 
and for $\chi_0 = 1$ $t_{\rm end} = 40/\omega_0$, where $\omega_0$ is the electromagnetic wave angular frequency ($\omega_0=2\pi c/\lambda_0$,
where we have considered $\lambda_0 = 1~{\rm \mu m}$).
 In all cases, 10 000 test particles were used.
  
\begin{figure*}
\begin{center}
\includegraphics[width=16cm]{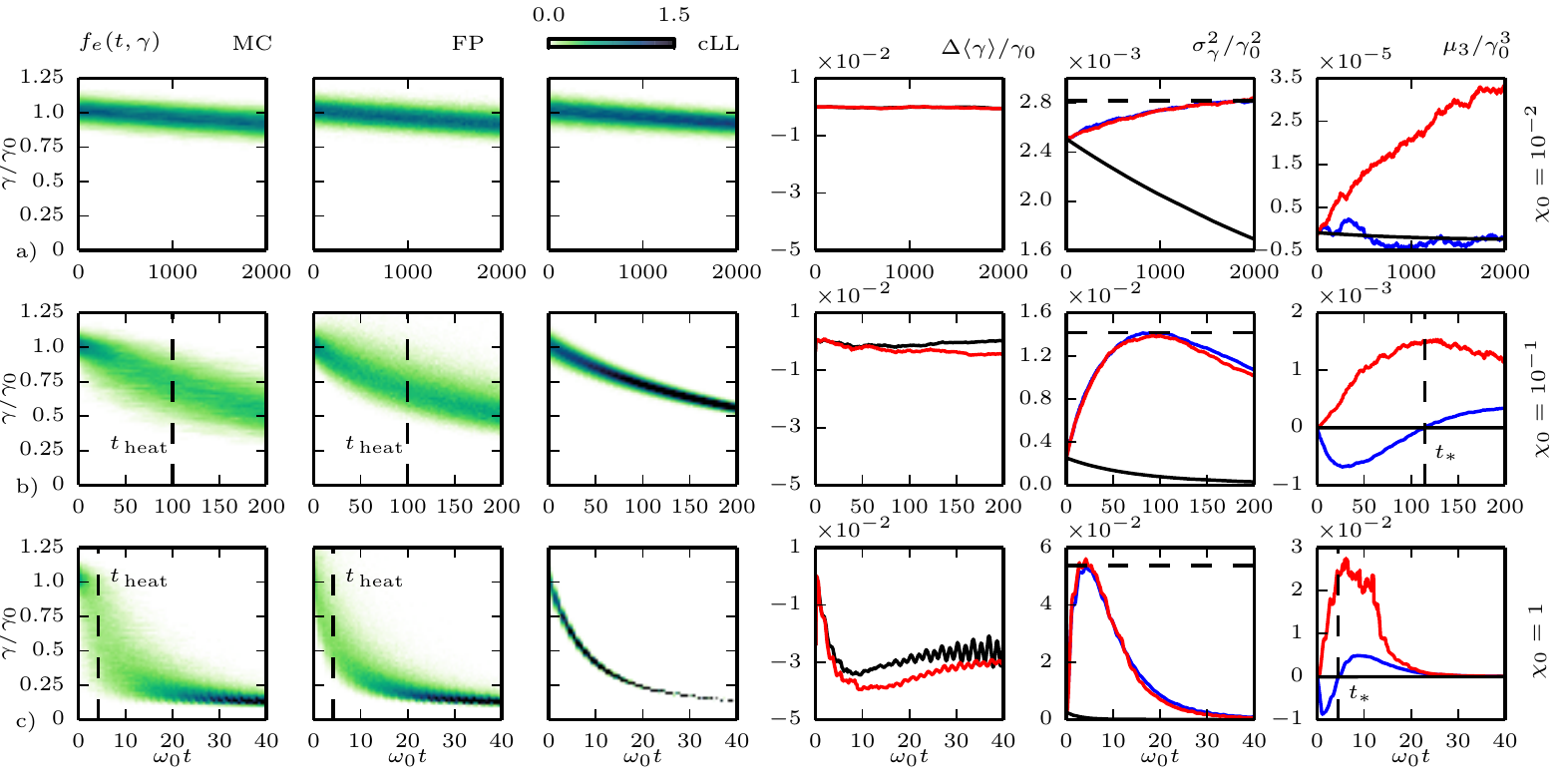}
\caption{Simulations of an ultra-relativistic electron beam in a counter-propagating linearly polarized plane wave for
a) $\chi_0 = 10^{-2}$, b) $\chi_0 =10^{-1}$, and c) $\chi_0 =1$.
The first three panels of each row show the electron distribution functions from the Monte-Carlo simulations (MC, first panels),
stochastic (Fokker-Planck) simulations (FP, second panels) and quantum-corrected deterministic simulations (cLL, third panels).
The fourth panel shows the difference in the prediction of the mean electron energy in between the MC simulation and
 the deterministic (black line) and FP (red line) simulations. 
  The two last panels (in each row) correspond to the moments of order 2 (energy variance) and 3 for the MC (blue line), 
  FP (red line) and deterministic (black line) simulations.} \label{figfePW}
\end{center}
\end{figure*}

 The simulation results are summarized in Fig.~\ref{figfePW}, 
following the same presentation than Fig.~\ref{fig:numResults:feB}.
 The first row a) corresponds to $\chi_0 = 10^{-2}$, 
 the second b) to $\chi_0 = 10^{-1}$, the third c) to $\chi_0 = 1$. 
 The  interpretation of these simulations follows the same lines than for the case of a constant magnetic field and,
as a result, the same conclusions can be drawn.
 
Let us start by considering the case $\chi_{\rm cl} < \chi_0 = 10^{-2} < \chi_{\rm qu}$.
As $\chi_0 \ll 1$ the $n \le 2$ higher order moments ($\propto \chi^{n+1}$) are small,
and the overall evolution of the electron distribution function $f_e$ is well reproduced,
as shown in Fig.~\ref{figfePW}a should one focus on the first four panels in Fig.~\ref{figfePW}a only.
Looking into more details, and in particular to the fifth panel (energy variance), one sees that here again the deterministic (cLL) description strongly
overestimate the electron beam cooling. This is explained again by this particular case lies outside of the regime of applicability of the deterministic (cLL) description (Fig.~\ref{fig:validity_mu3}a),
and well into that of the FP description. 
 
 \begin{figure}
\begin{center}
\includegraphics[width=8cm]{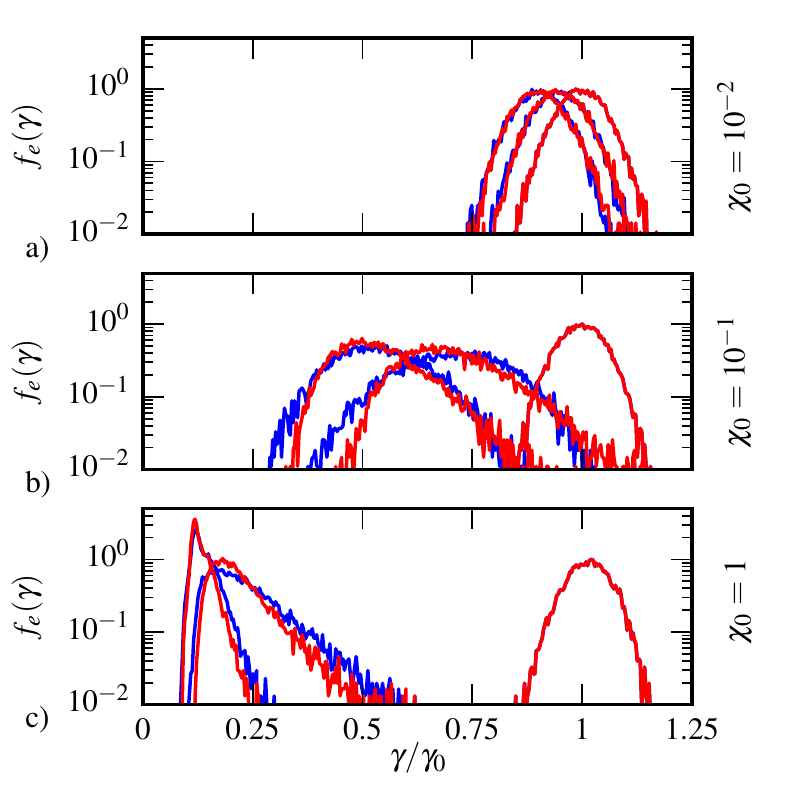}
\caption{Simulations of an ultra-relativistic electron beam in a counter-propagating linearly polarized plane wave for 
a) $\chi_0 =10^{-2}$, c) $\chi_0 =10^{-1}$ and d) $\chi_0 =1$ electron distribution functions  at  times $t = 0$, $t = t_{\rm end}/2$ and $t = t_{\rm end}$ (from right to left). 
The red lines correspond to FP simulations, the blue one to MC simulations.}\label{figfePWgaussian} 
\end{center}
\end{figure}

\begin{figure*}
\begin{center}
\includegraphics[width=16cm]{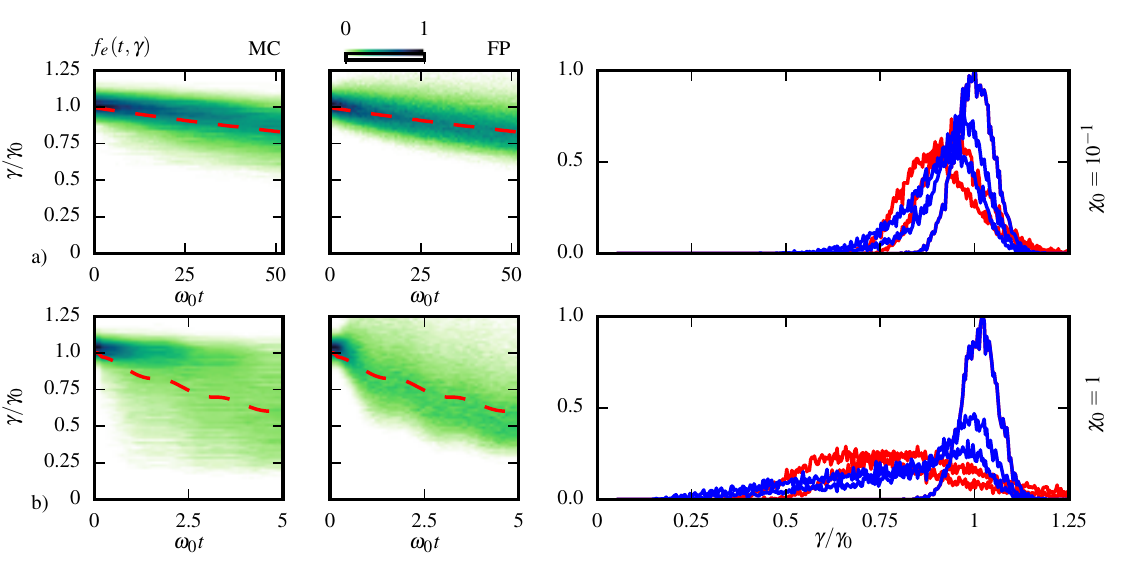}
\caption{Simulations of an ultra-relativistic electron beam in a counter-propagating linearly polarized plane wave for
a) $\chi_0 = 10^{-1}$, b) $\chi_0 =1$. This figure focuses on the early times of interaction ($t\le t_{\rm heating}$ during which 
the energy dispersion increases).
The first column corresponds to MC simulations, the second to the FP ones.
The last row shows snap-shots of the electron distribution functions at different times $t=0$, $t=t_{\rm heating}/4$ and $t=t_{\rm heating}/2$ (from right to left).
The red lines correspond to FP simulations, the blue one to MC simulations.}\label{figfePWheating}
\end{center}
\end{figure*}

Similarly, for $\chi_{\rm cl} < \chi_0 = 10^{-1} < \chi_{\rm qu}$, the stochastic (FP) scheme reproduces correctly the evolution 
of both the electron mean energy \textit{and} variance, and the global properties of the electron distribution function $f_e$ considering the two quantum approaches 
  are similar. In particular, the existence of a heating to cooling transition is recovered, and analytical
  predictions of Eq.~\eqref{eq:sigmaThreshold} are found to be in excellent agreement with our
  simulation results, see also Fig.~\ref{fig:averages:sigmamax} (green crosses).
As discussed previously for the case of an electron beam radiating a constant magnetic field, 
  and as predicted by our theoretical analysis (this particular simulation lies slightly outside of the regime of validity of the FP description, see Fig.~\ref{fig:validity_mu3}a),
  discrepancies in the third order moment appear on short time. For this case however, this error remains quite small, of the order of $\chi_0 = 10^{-1}$ times the normalized variance (second order moment).
Here again, our numerical results support the extended analysis of the domains of validity of the different approaches (Sec.~\ref{sec:averages:domainval}).
  
  Finally, when $\chi_0 = 1 > \chi_{\rm{qu}}$, important differences in the different approaches are observed, 
  in particular visible in the third momentum, and in the overall shape of the distribution function at early times as
   $\mu_3$ then assumes values of the same order than $\sigma_{\gamma}$.
  Once more, $\mu_3$ is negative during the heating phase, and quantum quenching sets in. 
Here again, this third order moment flip signs at $t_{\rm heating} \simeq 2\pi/\omega_0$, which thus provides a good
measure for the time up to which the discrete nature of photon emission has a noticeable effect on the overall
shape of the distribution function.

As previously stressed, this quenching is intimately linked to the discrete nature of photon emission,
and is here shown to greatly impact the physics at $\langle \chi \rangle \rightarrow 1$.
As it is associated to negative values of the third order moment $\mu_3$, the impact of {\it quenching} can be measured in forthcoming
experiments by analysing the skewness of the measured electron energy spectrum. 
This would provide a clear signature of the quantum (discrete) nature of radiation reaction.
Yet, unlike previously claimed in Ref.~\cite{harvey2017}, this does not translate in the electron bunch mean energy which is correctly 
described by the deterministic (radiation friction) description provided that it accounts for the quantum correction as previously underlined in Sec.~\ref{sec:averages}, 
and demonstrated in Fig.~\ref{figfePW} (fourth panels) where all three descriptions are shown to give similar results on the mean energy.

\subsection{Electron population with a broad energy dispersion}\label{sec:numResultsMJ}

For the sake of completeness, we finally consider the evolution of an electron population with 
an initially broad energy distribution radiating in a constant uniform external magnetic field. 
The electron energy distribution at the beginning of the simulation follows a (zero-drift) 3D Maxwell-J\"uttner distribution:
\begin{eqnarray}
f_e(t=0,\gamma) = \frac{\gamma \sqrt{\gamma^2 -1}}{\theta K_2(1/\theta)}\exp\left(-\frac{\gamma}{\theta} \right) \, ,
\end{eqnarray}
where $K_2$ is the modified Bessel function of second kind, and $\theta = T/(mc^2) = 600$ is the normalized temperature
corresponding to an initial electron mean Lorentz factor $\gamma_0 = \langle \gamma\rangle \simeq 1800$, 
and initial energy standard deviation $\sigma_{\gamma} \simeq 0.57~\gamma_0$.
Note also that such a broad distribution also presents a large initial asymmetry and $\hat{\mu}_3 \simeq 0.2$ at the beginning of the simulation.
Three magnetic field strengths have been considered corresponding to $\chi_0=\langle \chi \rangle(t=0) = 10^{-2}$, $\chi_0 = 10^{-1}$ and $\chi = 1$.

\begin{figure*}
\begin{center}
\includegraphics[width=16cm]{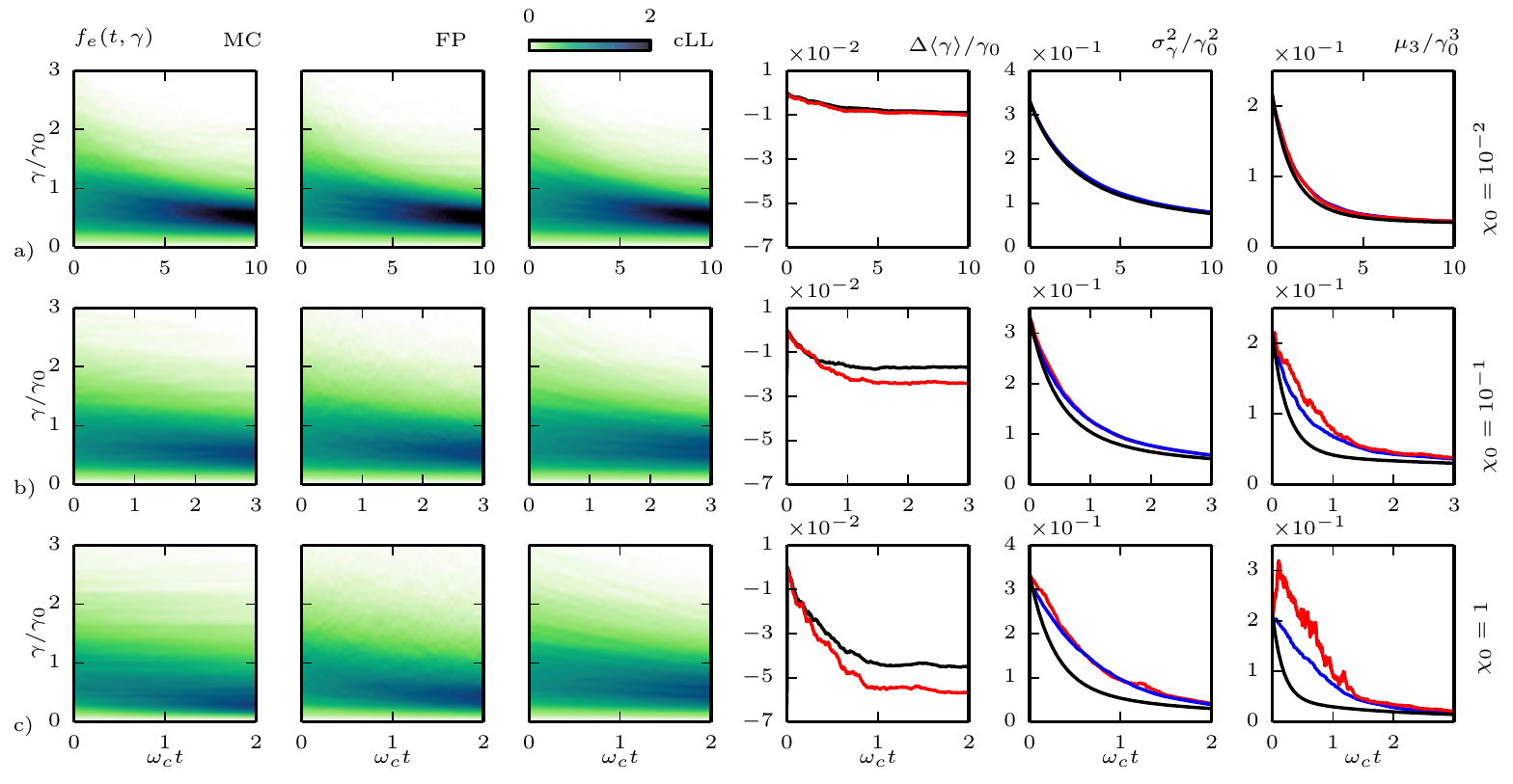}
\caption{Simulations of an electron bunch with initially broad (Maxwell-J\"uttner) energy distribution  in a constant, uniform magnetic field for 
a) $\chi_0 = 10^{-3}$, b) $\chi_0 =10^{-2}$, c) $\chi_0 =10^{-1}$ and d) $\chi_0 =1$. 
The first three panels of each row shows the electron distribution functions from the Monte-Carlo simulations (MC, first panels),
stochastic (Fokker-Planck) simulations (FP, second panels) and quantum-corrected deterministic simulations (cLL, third panels).
The fourth panels show the difference in the prediction of the mean electron energy in between the MC simulation and
 the deterministic (black line) and FP (red line) simulations. 
  The two last panels (in each row) correspond to the moments of order 2 (energy variance) and 3 for the MC (blue line), 
  FP (red line) and deterministic (black line) simulations.} \label{figfeBMJ}
\end{center}
\end{figure*}

The simulation results are summarized  in Fig.~\ref{figfeBMJ} following the same presentation than Figs.~\ref{fig:numResults:feB} and~\ref{figfePW}.

Let us first briefly focus on the case $\chi_0=10^{-2}$. 
This particular case is indeed particularly interesting as for such $\chi_0$,
both our extended analysis of the domain of validity (Sec.~\ref{sec:averages:domainval}) and previous simulations considering an initially narrow and symmetric electron beam showed that the on-set of stochasticity effects translate in an overestimated cooling by the deterministic (cLL) description.
In contrast, for this particular case of a broad energy distribution, and as predicted in Fig.~\ref{fig:validity_mu3}b, this particular case $\chi_0=10^{-2}$ now
lies exactly into the regime of the deterministic (cLL) approach. This is indeed confirmed by our simulations results (Fig.~\ref{figfeBMJ}a) where all three models predict the same evolution for all three moments. This finding further confirms our analysis that not only the average quantum parameter, but 
also the shape of the energy distribution of the electron population is important to assess the relative importance, and measurable aspect, of various effects of radiation reaction.

Furthermore, and for all values of $\chi_0$, the very large initial standard deviation, $\sigma_{\gamma}/\gamma_0 \simeq 0.57$, exceeds  
the predicted threshold $\sigma_{\gamma}^{\rm thr}/\gamma_0 \simeq 0.37$ [here computed for $\chi=1$ using Eq.~\eqref{eq:sigmaThreshold}].
As a result, and even at large $\chi$, all simulations (and in particular all quantum ones) predict only a cooling of the electron distribution
with $\sigma_{\gamma}$ continuously decreasing with time. 

Another remarkable point is that the FP description reproduces the MC results for $\chi_0=10^{-1}$ (and to some extent for $\chi_0 = 1$)
much more closely than for the initially narrow electron beam. Again, this is in good agreement with our extended analysis of the domain of valid of the
different descriptions of radiation reaction and is explained by the initially broad energy dispersion (variance) so that the deterministic terms in the equation of evolution of the moments are dominant, while the quantum terms only give a correction.  

Let us finally note that, beyond demonstrating the validity our approach and its capacity at being applied to general electron populations,
this particular case of a Maxwell-J\"uttner distribution is also interesting to broaden the spectrum of applications of our study.
Indeed, such relativistic distribution functions are typical of laser-solid interaction in the ultra-high intensity regime, as well as in relativistic astrophysics.

\section{Conclusion}\label{sec:conclusion}

The radiation reaction force acting on particles in strong electromagnetic fields has recently attracted increased attention 
as it may affect laser-plasma interaction under extreme light conditions, and for its impact on the particle dynamics in extreme astrophysical scenarios.
It is usually described using either a classical (potentially quantum corrected) friction force
or a full Monte-Carlo (MC) treatment, the latter allowing to account for the quantum and discrete nature of the photon emission. 

In the first part of the paper, we have revisited the basis of the classical treatment of radiation reaction.
The Landau and Lifshitz (LL) force was rewritten in the simple and intuitive form of a friction force for radiating ultra-relativistic particles.
Like the full LL equation, this reduced friction force also conserves the on-shell condition.
Its correction to account for the quantum reduction of the power radiated away by the emitting particle was then introduced heuristically
(as previously suggested in other works).\\

After briefly presenting the properties of high-energy photon emission as inferred from quantum electrodynamics (QED)  
in the non-linear moderately quantum regime, 
we then focused on a statistical description of photon emission and its back-reaction considering a population of ultra-relativistic electrons.
Starting from a linear Boltzmann (lB) equation with a {\it collision} operator describing incoherent photon emission in a quantum description, 
we performed a Fokker-Planck (FP) expansion of the {\it collision} operator in the parameter $\gamma_\gamma/\gamma$, 
the limit of which were discussed in details.
The resulting FP description, derived here for the first time under arbitrary particle and electromagnetic field configurations, is interesting for several reasons.

(i)~First, it takes the simple form of a partial differential equation, more easily handled in theoretical models than the lB description 
that relies on an integro-differential equation.

(ii)~The derived FP description is equivalent to a stochastic differential equation (SDE) for the electron momentum. 
In addition to the standard Lorentz force, this equation contains a deterministic drift term which is found to be leading term of the LL friction force 
with the quantum correction discussed above, hence justifying the heuristic treatment that consists in systematically correcting the LL friction force to account for the quantum reduction of the power-radiated away by the ultra-relativistic electron.
Note that throughout this work, this quantum-corrected LL friction force provides the deterministic treatment of radiation reaction.
Most importantly, this kinetic treatment also fully justifies - for the first time - the use of the quantum-corrected LL friction force to account for radiation reaction in Particle-In-Cell (PIC) codes.

(iii)~An additional diffusion term in the SDE is also derived for any $\chi \lesssim 1$.
It accounts for the stochastic nature of photon emission inherent to its quantum nature. 
The numerical implementation of the SDE accounting for this diffusion term is discussed, 
leading to the development of a new stochastic pusher for PIC codes that can be easily implemented by modifying the pusher accounting for the
(quantum-corrected) radiation-reaction force.
Its ability to correctly address various physical configurations is demonstrated, and it can be used in an intermediate regime where the lowest order term (deterministic friction force)  is not accurate enough, and the full MC procedure (equivalent to the lB description) is not yet necessary. 

(iv)~The formalism developed in this paper also allows to identify the limit of validity of the deterministic approach 
and to propose a first criterion, in terms of the quantum parameter $\chi$, for the transition between FP and lB/MC treatments,  
namely $\chi < \chi_{\rm qu} \simeq 0.25$. 
This criterium is not strict and has been extended, following considerations on the evolution successive moments of the distribution function,
to account for the general shape of the electron distribution function, as discussed at the end of this Section.
\\

Indeed, special attention was paid to the study of the successive moments of the electron distribution function.
Using both analytical and numerical modelling,  we evidence that the equation of evolution for the average energy of the particle ensemble is formally the same in all three models  (quantum-corrected LL/deterministic, FP, lB/MC). 
An estimate on the discrepancy on the mean electron energy predicted by the different models is derived, and it is shown to be small for most cases: 
that is all three descriptions lead to the same prediction for the electron mean energy.
This has serious implications for experiments, as the mean electron energy proves not to be, in general, a relevant measure to assess the importance of quantum effects beyond the simple reduction of the power radiated away by the particles.

The situation is different when considering the evolution of the higher order moments of the particles' energy distribution. 
In particular, only the FP and lB/MC descriptions provide the correct equation of evolution for the variance (energy dispersion).
Instead the third and higher order moment equations of evolution differ from one approach to another.

The study of the equation of evolution of the energy variance allows us to define an energy spreading ({\it heating}) time,  
when the variance of an initially narrow beam reaches a maximum value. 
The existence of this maximum energy spread follows from the competition between the deterministic part of the radiation reaction force and the stochastic nature of the quantum process of high-energy photon emission.
The former acts as a friction term and results in a cooling of the electron population, while the former leads to a natural spreading of the electron energy distribution function, associated to an increased effective temperature of the electron population.
This spreading can set in the moderately quantum regime of radiation reaction ($\chi \lesssim 1$), and is shown to be correctly modelled by both the FP and lB/MC treatments.
This has also interesting implications for experiments, as the stochastic nature of high-energy photons on the electron dynamics can then be diagnosed by a careful analysis of the resulting spread of the particle energy distribution.

The study of the evolution of the third order moment proved also to be particularly interesting as it reveals the discrepancy between all three models.
In particular, this study allows us to link negative values of the third moment to the so-called quenching of radiation losses.
This process is intimately linked to the discrete nature of high-energy photon emission and can be accounted for only using the lB/MC procedure.
We show that this quenching process can be more easily observed, and impact the electron distribution, when the electron initial distribution is quite narrow.
This has important implications for future experiments on radiation reaction, as it is here shown that "quenching", a purely quantum effect, can be diagnosed as a negative skewness of the resulting electron distribution function.
Let us further note that, unlike previously claimed, even though quenching affects the overall shape of the electron distribution function, it does not impact the mean electron energy that is correctly modelled using the deterministic radiation reaction force, provided it is corrected to account for the quantum reduction on the radiated power.

Finally, the analytical study of the successive moments of the electron energy distribution function allowed us a deeper insight into 
the regimes of validity of the various approaches considering more general distribution functions than usually investigated.
While previous studies have considered the average quantum parameter $\langle \chi \rangle$  as the only relevant parameter, we show that the initial electron spread in energy is also a key-parameter for determining the validity of the different (deterministic, FP or lB/MC) models. 
Our moment analysis thus allowed us to identify more rigorously the various regimes of validity of the different models as a function of both, 
the initial quantum parameter $\langle \chi \rangle$ and normalized electron energy spread $\sigma_{\gamma}/\langle\gamma\rangle$.
By doing so, we could also highlight under which conditions the various physical effects, and in particular the stochastic and/or discrete nature of high-energy photon emission may be more easily accessible in forthcoming experiments.

The theoretical findings of this work as well as the numerical tools developed here are important for future experiments on extreme light facilities.
They are also important to the relativistic astrophysics community as radiation reaction is known to strongly affect the particle dynamics in various extreme scenarios.

\section*{Acknowledgements}
The authors thank A. Grassi for useful discussions and help on the development of numerical tools. 
We also thank R. Capdessus, A. Di Piazza, T. Grismayer, M.~Lobet, P.~Mora, M. Tamburini, V.~T.~Tikhonchuk and M.~Vranic for fruitful discussions.
Financial support from PALM (ANR-10-LABX-0039-PALM, Junior Chair SimPLE) and 
Plas@Par (ANR-11-IDEX-0004-02-Plas@Par) is acknowledged.

\appendix

\section{Exact and approximate expression of $a_n(\chi)$ functions}\label{AppendixFunctions}

In what follows, we rewrite (in a single integral form) the $a_n(\chi)$ that appear when considering the successive moments of the kernel $w_{\chi}(\gamma,\gamma_{\gamma})$:

\begin{eqnarray}
\nonumber a_n(\chi) &=& \frac{\sqrt{3}}{2\pi} 3^{n+1} \chi^{n+1} \int_0^{+\infty} \!\!\! d\nu \left[ \frac{9\chi^2 \nu^{n+2}}{(2+3\nu \chi)^{n+3}}K_{2/3}(\nu) \right. \\
&+& \left. \frac{\nu^{n+1}}{(n+1)(2+3\nu\chi)^{n+1}}K_{5/3}(\nu) \right]\, .
\end{eqnarray}
For the reader's convenience, we also give their expansion in the limit $\chi \ll 1$:
\begin{eqnarray}
\nonumber a_n(\chi) \sim  \frac{\sqrt{3}}{4 \pi (n+1)} 3^{n+1}\,\Gamma\!\left(\frac{n}{2} + \frac{1}{6}\right)\,\Gamma\!\left(\frac{n}{2}+\frac{11}{6}\right) \, \chi^{n+1} \, ,
\end{eqnarray}
with $\Gamma$ the Gamma function.

\section{Conservation of the number of electrons and total energy, and photon production rate}\label{app:Conservation}

As shown in Sec.~\ref{sec:averages:CollOpMoments}, the integral over all possible electron energies of the {\it collision} operators for all three description is identically zero. 
It thus turns out that, for all three descriptions, the photon emission process does not impact the electron density distribution 
in the (${\bf x},{\bf\Omega}$) phase-space. 
As a result, all three descriptions conserve the total number of electrons ${N_e=\int\!d^3x\,d^2\Omega\,d\gamma f_e}$.\\

The situation is obviously different when looking at the total number of photons. Indeed, integrating the source term, rhs of Eq.~\eqref{eq:Master2},
over all possible photon energies and directions provides us with the local photon production rate:
\begin{eqnarray}
\label{eq:localProdRate}W_{\rm loc}(t,{\bf x}) = \int\!\!d^2\Omega\,d\gamma_{\gamma}\,\mathcal{S}[f_e]
=n_e(t,{\bf x})\big\langle W(\chi,\gamma)\big\rangle_{\alpha}\,,
\end{eqnarray}
where $n_e(t,{\bf x})$ is the electron density, and $W(\chi,\gamma) = a_0(\chi)/\gamma$ is the rate of photon emission 
[Eq.~\eqref{eq:PhotonRate}] introduced in Sec.~\ref{sec:stoch:PoissonProcess}.
As a result, the total number of photons $N_{\gamma}(t)=\int\!d^3x\,d^2\Omega_{\gamma}\,d\gamma_{\gamma} f_{\gamma}$ increases with time as:
\begin{eqnarray}
\frac{d}{dt}N_{\gamma} = \int\!d^3x\,W_{\rm loc}(t,{\bf x})\,,
\end{eqnarray}
the rhs of the previous equation denoting the total photon production rate.\\

Finally, the equations of evolution for the total energy $U_{e,\gamma}(t) = mc^2\!\!\int\!d^3x\,d^2\Omega\,d\gamma\,\gamma f_{e,\gamma}$ of the electron and photon population reads:
\begin{eqnarray}
\label{eq:TotalEnergy}\frac{d}{dt}U_{e} &=& \int\!\!d^3x\,{\bf J}_{e} \cdot {\bf E} - P_{\rm rad}^{\rm tot}(t)\,,\\
\frac{d}{dt}U_{\gamma} &=& P_{\rm rad}^{\rm tot}(t)\,,
\end{eqnarray}
where:
\begin{eqnarray}
P_{\rm rad}^{\rm tot}(t) = mc^2\!\!\!\int\!\!d^3\!x\,n_e(t,{\bf x})\,\langle S(\chi)\rangle_{\alpha},
\end{eqnarray}
is the total power radiated away by the electrons, and ${\bf J}_e(t,{\bf x}) = -e\,n_e(t,{\bf x}){\bf V}_e(t,{\bf x}$) is the electron current density 
[with ${\bf V}_e(t,{\bf x}) = c\,\langle u\,{\bf\Omega}\rangle_{\alpha}$ the electron mean velocity].
This confirms that all three descriptions conserve the total energy in the system, the only source of energy being the work of the external electromagnetic field.

\section{Monte-Carlo module}\label{AppendixMC}
Our implementation closely follows that presented in Refs.~\cite{duclous2011,arber2015,lobet2016}.
To treat the discontinuous process of high-energy photon emission each electron is first assigned a final optical depth $\tau_f$
sampled from $\tau_f = -\ln(r)$ with $0 < r \le 1$ a uniform random number. At the same time, a {\it current} optical depth $\tau_c$
is assigned to each electron, which is initialized at $0$ and updated in time (possibly using a sub-cycling with respect to the main, Lorentz,
loop at $\Delta t$) following:
\begin{eqnarray}
\frac{d\tau_c}{dt} = \int_0^{\chi} \frac{d^2N}{d\chi_{\gamma} dt} d\chi_{\gamma}\,.
\end{eqnarray}
When $\tau_c$ reaches the final optical depth $\tau_f$, the electron emits a photon. 
The emitted photon quantum parameter is computed inverting the cumulative distribution function:
\begin{eqnarray}\label{xi_sol}
{\rm CDF}(\chi_{\gamma}) = \frac{\int_0^{\chi_{\gamma}}G(\chi,\chi_{\gamma}')/\chi_{\gamma}'\,d\chi_{\gamma}'}{\int_0^{\chi}G(\chi,\chi_{\gamma}')/\chi_{\gamma}'\,d\chi_{\gamma}'}\, ,
\end{eqnarray}
with $\chi$ the electron quantum parameter at the time of emission.
From Eq.~\eqref{eq:linkChi}, this uniquely defines the energy of the emitted photon $\varepsilon_{\gamma}=mc^2\,\gamma\chi_{\gamma}/\chi$ 
(with $\gamma$ the energy of the radiating electron), and the electron momentum right after emission ${\bf p}^+$ is then updated considering 
forward emission $\bf\Omega={\bf p}/{\vert{\bf p}\vert}$:
\begin{eqnarray}
{\bf p}^+ = {\bf p} - \frac{\varepsilon_{\gamma}}{c}{\bf\Omega}\,.
\end{eqnarray}
Note that this implementation, which conserves momentum, does not exactly conserve energy.
The error made on the energy is however small for ultra-relativistic electrons ($\gamma \ll 1$)~\cite{lobetPhD}.

Finally, we note that, in between emission events, the electron dynamics governed by the Lorentz force is updated as in the previous pushers using the Boris approach.

\section{Approximated equations of evolution of the successive moments}\label{app:ApproximatedEqEvolutionMoments}

The equations of evolution of the successive moments of the distribution function, Eqs.~\eqref{eq:eonMeanEnergy} to~\eqref{eq:mun}, are exact. 
Yet, to evaluate the rhs of these equations, one needs to consider a particular distribution function. 
This would in general require to solve the full linear Boltzmann Eq.~\eqref{eq:Master1}.
Throughout this work, we have often adopted the approach that consists in developing the various functions of $\chi$ around the average value $\langle \chi \rangle$. 
In what follows, we list the resulting (approximated) equations of evolution [introducing $\bar{\tau} = 3\tau_e/(2\alpha^2)$].

\begin{eqnarray}
d_t \langle \gamma \rangle_{\alpha} \simeq -S(\langle \chi \rangle_{\alpha}) - \frac{1}{2} \hat{\sigma}_{\gamma}^2 \, \langle \chi \rangle_{\alpha}^2 S''(\langle \chi \rangle_{\alpha}) \, ,
\end{eqnarray}

\begin{eqnarray}
\left. d_t \sigma_{\gamma}^2 \right\vert_{\rm cLL}
 \simeq - 2 \sigma_{\gamma}^2 \frac{\langle \chi \rangle} {\langle \gamma \rangle} S'(\langle \chi \rangle) \, ,
\end{eqnarray}

\begin{eqnarray}
\bar{\tau} \left. d_t \sigma_{\gamma}^2 \right\vert_{\rm \alpha} &\simeq& \bar{\tau}  \left. d_t \sigma_{\gamma}^2 \right\vert_{\rm cLL} \\
\nonumber &+& \langle \gamma \rangle h(\langle \chi \rangle) + \sigma_{\gamma}^2\frac{\langle \chi \rangle}{\langle \gamma \rangle} h'(\langle \chi \rangle) \, ,
\end{eqnarray}

\begin{eqnarray}
\left. d_t \mu_3 \right\vert_{\rm cLL}
 \simeq - 3 \mu_3 \frac{\langle \chi \rangle} {\langle \gamma \rangle} S'(\langle \chi \rangle) \, ,
\end{eqnarray}

\begin{eqnarray}
\nonumber \bar{\tau}  \left. d_t \mu_3 \right\vert_{\rm FP} &\simeq& 
\bar{\tau} \left. d_t \mu_3 \right\vert_{\rm cLL} + 3 \sigma_{\gamma}^2 [ h(\langle \chi \rangle) \\
\!\!\!\! &+& \langle \chi \rangle h'(\langle \chi \rangle) ] + 3\mu_3 \frac{\langle \chi \rangle}{\langle \gamma \rangle} h'(\langle \chi \rangle) \, ,
\end{eqnarray}

\begin{eqnarray}
\nonumber \bar{\tau}  \left. d_t \mu_3 \right\vert_{\rm MC} &\simeq& 
\bar{\tau}  \left. d_t \mu_3 \right\vert_{\rm FP} - \langle \gamma \rangle^2 \, a_3(\langle \chi \rangle) \\
\nonumber &-& \sigma_{\gamma}^2 [a_3(\langle \chi \rangle) + 2 \, \langle \chi \rangle \,  a_3'(\langle \chi \rangle)] \\
&-& \mu_3 \frac{\langle \chi \rangle}{\langle \gamma \rangle} a_3'(\langle \chi \rangle) \, .
\end{eqnarray}

\begin{eqnarray}
d_t \mu_{n} \Big\vert_{\mbox{\tiny cLL}} \simeq  -n \mu_n \frac{\langle \chi \rangle}{\langle \gamma \rangle} S(\langle \chi \rangle) \, ,
\end{eqnarray}

\begin{eqnarray}
\nonumber \bar{\tau} d_t \mu_{n} \Big\vert_{\mbox{\tiny FP}} &\simeq& \bar{\tau} d_t \mu_n \Big\vert_{\mbox{\tiny cLL}} + \frac{n(n-1)}{2}\mu_{n-2} \langle \gamma \rangle h(\langle \chi \rangle) \\
\nonumber &+& 
\frac{n(n-1)}{2} \mu_{n-1} [h(\langle \chi \rangle) + \langle \chi \rangle h'(\langle \chi \rangle)] \\
&+& \frac{n(n-1)}{2} \mu_n \frac{\langle \chi \rangle}{\langle \gamma \rangle} h'(\langle \chi \rangle) \, ,
\end{eqnarray}

\begin{eqnarray}\label{eq:approx_mun}
\bar{\tau}\nonumber d_t \mu_{n} \Big\vert_{\mbox{\tiny MC}} &\simeq& \sum_{k = 0}^{n-1} \sum_{l = 0}^{n-k} (-1)^{n-k} \dbinom{n}{k} \dbinom{n-k}{l} \langle \gamma \rangle^{n-k-l} \\ 
&\Big[& \mu_{k+l}\,  a_n(\langle \chi \rangle) + \mu_{k+l+1} \frac{\langle \chi \rangle}{\langle \gamma \rangle} \,  a_n'(\langle \chi \rangle) \Big] \, .
\end{eqnarray}



\begin{thebibliography}{10}
\bibitem{footnote1}
Note that for the first two terms in Eq.~\eqref{eq_LL} to cancel upon
  integration requires to rewrite $dp^{\mu}\!/d\tau$ using the Lorentz force
  [Eq.~\eqref{eq_Lorentz}] only. This requires the radiation force to be
  smaller than the Lorentz force in the electron proper-frame (not in all
  reference frames), which is exactly the condition upon which the LL force is
  derived, see Ref~\cite{Landau_CED}.
\bibitem{footnote3}
As discussed in Ref.~\cite{Di_Piazza_RMP}, pair production as well as
  higher-order coherent QED processes (such as coherent multiple photon
  emission) are neglected.
\bibitem{footnote5}
Some pathological situations can be found where it is actually positive, their
  physical significance is however questionable and these scenarios will not be
  considered here.
\bibitem{footnote6}
There is of course a degree of arbitrariness here, and the domain of validity
  can be adjusted depending on the required degree of precision.
\end{thebibliography}
\end{document}